\documentclass{article}

\usepackage{arxiv}

\usepackage[utf8]{inputenc} 
\usepackage[T1]{fontenc}    
\usepackage{hyperref}       
\usepackage{url}            
\usepackage{booktabs}       
\usepackage{amsfonts}       
\usepackage{nicefrac}       
\usepackage{microtype}      
\usepackage{graphicx}
\usepackage{natbib}
\usepackage{doi}

\usepackage{graphicx}
\usepackage{tikz}
\usetikzlibrary{shapes,positioning}
\usepackage{colortbl}
\usepackage{multicol}
\usepackage{epstopdf}
\usepackage{caption}
\usepackage{siunitx}
\usepackage{mathtools}

\makeatletter
\newcommand\resetstackedplots{
\makeatletter
\pgfplots@stacked@isfirstplottrue
\makeatother
\addplot [forget plot,draw=none] coordinates{(1,0)(2,0)(3,0)(4,0)(5,0)};
}
\makeatother

\usepackage{amsmath}
\usepackage[colorinlistoftodos]{todonotes}
\usepackage{bm}
\usepackage{multirow}
\definecolor{dark_magenta}{HTML}{8B008B}
\definecolor{forest_green}{HTML}{228B22}
\definecolor{dark_gray}{HTML}{373737}
\definecolor{aquamarine}{HTML}{13B3AC}
\definecolor{light_gray}{HTML}{BEBEBE}
\definecolor{lime}{HTML}{7FFF00}

\usepackage[rightcaption]{sidecap}
\sidecaptionvpos{figure}{c}

\usepackage{pgfplots}
\usetikzlibrary{patterns}

\makeatletter
\tikzset{
        hatch distance/.store in=\hatchdistance,
        hatch distance=5pt,
        hatch thickness/.store in=\hatchthickness,
        hatch thickness=5pt
        }
\pgfdeclarepatternformonly[\hatchdistance,\hatchthickness]{north east hatch}
    {\pgfqpoint{-1pt}{-1pt}}
    {\pgfqpoint{\hatchdistance}{\hatchdistance}}
    {\pgfpoint{\hatchdistance-1pt}{\hatchdistance-1pt}}%
    {
        \pgfsetcolor{\tikz@pattern@color}
        \pgfsetlinewidth{\hatchthickness}
        \pgfpathmoveto{\pgfqpoint{0pt}{0pt}}
        \pgfpathlineto{\pgfqpoint{\hatchdistance}{\hatchdistance}}
        \pgfusepath{stroke}
    }
\pgfdeclarepatternformonly[\hatchdistance,\hatchthickness]{north west hatch}
	{\pgfqpoint{0pt}{0pt}}
	{\pgfqpoint{\hatchdistance}{\hatchdistance}}
	{\pgfpoint{\hatchdistance-1pt}{\hatchdistance-1pt}}%
	{
    		\pgfsetcolor{\tikz@pattern@color}
  	  	\pgfsetlinewidth{\hatchthickness}
    		\pgfpathmoveto{\pgfqpoint{\hatchdistance}{0pt}}
    		\pgfpathlineto{\pgfqpoint{0pt}{\hatchdistance}}
    		\pgfusepath{stroke}
}
\pgfdeclarepatternformonly[\hatchdistance,\hatchthickness]{horizontal hatch}
	{\pgfqpoint{-2pt}{-2pt}}
	{\pgfqpoint{\hatchdistance}{\hatchdistance}}
	{\pgfpoint{\hatchdistance-1pt}{\hatchdistance-1pt}}%
	{
    		\pgfsetcolor{\tikz@pattern@color}
  	  	\pgfsetlinewidth{2*\hatchthickness}
    		\pgfpathmoveto{\pgfqpoint{\hatchdistance}{\hatchdistance}}
    		\pgfpathlineto{\pgfqpoint{0pt}{\hatchdistance}}
    		\pgfusepath{stroke}
}
\pgfdeclarepatternformonly[\hatchdistance,\hatchthickness]{vertical hatch}
	{\pgfqpoint{-2pt}{-2pt}}
	{\pgfqpoint{\hatchdistance}{\hatchdistance}}
	{\pgfpoint{\hatchdistance-1pt}{\hatchdistance-1pt}}%
	{
    		\pgfsetcolor{\tikz@pattern@color}
  	  	\pgfsetlinewidth{2*\hatchthickness}
    		\pgfpathmoveto{\pgfqpoint{\hatchdistance}{\hatchdistance}}
    		\pgfpathlineto{\pgfqpoint{\hatchdistance}{0pt}}
    		\pgfusepath{stroke}
}
\makeatother

\title{p-adaptive discontinuous Galerkin method for the shallow water equations on heterogeneous computing architectures}

\date{November 2023}

\newif\ifuniqueAffiliation

\ifuniqueAffiliation 
}
\else
\usepackage{authblk}

\author[1,2]{%
	Sara Faghih-Naini 
}
\author[1]{%
	Vadym Aizinger\thanks{\texttt{vadym.aizinger@uni-bayreuth.de}}%
}
\author[3,2]{%
	Sebastian Kuckuk
}
\author[2]{%
	Richard Angersbach
}
\author[3,2]{%
	Harald K{\"o}stler
}
\affil[1]{Chair of Scientific Computing, University of Bayreuth,              95440 Bayreuth, Germany}
\affil[2]{Chair of Computer Science 10, Friedrich-Alexander Universit{\"a}t Erlangen-N{\"u}rnberg, 91058 Erlangen, Germany}
\affil[2]{Erlangen National High Performance Computing Center (NHR@FAU), Friedrich-Alexander Universit{\"a}t Erlangen-N{\"u}rnberg, 91058 Erlangen, Germany}
\fi

\renewcommand{\shorttitle}{p-adaptive DG method for the SWE on heterogeneous computing architectures}

\begin{document}
\maketitle

\begin{abstract}
Heterogeneous computing and exploiting integrated CPU-GPU architectures has become a~clear current trend since the flattening of Moore's Law. In this work, we propose a~numerical and algorithmic re-design of a~p-adaptive quadrature-free discontinuous Galerkin method (DG) for the shallow water equations (SWE). Our new approach separates the computations of the non-adaptive (lower-order) and adaptive (higher-order) parts of the discretization form each other. Thereby, we can overlap computations of the lower-order and the higher-order DG solution components. Furthermore, we investigate execution times of main computational kernels and use automatic code generation to optimize their distribution between the CPU and GPU. Several setups, including a~prototype of a~tsunami simulation in a~tide-driven flow scenario, are investigated, and the results show that significant performance improvements can be achieved in suitable setups.
\end{abstract}


\section{Introduction}
\label{introduction}
One of the key factors limiting the accuracy and the physical relevance of climate models is the computational performance of the hardware those models are executed on. The current computational paradigm for numerical models of ocean and atmosphere mostly relies on massively parallel and, in part, hybrid platforms. However, new ways are required in order to achieve improvements in computational performance in the time of failing Moore's Law and more heterogeneous landscape of relevant computing architectures.
\par
There are some approaches to heterogeneous shallow water simulations. In~\citep{Echeverribar2020}, for example, a coupled 1D-2D model for real flood cases is hybridized using a heterogeneous CPU-GPU architecture. A different approach is taken in~\citep{Chaplygin2022} for a 2D shallow water model, where the domain is partitioned into subdomains distributed between CPUs and GPUs. Furthermore, in~\citep{Fu2017}, the global SWE are solved heterogeneously by dividing subblocks of patches into a CPU and an accelerator part. However, to the best of our knowledge, no works attempt to adapt the original algorithm's numerics to parallelize it between different architectures.
\par
Among the most promising numerical methodologies aiming to optimize the computational performance of PDE discretizations are adaptive schemes. They are particularly attractive in combination with finite element and finite volume methods and rely on adaptive mesh refinement (h-adaptivity) or local adjustment of the polynomial order of the discretization (p-adaptivity).
However, adaptive numerical schemes for time dependent problems have a~critical limitation when used in combination with massively parallel (usually based on distributed memory parallelization) computing, namely the issue of obtaining a~good load balance throughout the simulation.  
Several strategies of load-balancing have been proposed, e.g.,~\citep{Hendrickson2000,Biswas2000,TerescoDF2006,Baiges2017}, but they increase the code complexity and result in additional computational overhead. This is one of the main reasons why adaptive numerical schemes declined in popularity with in some communities like in numerical weather prediction or ocean modeling in the last decade. 
\par
The current work attempts to take a~new approach to the load balancing issue in connection with a~p-adaptive DG method: by separating the numerical scheme and the solution algorithm into a~lower-order non-adaptive (base computation) and a~higher-order adaptive (correction computation) parts, we can execute the correction computation on a~separate hardware without influencing the load balance of the base computation. This strategy of encapsulating the adaptive part of the numerical method in a~separate kernel offers, in addition, a~meaningful way to map time-dependent adaptive finite element schemes to task-based programming models particularly attractive for heterogeneous hardware architectures (see, e.g., \citep{Bosilca2013,Garcia-Gasulla2019}).
\par
The tight coupling between the base and correction computations favors hardware architectures minimizing the performance impact of communication between their parts (e.g. CPU and GPU). This motivates the focus of our current work on integrated CPU-GPU architectures represented by Systems-on-a-Chip (SoCs). They are well suited for heterogeneous computing environments, which is a~clear current trend, offer low-latency data transfer between the CPU and GPU, and tend to be energy efficient. In order to fully exploit these systems, hardware-driven algorithm design, i.e., fundamental changes in the algorithm are necessary, and the multiple instruction multiple data (MIMD) level of parallelism in Flynn's taxonomy \citep{Flynn1972} may become beneficial. 
\par
We begin in the next section, by introducing the mathematical model and its discretization by a~quadrature-free p-adaptive DG method. Sec.~\ref{order_separation} details our new approach of separating the lower-order degrees of freedom from the remainder of the discrete solution. In Sec.~\ref{code_generation}, we explain the implementation of this approach within our code generation framework and discuss the necessary adaptations. Numerical results used for the evaluation of the computational performance of the proposed scheme are presented in Sec.~\ref{results}. There we also detail individual kernel execution times for a realistic simulation setup. A short conclusions and outlook section wraps up the manuscript.

\section{Quadrature-free DG formulation of the shallow water equations}
\label{quadfree-dg}
The presentation in this section closely follows \citep{FaghihNainiKAZGK2020, FaghihNainiKZKA2022} and is included here for completeness. The discontinuous Galerkin discretization of the 2D SWE used in this work was originally proposed in~\citep{AizingerDawson2002} and further developed to simulate the 3D primitive ocean equations with free surface in~\citep{DawsonAizinger2005,AizingerPDPN2013}. 
\par
The 2D SWE in conservative form on some domain $\Omega$ are given by
\begin{align}
\label{mass}
&\partial_t \xi+\nabla \cdot \boldsymbol{q}=0,\\
\label{momentum}
&\partial_t \boldsymbol{q}
+\nabla \cdot \left(\boldsymbol{qq}^T /H \right)
+\tau_{\textrm{bf}} \boldsymbol{u}
+\left( \begin{smallmatrix} 0 & -f_c\\ f_c & 0 \end{smallmatrix} \right) \boldsymbol{q}
+gH\nabla\xi
=\boldsymbol{F}
\end{align}
where $\xi$ represents the surface elevation with respect to some datum (e.g., the mean sea level), $H = h_b + \xi$ is the total fluid depth ($h_b$ denotes here the bathymetry with respect to the same datum), and $\boldsymbol{q} \equiv (U,V)^T$ is the depth integrated horizontal velocity. Further physical quantities are the bottom friction coefficient $\tau_\textrm{bf}$, the Coriolis coefficient $f_c$, the gravitational acceleration $g$ and the body Force $F$. Their values used in the simulations are summarized in Sec.~\ref{results}. 

For the quadrature-free formulation we also need the auxiliary equation $\boldsymbol{q} = \boldsymbol{u}H$ with depth averaged velocity $\boldsymbol{u}=(u,v)^T$. Using the notation $\bm{c} := (\xi, U,V)^T$, system~\eqref{mass}--\eqref{momentum} can be written in the following compact form (cf.~\citep{FaghihNainiKAZGK2020}):
\begin{align}
\label{compact_qf_1}
\partial_t \boldsymbol{c}+ \nabla \cdot \boldsymbol{{A}}= \boldsymbol{r},\\
\label{compact_qf_2}
\boldsymbol{u} H = \boldsymbol{q},
\end{align}
where
\begin{equation}
\boldsymbol{{A}}(\boldsymbol{c}, \boldsymbol{u})=\begin{pmatrix}
U & V\\
Uu+\frac{g\xi(H+h_b)}{2}&U v\\
V u& V v+\frac{g\xi(H+h_b)}{2}
\end{pmatrix}\quad
\text{ and } \quad
\boldsymbol{r}(\boldsymbol{c}, \boldsymbol{u})=\begin{pmatrix}
0\\
-\tau_{\textrm{bf}}u+f_cV+g\xi \partial_x h_b+F_x\\
-\tau_{\textrm{bf}}v-f_cU+g\xi \partial_y h_b+F_y\\
\end{pmatrix}. 
\label{def_A_h_qf}
\end{equation}

The examples in this work use the following types of boundary conditions for the SWE: 
\par
{\em Land boundary:} $\boldsymbol{q} \cdot \boldsymbol{n} = 0$. 
\par
{\em Open-sea boundary:} $\xi = \hat{\xi}$, 
where $\hat{\xi}$ is a specified water elevation.
\par
\noindent
Initial conditions for the elevation and velocity are provided as $\xi(\boldsymbol{x},0) = \xi_0(\boldsymbol{x}) \text{ and } {\boldsymbol{q}}(\boldsymbol{x}, 0) = {\boldsymbol{q}}_0(\boldsymbol{x})$.
\par
Given a triangulation $\Omega = \bigcup  \Omega_e$, we obtain the local variational formulation of system \eqref{compact_qf_1}--\eqref{compact_qf_2} on an~element $\Omega_e$ by multiplying with sufficiently smooth test functions $\boldsymbol{\phi}$ and $\boldsymbol{\psi}$, followed by the integration by parts. We use the notation $(\cdot,\cdot)_{\Omega_e}$ and $\langle\cdot,\cdot\rangle_{\partial \Omega_e}$ for the $L^2$-scalar products on elements and edges, respectively, and denote by $\boldsymbol{n}_e$ an~exterior unit normal to $\partial \Omega_e$
\begin{align}
\label{variational_1}
&\left(\partial_t \boldsymbol{c},\boldsymbol{\phi}\right)_{\Omega_e}-\left(\boldsymbol{{A}}, \nabla \boldsymbol{\phi}\right)_{\Omega_e}+\langle\boldsymbol{{A}}\cdot\boldsymbol{n}_e,  \boldsymbol{\phi}\rangle_{\partial\Omega_e}= \left(\boldsymbol{r},\boldsymbol{\phi}\right)_{\Omega_e},\\
\label{variational_2}
&\left(\boldsymbol{u} H, \boldsymbol{\psi} \right)_{\Omega_e}
= \left(\boldsymbol{q}, \boldsymbol{\psi} \right)_{\Omega_e}.
\end{align}
\par
Let $\mathbb{P}^p(\Omega_e)$ be the polynomial space of order (i.e., the highest polynomial degree) $p$ on $\Omega_e$. We derive the semi-discrete formulation from~\eqref{variational_1}--\eqref{variational_2} by replacing $\boldsymbol{c}$ and $\boldsymbol{u}$ with the discrete solution $\boldsymbol{c}_\Delta, \boldsymbol{u}_\Delta$ and utilizing test functions $\boldsymbol{\phi}_\Delta \in \mathbb{P}^p(\Omega_e)^3, \text{ and }\boldsymbol{\psi}_\Delta \in \mathbb{P}^p(\Omega_e)^2$
\begin{align}
\label{variational_disc_1}
&\left(\partial_t \boldsymbol{c}_\Delta,\boldsymbol{\phi}_\Delta\right)_{\Omega_e}
-\left(\boldsymbol{{A}}, \nabla \boldsymbol{\phi}_\Delta\right)_{\Omega_e}
+\langle\boldsymbol{\widehat{A}},\boldsymbol{\phi}_\Delta \rangle_{\partial\Omega_e}
= \left(\boldsymbol{r},\boldsymbol{\phi}_\Delta\right)_{\Omega_e},\\
\label{variational_disc_2}
&\left(\boldsymbol{u}_\Delta H_\Delta, \boldsymbol{\psi}_\Delta \right)_{\Omega_e}
= \left(\boldsymbol{q}_\Delta, \boldsymbol{\psi}_\Delta \right)_{\Omega_e}.
\end{align}
The edge flux $\boldsymbol{{A}}(\boldsymbol{c}_\Delta, \boldsymbol{u}_\Delta) \cdot\boldsymbol{n}_e$ is approximated on $\partial\Omega_e$ by a~numerical flux $\boldsymbol{\widehat{{A}}}(\boldsymbol{c}_\Delta, \boldsymbol{u}_\Delta,\boldsymbol{c}^+_\Delta, \boldsymbol{u}^+_\Delta, \boldsymbol{n}_e)$ that depends on discontinuous values of the solution on element $\Omega_e$ (i.e., $\boldsymbol{c}_\Delta, \boldsymbol{u}_\Delta$) and its edge neighbor (i.e., $\boldsymbol{c}^+_\Delta, \boldsymbol{u}^+_\Delta$). On exterior domain boundaries, the specified boundary conditions for the elevation or velocity are utilized in the flux computation. In this work, we rely on the Lax--Friedrichs flux \citep{HajdukHAR2018} modified (see~\citep{FaghihNainiKAZGK2020}) for our quadrature-free integration scheme, that is
\small
\begin{equation*}
\boldsymbol{\widehat{{A}}}(\boldsymbol{c}_\Delta, \boldsymbol{u}_\Delta,\boldsymbol{c}^+_\Delta, \boldsymbol{u}^+_\Delta, \boldsymbol{n}_e)
=\frac{1}{2}\left(\left(\boldsymbol{A}(\boldsymbol{c}_\Delta, \boldsymbol{u}_\Delta)+\boldsymbol{A}(\boldsymbol{c}_\Delta^+, \boldsymbol{u}_\Delta^+)\right) \cdot \boldsymbol{n}_e
+\lambda(\boldsymbol{c}_\Delta-\boldsymbol{c}_\Delta^+)\right),
\end{equation*}
\normalsize
with the following approximation of $\lambda|_E$ for each edge $E$ of $\Omega_e$:
\begin{equation}\label{eq:lambda}
\lambda_{|_E} \coloneqq \max\limits_{\Omega_e : \bm{x}_E \in \partial \Omega_e} \left|{\bm{u}_\Delta}_{|_{\Omega_e}}(\bm{x}_E) \cdot\boldsymbol{n}_e\right|+ \max \limits_{\Omega_e : \bm{x}_E \in \partial \Omega_e} \sqrt{g {H_\Delta}_{|_{\Omega_e}(\bm{x}_E)}}\,,
\end{equation}
where $\boldsymbol{x}_E$ denotes the midpoint of edge $E$.
The main computational kernels of the above discretizations are the evaluations of the element and edge integrals in \eqref{variational_disc_1} and \eqref{variational_disc_2}.
\par
As in~\citep{FaghihNainiKAZGK2020}, given  a~basis $\varphi_{ei}(\bm{x}), \,i=1, \ldots, K(p)$ of $\mathbb{P}^p(\Omega_e)$, $\bm{c}_\Delta$ and $\bm{u}_\Delta$ can be represented as
\begin{align}
\bm{c}_\Delta(t,\bm{x})_{|_{\Omega_e}} &=(\xi_\Delta, U_\Delta, V_\Delta)^T(t,\bm{x})=\sum_{j=1}^3\sum_{i=1}^{K(p)} c_{ei}^j\varphi_{ei}(\bm{x})\,\bm{e}_j, \label{c_h}\\
 \bm{u}_\Delta(t,\bm{x})_{|_{\Omega_e}} &=(u_\Delta, v_\Delta)^T(t,\bm{x})=\sum_{j=1}^2\sum_{i=1}^{K(p)} u_{ei}^j\varphi_{ei}(\bm{x})\,\bm{e}_j\label{c_h_tilde}
\end{align}
with $\bm{e}_j$ denoting the $j$-th unit vector in $\mathbb{R}^3$ in~\eqref{c_h} or $\mathbb{R}^2$ in~\eqref{c_h_tilde}. 
\par
Our implementation employs triangles. The number of basis functions $K(p)$ depends on the chosen polynomial approximation space; it has the following values in $\mathbb{R}^2$: $K(0)\!=\!1, \, K(1)\!=\!3, \, K(2)\!=\!6$, and $K(3)\!=\!10$.
The basis functions employed in our implementation are hierarchical and orthonormal with respect to the $L^2$-scalar product on ${\Omega_e}$. 
\par
The temporal discretization of system \eqref{variational_disc_1}--\eqref{variational_disc_2} is performed using a~strong stability preserving (SSP) Runge--Kutta method \citep{GottliebShu1998}. In the test cases presented in Sec.~\ref{results}, we use a~two-stage SSP Runge--Kutta method given in~\citep{ReuterAWFK2016}.
\par
Our scheme relies on a dynamic p-adaptive algorithm which adjusts and limits, if necessary, the local approximation order using an adaptivity indicator. For DG discretizations which rely on hierarchical bases, p-adaptive (as opposed to h- and hp-adaptive) schemes are particularly attractive due to simplicity of implementation. Numerical results in the literature and our previous studies for p-adaptive schemes, together with the performance measurements in Sec.~\ref{results} show savings of computational time compared to non-adaptive schemes of similar accuracy \citep{Kubatko2009,FaghihNainiA2022}. All our dynamically p-adaptive numerical runs use a slightly modified Jump-Reconstruction-Limiting (JRL) indicator from~\citep{FaghihNainiA2022}.

\section{Algorithmic adaptation: order separation}
\label{order_separation}
The quadrature-free formulation utilized in our solver -- when combined with hierarchical bases -- naturally separates the discrete equations for different polynomial orders. The main idea is to evaluate the update for the non-adaptive degrees of freedom of the DG approximation independently from the adaptive higher-order part of the solution. Since the quadrature-free formulation only contains product-type nonlinearities, a~p-adaptive scheme for such a discretization boils down to adding and removing terms without affecting the rest of the DG approximation. The non-adaptive lower-order DG solution is computed for all elements, and a higher-order correction is applied where necessary.

\begin{SCfigure}[][h]
\centering
\includegraphics[width=0.36\textwidth, trim=65 100 295 170, clip]{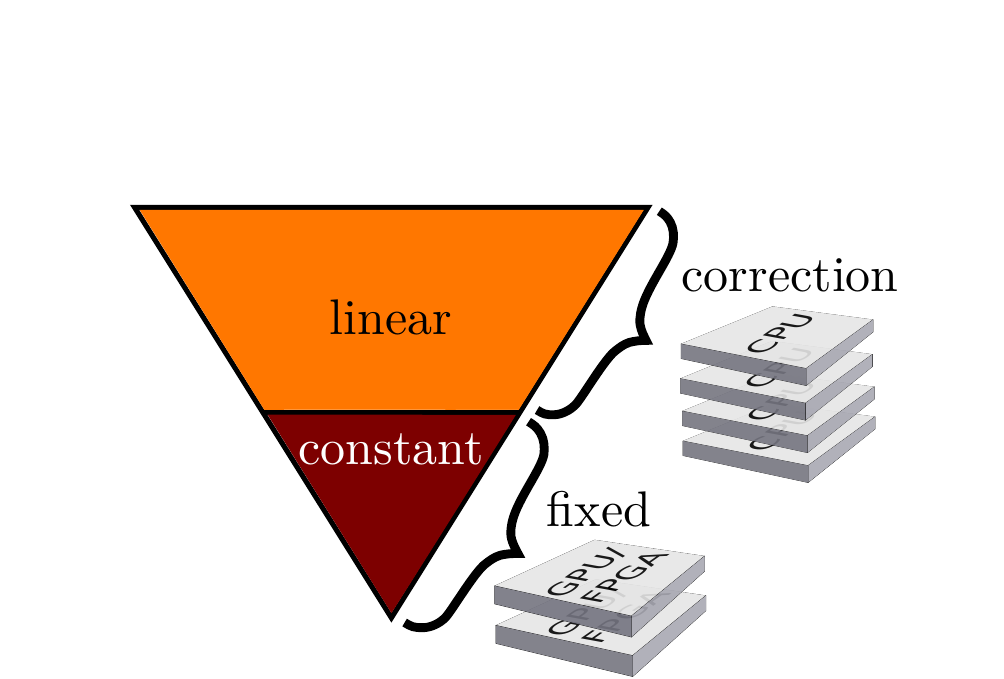}
\includegraphics[width=0.36\textwidth, trim=65 100 295 170, clip]{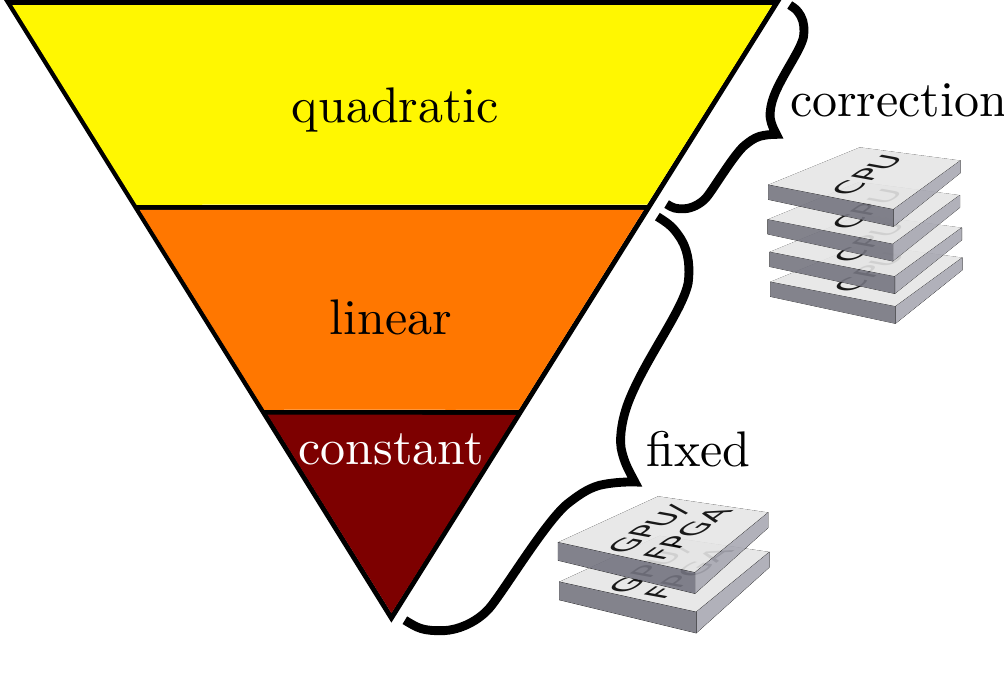}
  \caption{Schematic idea of separation approach: The solution for the non-adaptive part (left: piecewise constant, right: piecewise linear) is computed for all elements on one hardware, e.g a GPU. An~adaptive correction (left: linear, right: quadratic contributions) is then applied to some elements. The latter computation can use a different hardware, e.g., a~CPU.}
  \label{fig:layers}
\end{SCfigure}

The setups evaluated in Sec. \ref{results} include a constant non-adaptive part with a linear correction and a linear non-adaptive part with a quadratic correction as shown in Fig.~\ref{fig:layers}.
This approach is naturally extendable to any higher order DG discretization.
Our presentation of the separation methodology needs the algebraic representation of element integrals, thus we repeat equation (20) from~\citep{FaghihNainiKAZGK2020} (adapted to the notation in this work).
Inserting the basis representations \eqref{c_h} and \eqref{c_h_tilde} into \eqref{variational_disc_1} and testing the first equation with $\bm{\phi}_\Delta = \varphi_{eq} \bm{e}_1$ we obtain for $q \in  {1,\dots,K(p)}$:
\begin{align}
\left(\bm{{A}}(\bm{c}_\Delta, \bm{u}_\Delta), \nabla (\varphi_{eq} \bm{e}_1) \right)_{\Omega_e}
= \sum_{i=1}^{K(p)}\left[ c_{ei}^{2}\int_{\Omega_e} \frac{\partial\varphi_{eq}} {\partial x}\varphi_{ei}\mathrm{d}\bm{x}+ c_{ei}^{3}\int_{\Omega_e} \frac{\partial\varphi_{eq}} {\partial y}\varphi_{ei}\mathrm{d}\bm{x}\right].
\label{mass_basis}
\end{align}
\par
We assume that up to order $b$, the computation is done for all elements, and the correction for order $b+1$ is only applied to some elements. Then the base computation would involve the following: 
\begin{align*}
\left(\bm{A}(\bm{c}_\Delta, \bm{u}_\Delta), \nabla (\varphi_{eq} \bm{e}_1) \right)^{base}_{\Omega_e}
=\sum_{i=1}^{K(b)}\left[ c_{ei}^{2}\int_{\Omega_e} \frac{\partial\varphi_{eq}} {\partial x}\varphi_{ei}\mathrm{d}\bm{x}+ c_{ei}^{3}\int_{\Omega_e} \frac{\partial\varphi_{eq}} {\partial y}\varphi_{ei}\mathrm{d}\bm{x}\right]
\end{align*} 
for $q \in  {1,\dots,K(b)}$. \\
The correction consists of 
\begin{align*}
\left(\bm{A}(\bm{c}_\Delta, \bm{u}_\Delta), \nabla (\varphi_{eq} \bm{e}_1) \right)^{correction}_{\Omega_e}
=\sum_{i=K(b)+1}^{K(p)}\left[ c_{ei}^{2}\int_{\Omega_e} \frac{\partial\varphi_{eq}} {\partial x}\varphi_{ei}\mathrm{d}\bm{x}+ c_{ei}^{3}\int_{\Omega_e} \frac{\partial\varphi_{eq}} {\partial y}\varphi_{ei}\mathrm{d}\bm{x}\right]
\end{align*} 
for $q \in  {1,\dots,K(b)}$ and 
\begin{align*}
\left(\bm{A}(\bm{c}_\Delta, \bm{u}_\Delta), \nabla (\varphi_{eq} \bm{e}_1) \right)^{correction}_{\Omega_e}
&=\sum_{i=1}^{K(p)}\left[ c_{ei}^{2}\int_{\Omega_e} \frac{\partial\varphi_{eq}} {\partial x}\varphi_{ei}\mathrm{d}\bm{x}+ c_{ei}^{3}\int_{\Omega_e} \frac{\partial\varphi_{eq}} {\partial y}\varphi_{ei}\mathrm{d}\bm{x}\right]
\end{align*}
for $q \in  {K(b)+1,\dots,K(p)}$.
\par
The edge integrals are separated in a similar manner, of course taking into account the local approximation order of the current element. Note that, when computing the coefficient $\lambda$, i.e., \eqref{eq:lambda}, in front of the penalty term for the Lax-Friedrichs flux, regardless of the approximation order, exclusively the constant part of the solution is used for evaluation of the velocities and the surface elevation fields. In a~similar manner, the nonlinear bottom friction on the right-hand side uses the piecewise constant solution part, instead of using the full-order approximation for computing the velocity magnitude. This approach showed no loss in solution quality and is used in our implementation to avoid special treatment when applying a correction.
\par
The solution $\bm{u}_\Delta(t,\bm{x})|_{\Omega_e}=:\bm{u}_e$ of \eqref{variational_disc_2} involves solving an element-local linear system which looks differently for different approximation orders, i.e., the higher-order degrees of freedom can actually affect the lower-order ones. However, our tests (not shown here) indicate an equal solution quality when omitting higher-order contributions in the lower-order computations, so that, in order to compute $\bm{u}_e$, using an LU-factorization, we are now generally solving 
\begin{align*}
\left(H_{i,j}\right)_{i,j \in \{1,\dots, K(p)\}} \left({u}^{l}_j\right)_{j \in \{1,\dots, K(p)\}} =  \left(c^{l+1}_i\right)_{i \in \{1,\dots, K(p)\}} \quad  l=1,2
\end{align*} 
with\\
\begin{minipage}{0.55\textwidth}
\begin{align*}
&\left(H_{i,j}\right)_{i,j \in \{1,\dots, K(p)\}} :=  \int_{\Omega_{e}} \left(\sum_{n=1}^{K(p,i,j)}c^1_{en}\varphi_{en} +h_b\right) \varphi_{ej} \varphi_{ei}\\ 
&\left(u^{l}_j\right)_{j \in \{1,\dots, K(p)\}} := u^{l}_{ej}\\
&\left(c^{l+1}_i\right)_{i \in \{1,\dots, K(p)\}} :=\int_{\Omega_{e}} \sum_{n=1}^{K(p)}c^{l+1}_{en}\varphi_{en}\varphi_{ei}, \quad l=1,2
\end{align*} 
\end{minipage}
\begin{minipage}{0.4\textwidth}
$K(p,i,j)=\begin{cases}1, &\text{if } i=j=1\\
3,& \text{if }  p  <= 1 \land !(i = j =1)  \\
6,& \text{if }  p  <= 2 \land !(i <= 3 \land j <=3 ) \\
10,& \text{if } p  <= 3 \land !(i <= 6 \land j <=6 )
\end{cases}
$ 
\end{minipage}\hfill\\
Since the lower-order terms in $\left(c^{l+1}_i\right)_{i \in \{1,\dots, K(p)\}}$ are not affected by the higher-order ones due to orthonormality property, the lower-order system does not have to be re-assembled when applying the higher-order correction.

\par

\section{Code generation}
\label{code_generation}
The new separation approach was implemented in GHODDESS\footnote{https://i10git.cs.fau.de/ocean/ghoddess-release} (Generation of Higher-Order Discretizations Deployed as ExaSlang Specifications), a~Python frontend to the \mbox{ExaStencils} code generation framework~\citep{FaghihNainiKAZGK2020, LengauerABCRTGHKCGGKKRSS2020, AltKFFOPAHK2023}. GHODDESS uses SymPy\footnote{https://www.sympy.org}~\citep{MeurerSPCKRKIMSRVGMBGVJPCTRSFKCS2017}  to perform symbolic differentiation and integration. It implements the complete program specification, including optimizations such as buffering geometric information. 

\par
\mbox{ExaStencils} is a~highly advanced framework for generating C++ stencil codes on block-structured grids. %
The parallelization and communication routines of the code are automatically generated using OpenMP, MPI, and CUDA as backends. %
For automatic optimizations, \mbox{ExaStencils} supplies code transformations such as common subexpression elimination, polyhedral loop transformations \citep{KronawitterLengauer2018}, explicit single instruction multiple data (SIMD) vectorization, and address pre-calculation. %
\mbox{ExaStencils} provides its external domain specific language (DSL)  \mbox{ExaSlang}~\citep{LengauerABCRTGHKCGGKKRSS2020} consisting of four language layers with different abstraction levels. %
Each layer is designed to provide a~tailored language for the different aspects of a problem and its corresponding solution methods. %
ExaSlang~4 is the most comprehensive layer of our DSL, as it can hold the whole program specification and makes concepts such as parallelization, domain partitioning, and data I/O available to users. %
For this reason, it was chosen as an intermediate target for the mapping from the symbolic description in GHODDESS. %
\mbox{ExaStencils'} source-to-source compiler, written in Scala, is then responsible for parsing the \mbox{ExaSlang} input, applying code transformations, and emitting the target C++ code. %
\par
For the hardware-driven algorithm design, we extend our code generation pipeline on different abstraction layers. %
One is on the algorithmic side within GHODDESS and incorporates the separated kernels described in Sec.~\ref{order_separation}. %
The symbolic program description also contains the control flow for distributing individual kernels to specific architecture components. %
Executing kernels in this heterogeneous manner requires data synchronization and bookkeeping concepts. %
Automating these rather technical steps can be highly beneficial for productivity, making them an ideal target for code generation with ExaStencils. %
\par
By default, ExaStencils supplies a standard data migration method for architectures with discrete memory locations for the host and device. %
Explicit memory transfer statements between the discrete memory locations and data structures for their version tracking are generated. %
These transfer operations, however, incur large latencies and can impact the execution time significantly when performed at a high frequency. %
Therefore integrated architectures such as the NVIDIA Jetson systems appear particularly promising for our heterogeneous approach and excel, in addition, in the energy-to-solution metric~\citep{GevelerRAGT2016}. %
Since the CPU and the GPU share the same die and the system memory, no distinct memory locations and transfer operations are needed -- thus allowing for a low-latency communication between the CPU and the GPU. %
Supporting the NVIDIA Jetson systems requires specialized extensions within ExaStencils. %
We implemented them as distinct building blocks so that all future users can benefit from the new code generation capabilities. %
The current Jetson systems employ ARM-based CPUs, for which we replenished an existing building block for an automatic SIMD vectorization with NEON intrinsics~\citep{LengauerABCRTGHKCGGKKRSS2020}. %
The other building blocks are memory management techniques of the CUDA platform, namely pinned memory, unified memory access, and zero-copy memory explained in the following. %
%
\begin{itemize}
\item
\textbf{Pageable} memory is referred to as memory which can be automatically swapped (paged) by the operating system between the primary (usually RAM) and secondary (e.g., external drive) storage. Since GPUs cannot directly access data from pageable host memory,
data transfers to and from GPU often incur overhead from internal copy operations to page-locked or pinned host buffers issued by the CUDA runtime. %
For this purpose, CUDA offers the (de-)allocation of \textbf{pinned} host memory to avoid the additional copy and increase the transfer bandwidth. %
\item
\textbf{Unified} memory bundles the previously separate host and device memory allocations into a single allocation. %
It also obviates explicit memory transfers with an automatic on-demand migration determined by the CUDA runtime via a page-fault mechanism. %
While this model significantly simplifies the development of heterogeneous codes, it often performs poorly due to the overhead from the fault handling. %
Explicit prefetching of data can mitigate this performance penalty and allows for fine-grained overlapping with kernel executions at the cost of additional code complexity. %
Still, the complexity can be overcome by generating automated prefetching routines. %
\item
\textbf{Zero-copy} memory allows GPU threads to access host memory directly. %
Users are provided with a shared virtual memory space for host and device data given by mapping the allocated host memory to the CUDA address space. %
This method is especially beneficial for systems with integrated GPUs such as the NVIDIA Jetson architectures. %
While this approach does not need explicit migration requests, synchronization between CPU and GPU execution is necessary for critical regions. %
The required bookkeeping for this purpose is automatically generated by the ExaStencils compiler. %
\end{itemize}
%
%
%
%
%

\section{Performance results}
\label{results}

The performance measurements were carried out on two test platforms. The first one is an NVIDIA Jetson AGX Xavier SoC (called in the following ARM-AGX), which is a~part of the ICARUS\footnote{http://www.mathematik.tu-dortmund.de/sites/icarus-green-hpc} cluster at TU Dortmund, containing an NVIDIA Carmel Armv8.2 CPU with eight cores and an NVIDIA Volta GPU. The CPU's frequency was fixed at 2100\,MHz in our test runs. The second test platform is a server with two AMD Epyc 7742 processors with 64 cores each and one NVIDIA Quadro RTX 6000 GPU (called in the following AMD-RTX). There, the CPU frequency was fixed at 2250\,MHz. We used OpenMP for the CPU parallelization and chose the number of threads to get similar execution times between the pure CPU and pure GPU versions of our code: three threads on the ARM-AGX and 64 threads on the AMD-RTX turned out to produce the best match. For the measurements presented in the following sections, based on our exhaustive testing, we used the fastest memory management techniques for each specific setup. On the ARM-AGX, the pure GPU code was narrowly fastest with pageable memory, the heterogeneous one clearly with zero-copy memory. On the AMD-RTX, pinned memory turned out to be fastest for all code variants. 
\par
The main computational kernels in our SWE code are as follows (cf. Figs.~\ref{flow_non-separated}, \ref{flow_adaptive-separated}): edge computation (cf. \eqref{variational_disc_1} in Sec.~\ref{quadfree-dg}), element and right-hand-side (RHS) computation (cf. \eqref{variational_disc_1} in Sec.~\ref{quadfree-dg}), auxiliary computation $\boldsymbol{u}H = \boldsymbol{q}$ (cf. \eqref{variational_disc_2} in Sec.~\ref{quadfree-dg}), minimum depth control to avoid negative depths, boundary condition (BC) evaluation, the Runge-Kutta step update, and, in dynamically p-adaptive runs, the adaptivity indicator.
\par

We investigate the computational performance using two simulation scenarios. The first one, a radial dam break on a randomly perturbed uniform mesh, was chosen because of the domain simplicity and easy problem customizability. The goal of the second test setup, a tide-driven flow in a realistic domain with a block-structured grid~\citep{FaghihNainiKZKA2022} consisting of several blocks, is to demonstrate the applicability of the new approach for more complex problems. The first test problem is used to evaluate the performance of main code kernels on ARM-AGX and to quantify the effect of separation on the total execution time. In addition, we designed a range of statically adaptive setups with varying fractions of higher-order elements to compare the overhead of using a~p-adaptive scheme vs. a higher-order scheme without adaptivity. The same test -- carried out on the AMD-RTX -- is also employed to illustrate the latency effect of a~discrete GPU on the execution time. Finally, the computational performance of a dynamic p-adaptive simulation is evaluated in detail using various configurations (separated, i.e. adaptive and non-adaptive part computed separately, unseparated, i.e. the standard approach, only CPU, only GPU, hybrid CPU and GPU). Here we also specifically study the overhead caused by our separation approach, which mostly boils down to transferring the solution parts between non-adaptive and adaptive kernels. Based on the insights originating from this detailed performance evaluation, we employ the second simulation scenario with the optimal settings to run a more complex test problem on a realistic domain.

Measuring the runtime contributions of individual kernels in different discretization spaces already provides valuable insights for practical application tuning.
In future work, however, developing a detailed performance model to guide this optimization process could prove to be a worthwhile endeavor.
Setting up such a performance model presents several challenges that extend beyond the scope of our current work.
The foremost challenge stems from the complexity of the kernels, which can be substantial, especially for higher orders or intricate operations like indicator evaluation.
This complexity also limits the utility of automated tools to some extent.
After establishing an execution model for each kernel, these models must then be mapped to hardware models for both CPU and GPU, including their interconnects and potentially shared memory spaces.
Ideally, this modeling effort should also encompass other factors such as CPU and GPU caching, extending beyond the scope of individual kernel models.
While there is a substantial body of research on CPU execution modeling, literature pertaining to the remaining aspects is comparatively sparse.
Lastly, it is crucial to account for the impact of variations that occur at execution time, such as changes in the adapted elements.
This includes quantifying these effects and adjusting the tuning approach accordingly.

The values of the physical parameters used in the simulations in this section are listed in Tab.~\ref{physical_quantities}. In Sec.~\ref{dam_break}, the test problem uses a~linear friction law, whereas the setup in Sec.~\ref{bahamas} uses the quadratic one.

\begin{table}[h]
	\centering
	\renewcommand{\arraystretch}{1.55}
	\setlength{\tabcolsep}{4pt}
\begin{tabular}{c|lp{4.5cm}rr}
\textbf{variable name}& \textbf{unit}& \textbf{meaning} & \textbf{value in setup Sec.~\ref{dam_break}} & \textbf{value in setup Sec.~\ref{bahamas}} \\
\hline
 $\tau_{\textrm{bf}}$ & $\frac{\text{m}}{\text{s}}$ & bottom friction coefficient & $0.0001 \cdot H$ & $0.009 \cdot |\boldsymbol{u}|$   \\
 $f_c$ & $\frac{1}{\text{s}}$ & Coriolis coefficient & $1.0\cdot 10^{-5}$ & $3.19\cdot 10^{-5}$ \\
 $g$ & $\frac{\text{m}}{\text{s}^2}$ & gravitational acceleration & $1$ & $9.81$  \\
 $\boldsymbol{F}$ &$\frac{\text{m}^2}{\text{s}^2}$ & body force (variable atmospheric pressure, tidal potential) & - & -
\end{tabular}
	\caption{Overview of physical quantities and their values in simulation setups.}  
	\label{physical_quantities}
\end{table}
\subsection{Radial dam break}
\label{dam_break}

\begin{SCfigure}[][h]
\centering
\centering
    \includegraphics[width=0.4\textwidth, trim=110 140 140 180, clip]{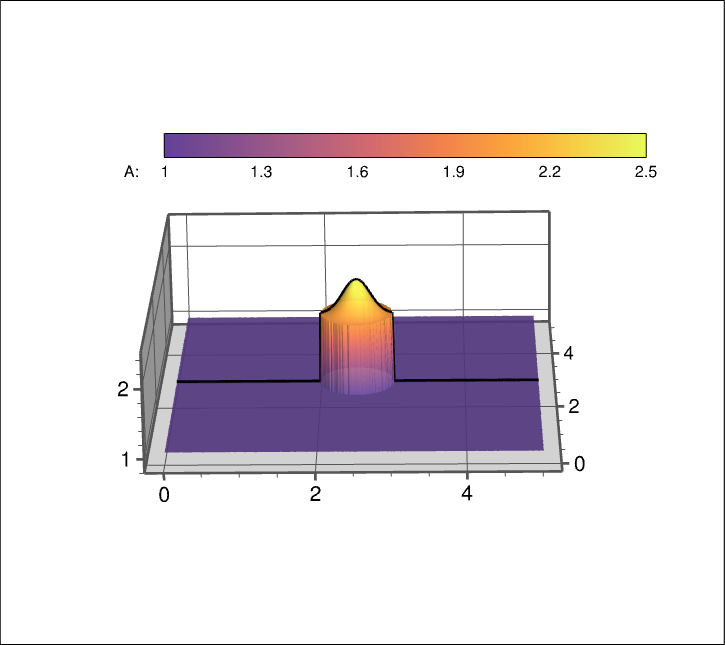}	\includegraphics[angle=0,width=0.13\textwidth, trim=1650 570 20  300, clip]{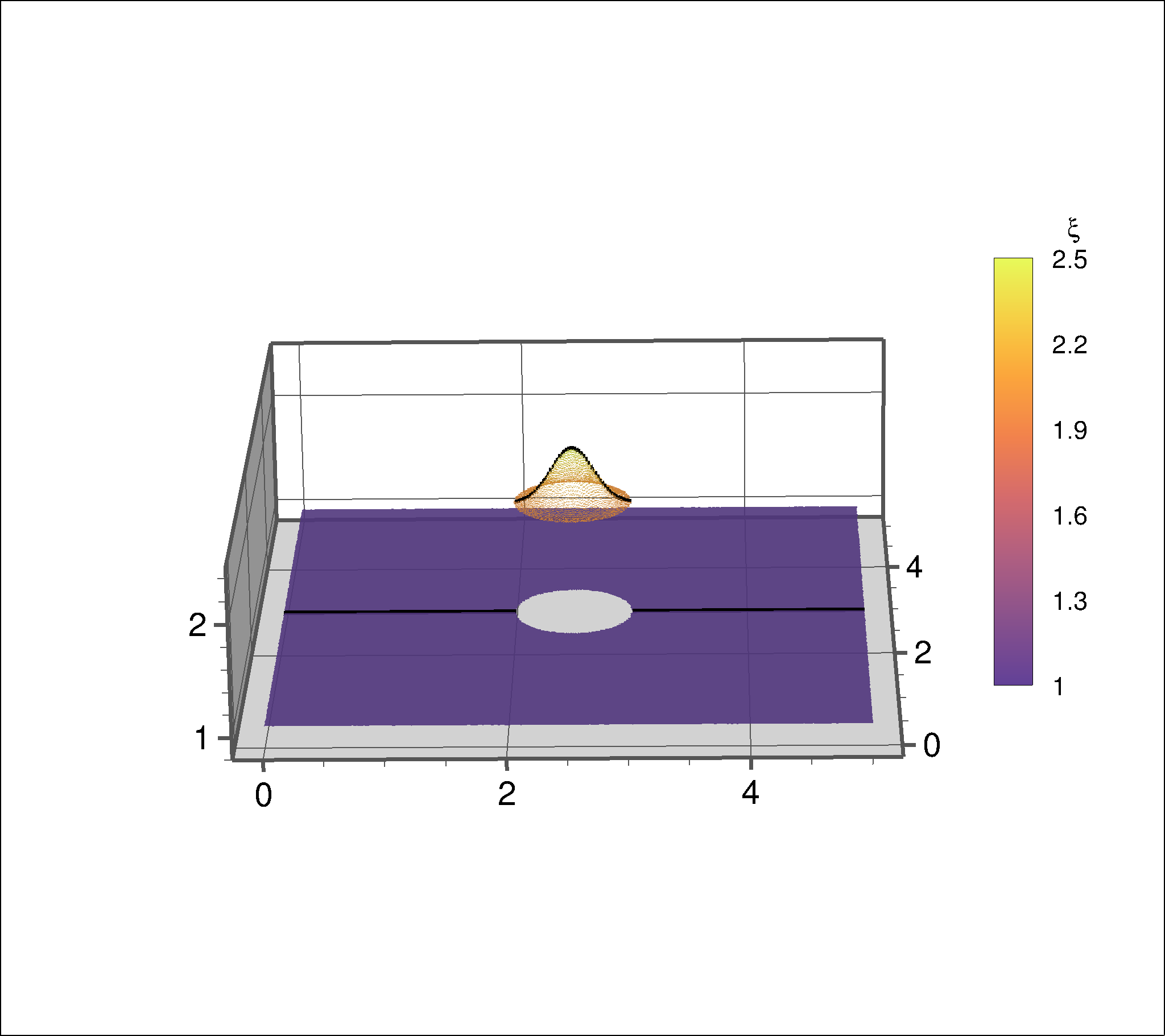}  
  \caption{Radial dam break: Analytic initial condition for surface elevation with slice at $y=2.5$.}
  \label{fig:dambreak_IC}
\end{SCfigure}

Here we consider a slightly modified dam break example from~\citep{FaghihNainiA2022}, which was based in turn on~\citep{Leveque2002,Hajduk2021}. We set $\Omega = [0,5]\times [0,5]$ and $g=1$; however, contrary to~\citep{FaghihNainiA2022}, a~constant bathymetry $h_b =0.5$ is used. On the exterior boundaries, we impose no normal flow boundary conditions and the initial condition (see also Fig.~\ref{fig:dambreak_IC}) is given by 
\begin{align*}
&\xi(x, y ,t)=\begin{cases}2+0.5\, e^{-15((x-2.5)^2+(y-2.5)^2)}, &\text{if } (x-2.5)^2+(y-2.5)^2<0.25,\\
1,& \text{otherwise},
\end{cases}\\
&U(x,y,t)= 0, \; V(x,y,t) = 0.
\end{align*}

\begin{figure}[h!]
\centering
    \includegraphics[width=0.325\textwidth, trim=100 140 140 230, clip]{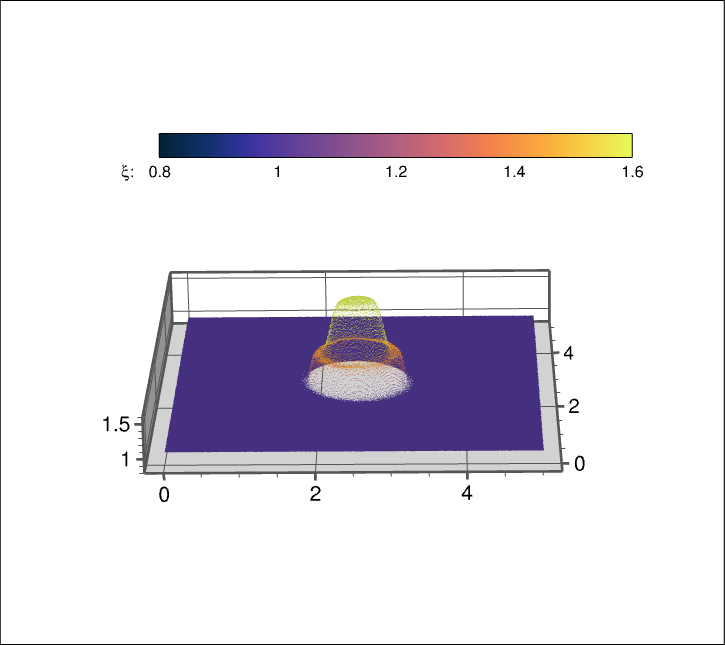}
     \includegraphics[width=0.325\textwidth, trim=100 140 140 230, clip]{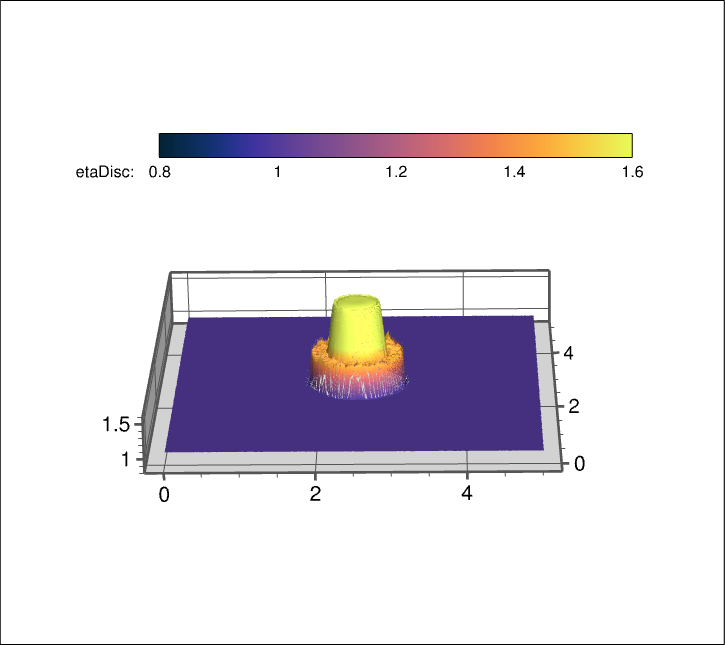}
     \includegraphics[width=0.325\textwidth, trim=100 140 140 230, clip]{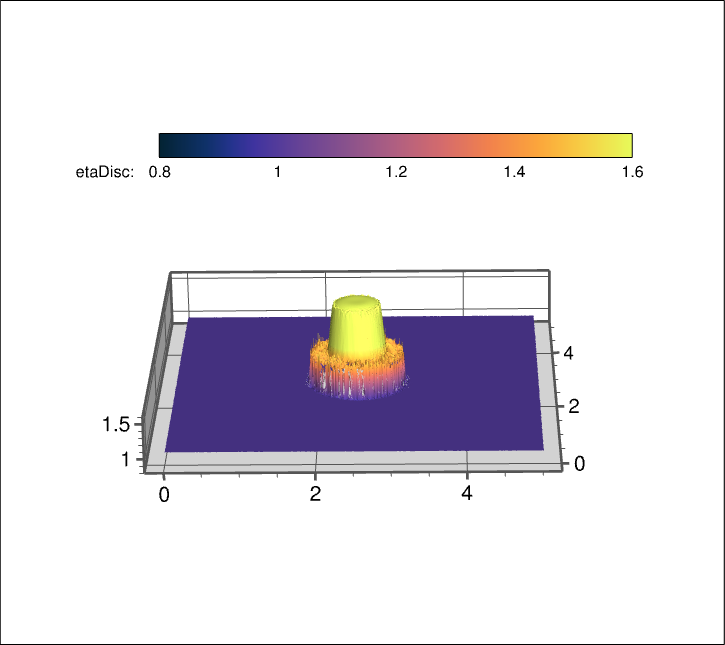}
     \includegraphics[width=0.49\textwidth, trim=2 200 2 260, clip]{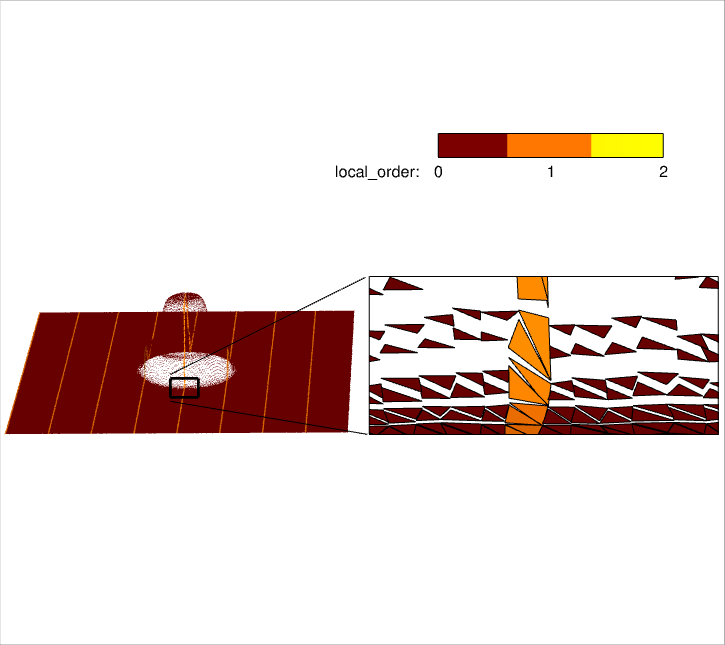}
     \includegraphics[width=0.49\textwidth, trim=2 200 2 260, clip]{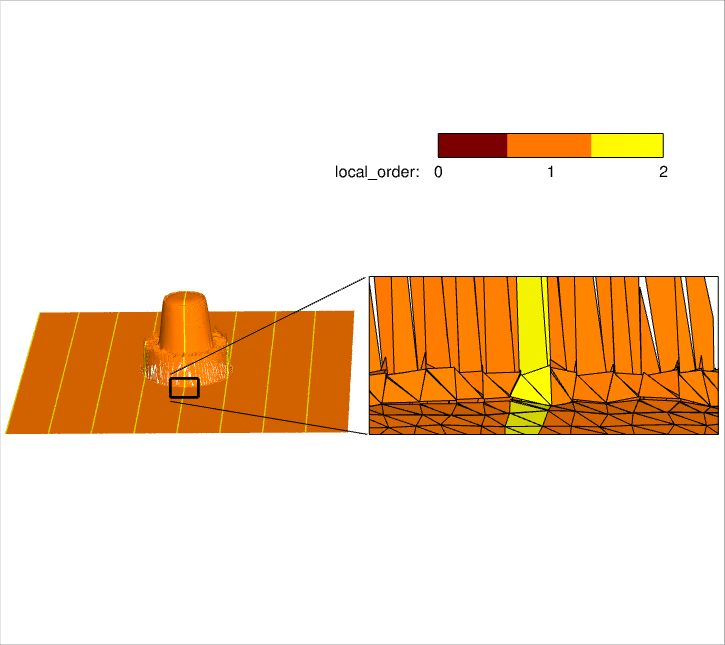}
  	\includegraphics[angle=0,width=0.65\textwidth, trim=300 1300 200  300, clip]{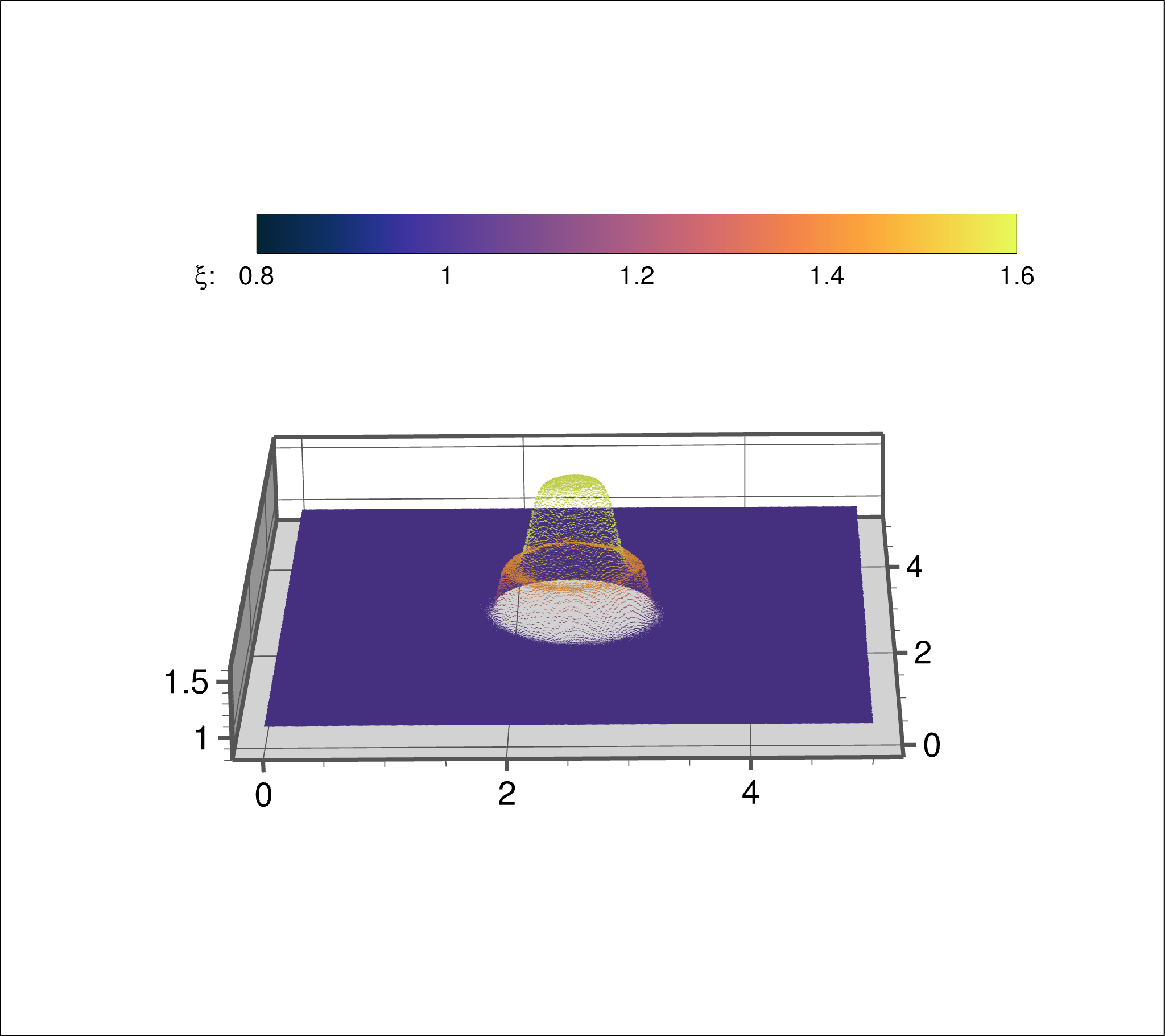} \hfill 
\includegraphics[width=0.34\textwidth, trim=1090 1300 150  300, clip]{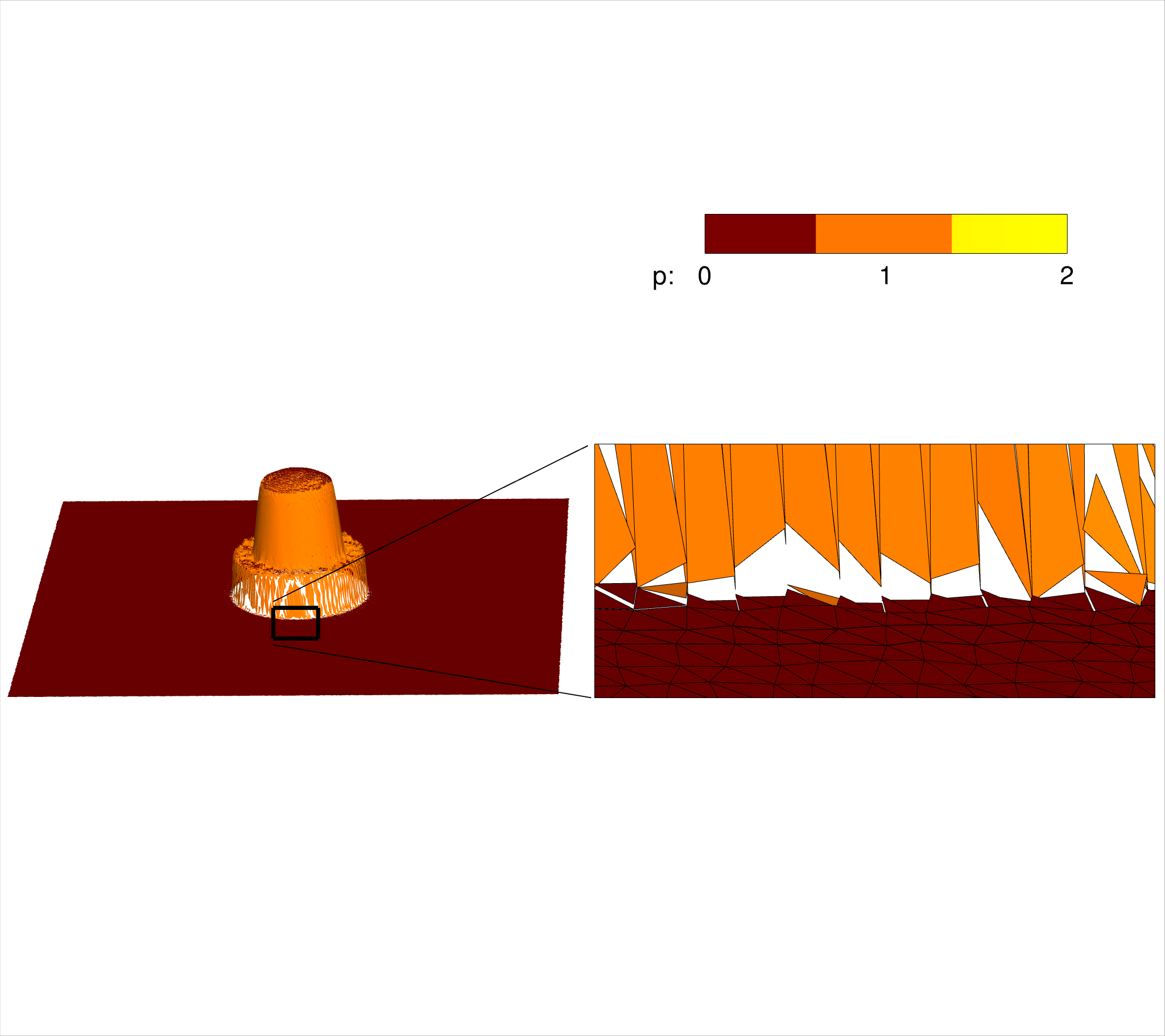} 
  \caption{Radial dam break: Surface elevation at $t=0.1\,\mathrm{s}$. Top row: constant (p0, left), linear (p1, middle), and quadratic solution (p2, right). Bottom row: statically adaptive solution with every 32$^{nd}$ element using the higher approximation order: constant-linear (p0-1, left) and linear-quadratic (p1-2, right), color-coding shows the local approximation order.}
  \label{fig:dambreak_static}
\end{figure}

For all performance measurements, we use a randomly perturbed uniform mesh with 2\,097\,152 triangles.
For illustration purposes, Fig.~\ref{fig:dambreak_static} (top) shows the surface elevation at $t=0.1\,\mathrm{s}$ on a mesh with 131\,072 triangles for different approximation orders and Fig.~\ref{fig:dambreak_static} (bottom) the local approximation order for the statically adaptive setup, in which every 32$^{nd}$ element is forced to use a~higher order. This particular test was selected because it is challenging for both CPU and GPU to achieve efficient vectorization and memory accesses.
In the static setups, we compute 100 time steps with $\Delta t=0.00001\,s$ and two substeps (Runge-Kutta stages) each. The execution time is then averaged over the 200 substeps. 

\begin{figure}[h!]
\centering
     \includegraphics[width=0.32\textwidth, trim=160 190 170 280, clip]{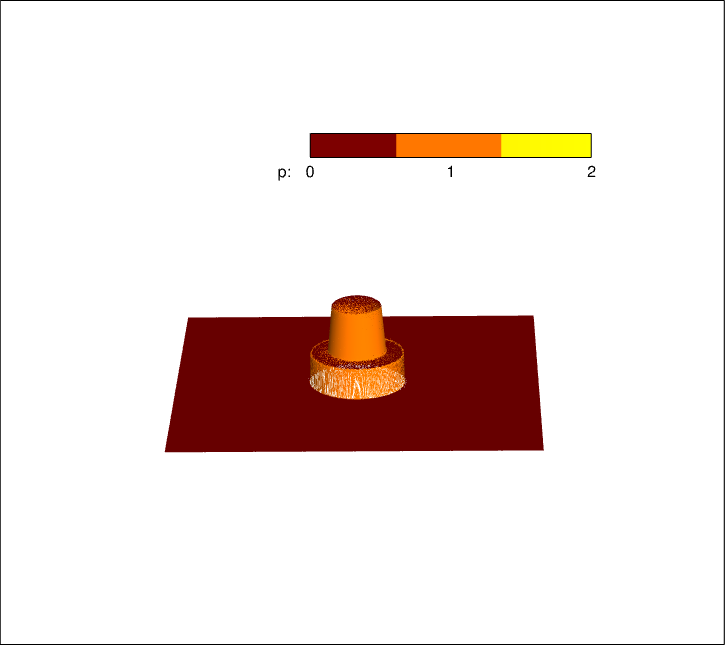}
          \includegraphics[width=0.32\textwidth, trim=160 190 170 280, clip]{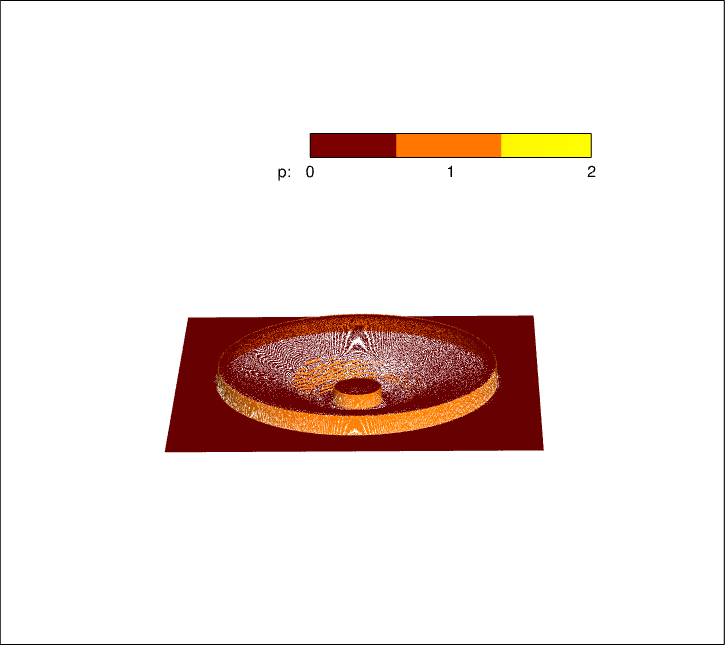}
               \includegraphics[width=0.32\textwidth, trim=160 180 170 280, clip]{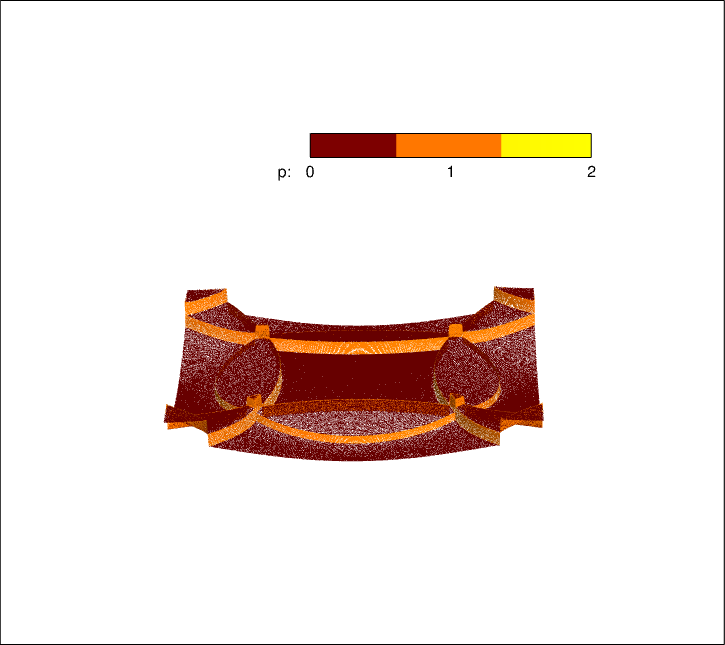}
                    \includegraphics[width=0.32\textwidth, trim=160 190 170 280, clip]{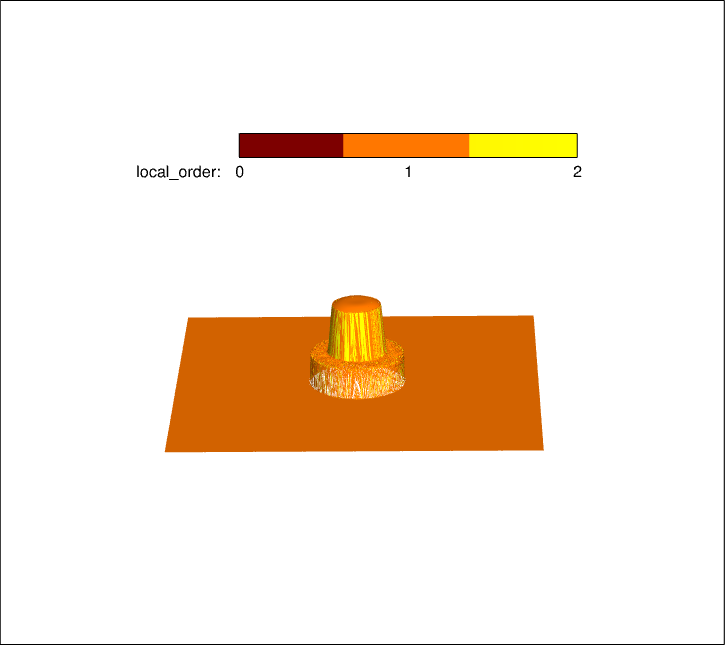}
          \includegraphics[width=0.32\textwidth, trim=160 190 170 280, clip]{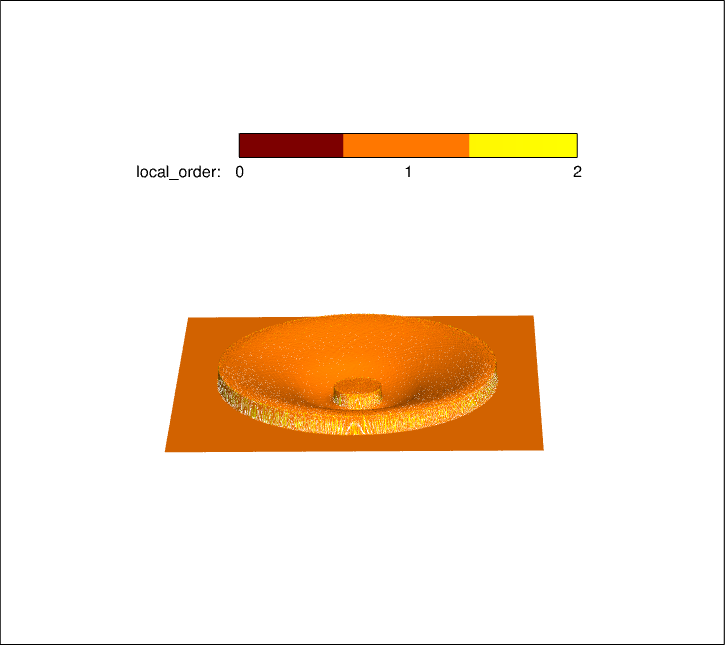}
               \includegraphics[width=0.32\textwidth, trim=160 180 170 280, clip]{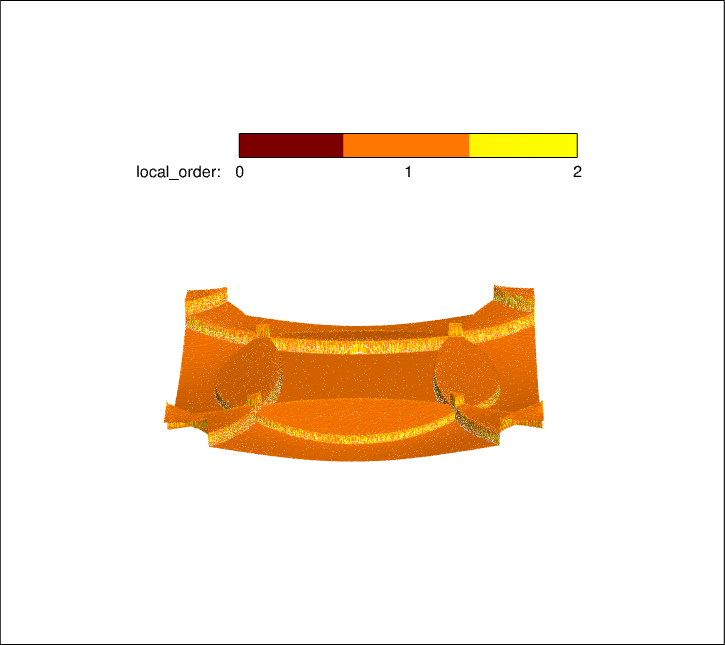}
\includegraphics[width=0.34\textwidth, trim=1090 1300 150  300, clip]{figures/dambreak/dambreak_eta_p_l8_o0-1_1_hr_colorbar.png} 
  \caption{Radial dam break: dynamic p-adaptive test. Surface elevation at $t=0.1\,\mathrm{s}$ (left), $t=1.0\,\mathrm{s}$ (middle) and $t=2.5\,\mathrm{s}$ (right). Top row: p0-1, bottom row: p1-2. Color-coding shows the local approximation order.}
  \label{fig:dambreak}
\end{figure}

The surface elevation and the local approximation order for the dynamic p-adaptive cases are shown in Fig.~\ref{fig:dambreak} at $t=0.1\,\mathrm{s}$, $t=1.0\,\mathrm{s}$ and $t=2.5\,\mathrm{s}$. These simulations were run for the total of 12\,500 time steps with $\Delta t=0.0002\,s$. Kernel timings were averaged over all substeps to capture the variations in the adaptive part of the solution algorithm.

In Fig.~\ref{flow_non-separated}, we compare the unseparated setups for the piecewise constant, linear, and quadratic solutions without adaptivity to the statically adaptive ones with 1/32 of the elements fixed at the higher order and to the dynamically p-adaptive results. We clearly see that, for the adaptive setups, the total execution times are much lower than those of the non-adaptive higher-order version.

\setlength{\tabcolsep}{4pt}
\renewcommand{\arraystretch}{1.05}
\tikzset{c-rectangle2/.style={rectangle,  thick, rounded corners, minimum
width=2cm, minimum height=1.0cm,text centered, text width = 4.45cm, draw=black,
fill=white},
circle/.style={ellipse,  thick, minimum
width=2cm, minimum height=0.5cm,text centered, text width = 1.3cm, draw=black,
fill=white},
rectangle_white/.style={rectangle,  thick, minimum
width=2cm, minimum height=0.5cm,text centered, text width = 2.8cm, draw=white,
fill=white},
circle_large/.style={ellipse,  thick, minimum
width=2cm, minimum height=0.5cm,text centered, text width = 2.7cm, draw=black,
fill=white},
arrow/.style={ thick,->,>=stealth}}
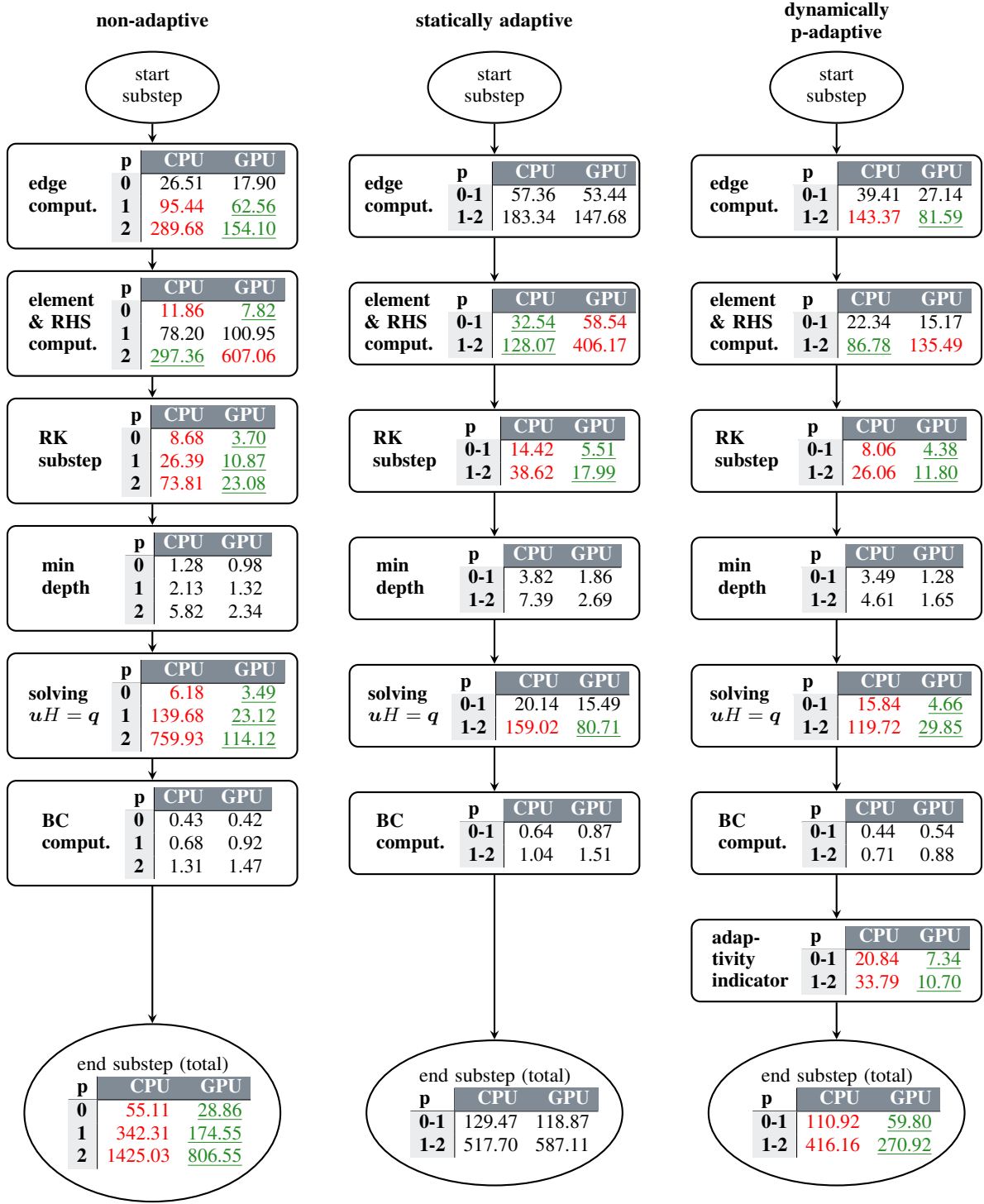
\begin{figure}[h!]
\centering
\begin{tikzpicture}[node distance = 2.05cm]
\footnotesize
	\node (headline) [rectangle_white] {\textbf{non-adaptive}};
	\node (start) [circle, below of=headline, yshift=1cm] {start substep};
    \node (edge) [c-rectangle2, below of=start, yshift=0.3cm] { \begin{tabular}{p{1.2cm}l|rr}
    \multirow{4}{\linewidth}{\textbf{edge comput.}} &\textbf{p} &\textcolor[HTML]{FFFFFF} 
     { \cellcolor[HTML]{808B96 }\textbf{CPU}} &  \cellcolor[HTML]{808B96 }\textcolor[HTML]{FFFFFF}{\textbf{GPU}} \\ \cline{2-4}
&\cellcolor[HTML]{EAECEE }\textbf{0} & 26.51 & 17.90\\ 

&\cellcolor[HTML]{EAECEE }\textbf{1} & \textcolor{red}{95.44} & \textcolor{forest_green}{\underline{62.56}}\\ 

&\cellcolor[HTML]{EAECEE }\textbf{2} & \textcolor{red}{289.68} & \textcolor{forest_green}{\underline{154.10}}\\ 
      \end{tabular}};
    \node (el_rhs) [c-rectangle2, below of=edge] { \begin{tabular}{p{1.2cm}l|rr}
    \multirow{4}{\linewidth}{\textbf{element \& RHS comput.}} &\textbf{p} &\textcolor[HTML]{FFFFFF} 
     { \cellcolor[HTML]{808B96 }\textbf{CPU}} &  \cellcolor[HTML]{808B96 }\textcolor[HTML]{FFFFFF}{\textbf{GPU}} \\ \cline{2-4}
&\cellcolor[HTML]{EAECEE }\textbf{0} & \textcolor{red}{11.86} & \textcolor{forest_green}{\underline{7.82}}\\ 

&\cellcolor[HTML]{EAECEE }\textbf{1} & 78.20 & 100.95\\ 

&\cellcolor[HTML]{EAECEE }\textbf{2} & \textcolor{forest_green}{\underline{297.36}} & \textcolor{red}{607.06}\\ 
 \end{tabular}};
    \node (rk) [c-rectangle2,below of=el_rhs] { \begin{tabular}{p{1.2cm}l|rr}
    \multirow{4}{\linewidth}{\textbf{RK substep}} &\textbf{p} & \textcolor[HTML]{FFFFFF} 
     { \cellcolor[HTML]{808B96 }\textbf{CPU}} &  \cellcolor[HTML]{808B96 }\textcolor[HTML]{FFFFFF}{\textbf{GPU}} \\ \cline{2-4}
&\cellcolor[HTML]{EAECEE }\textbf{0} & \textcolor{red}{8.68} & \textcolor{forest_green}{\underline{3.70}}\\ 

&\cellcolor[HTML]{EAECEE }\textbf{1} & \textcolor{red}{26.39} & \textcolor{forest_green}{\underline{10.87}}\\ 

&\cellcolor[HTML]{EAECEE }\textbf{2} & \textcolor{red}{73.81} & \textcolor{forest_green}{\underline{23.08}}\\  
 \end{tabular}};
    \node (min_depth) [c-rectangle2, below of=rk] { \begin{tabular}{p{1.2cm}l|rr}
    \multirow{4}{\linewidth}{\textbf{min depth}} & \textbf{p}  & \textcolor[HTML]{FFFFFF} 
     { \cellcolor[HTML]{808B96 }\textbf{CPU}} &  \cellcolor[HTML]{808B96 }\textcolor[HTML]{FFFFFF}{\textbf{GPU}} \\ \cline{2-4}
&\cellcolor[HTML]{EAECEE }\textbf{0} & 1.28 & 0.98\\ 

&\cellcolor[HTML]{EAECEE }\textbf{1} & 2.13 & 1.32\\ 

&\cellcolor[HTML]{EAECEE }\textbf{2} & 5.82 & 2.34\\
  \end{tabular}};
    \node (ctilde) [c-rectangle2, below of=min_depth] {\begin{tabular}{p{1.2cm}l|rr}
    \multirow{4}{\linewidth}{\textbf{solving $\boldsymbol{u} H = \boldsymbol{q}$}} & \textbf{p} & \textcolor[HTML]{FFFFFF} 
     { \cellcolor[HTML]{808B96 }\textbf{CPU}} &  \cellcolor[HTML]{808B96 }\textcolor[HTML]{FFFFFF}{\textbf{GPU}} \\ \cline{2-4}
&\cellcolor[HTML]{EAECEE }\textbf{0} & \textcolor{red}{6.18} & \textcolor{forest_green}{\underline{3.49}}\\ 

&\cellcolor[HTML]{EAECEE }\textbf{1} & \textcolor{red}{139.68} & \textcolor{forest_green}{\underline{23.12}}\\ 

&\cellcolor[HTML]{EAECEE }\textbf{2} & \textcolor{red}{759.93} & \textcolor{forest_green}{\underline{114.12}}\\ 
  \end{tabular}};
    \node (bc) [c-rectangle2, below of=ctilde] {\begin{tabular}{p{1.2cm}l|rr}
    \multirow{4}{\linewidth}{\textbf{BC comput.}} &  \textbf{p}  & \textcolor[HTML]{FFFFFF} 
     { \cellcolor[HTML]{808B96 }\textbf{CPU}} &  \cellcolor[HTML]{808B96 }\textcolor[HTML]{FFFFFF}{\textbf{GPU}} \\ \cline{2-4}
&\cellcolor[HTML]{EAECEE }\textbf{0} & 0.43 & 0.42\\ 

&\cellcolor[HTML]{EAECEE }\textbf{1} & 0.68 & 0.92\\ 

&\cellcolor[HTML]{EAECEE }\textbf{2} & 1.31 & 1.47\\   
  \end{tabular}};
    \node (end) [circle_large, below of=bc, yshift=-2.45cm] {end substep (total) \begin{tabular}{l|rr}
       \textbf{p}& \textcolor[HTML]{FFFFFF} 
     { \cellcolor[HTML]{808B96 }\textbf{CPU}} &  \cellcolor[HTML]{808B96 }\textcolor[HTML]{FFFFFF}{\textbf{GPU}} \\ \hline
\cellcolor[HTML]{EAECEE }\textbf{0} & \textcolor{red}{55.11} & \textcolor{forest_green}{\underline{28.86}}\\ 

\cellcolor[HTML]{EAECEE }\textbf{1} & \textcolor{red}{342.31} & \textcolor{forest_green}{\underline{174.55}}\\ 

\cellcolor[HTML]{EAECEE }\textbf{2} & \textcolor{red}{1425.03} & \textcolor{forest_green}{\underline{806.55}}\\   
  \end{tabular}};
    \draw[arrow] (start) -- (edge);
    \draw[arrow] (edge) -- (el_rhs);
    \draw[arrow] (el_rhs) -- (rk);
    \draw[arrow] (rk) -- (min_depth);
    \draw[arrow] (min_depth) -- (ctilde);
    \draw[arrow] (ctilde) -- (bc);
    \draw[arrow] (bc) -- (end);
    
	\node (headline2) [rectangle_white, xshift=5.5cm] {\textbf{statically adaptive}};
\node (start2) [circle, below of=headline2, yshift=1cm] {start substep };
    \node (edge2) [c-rectangle2, below of=start2, yshift=0.3cm] { \begin{tabular}{p{1.2cm}l|rr}
    \multirow{3}{\linewidth}{\textbf{edge comput.}} & \textbf{p} &\textcolor[HTML]{FFFFFF} 
     { \cellcolor[HTML]{808B96 }\textbf{CPU}} &  \cellcolor[HTML]{808B96 }\textcolor[HTML]{FFFFFF}{\textbf{GPU}} \\ \cline{2-4}
&\cellcolor[HTML]{EAECEE }\textbf{0-1} & 57.36 & 53.44\\ 

&\cellcolor[HTML]{EAECEE }\textbf{1-2} & 183.34 & 147.68\\  
      \end{tabular}};
    \node (el_rhs2) [c-rectangle2, below of=edge2] { \begin{tabular}{p{1.2cm}l|rr}
    \multirow{3}{\linewidth}{\textbf{element \& RHS comput.}} &\textbf{p} &\textcolor[HTML]{FFFFFF} 
     { \cellcolor[HTML]{808B96 }\textbf{CPU}} &  \cellcolor[HTML]{808B96 }\textcolor[HTML]{FFFFFF}{\textbf{GPU}} \\ \cline{2-4}
&\cellcolor[HTML]{EAECEE }\textbf{0-1} & \textcolor{forest_green}{\underline{32.54}} & \textcolor{red}{58.54}\\ 

&\cellcolor[HTML]{EAECEE }\textbf{1-2} & \textcolor{forest_green}{\underline{128.07}} & \textcolor{red}{406.17}\\ 
 \end{tabular}};
    \node (rk2) [c-rectangle2,below of=el_rhs2] { \begin{tabular}{p{1.2cm}l|rr}
    \multirow{3}{\linewidth}{\textbf{RK substep}} &\textbf{p} & \textcolor[HTML]{FFFFFF} 
     { \cellcolor[HTML]{808B96 }\textbf{CPU}} &  \cellcolor[HTML]{808B96 }\textcolor[HTML]{FFFFFF}{\textbf{GPU}} \\ \cline{2-4}
&\cellcolor[HTML]{EAECEE }\textbf{0-1} & \textcolor{red}{14.42} & \textcolor{forest_green}{\underline{5.51}}\\ 

&\cellcolor[HTML]{EAECEE }\textbf{1-2} & \textcolor{red}{38.62} & \textcolor{forest_green}{\underline{17.99}}\\ 
 \end{tabular}};
    \node (min_depth2) [c-rectangle2, below of=rk2] { \begin{tabular}{p{1.2cm}l|rr}
    \multirow{3}{\linewidth}{\textbf{min depth}} & \textbf{p}  & \textcolor[HTML]{FFFFFF} 
     { \cellcolor[HTML]{808B96 }\textbf{CPU}} &  \cellcolor[HTML]{808B96 }\textcolor[HTML]{FFFFFF}{\textbf{GPU}} \\ \cline{2-4}
&\cellcolor[HTML]{EAECEE }\textbf{0-1} & 3.82 & 1.86\\ 

&\cellcolor[HTML]{EAECEE }\textbf{1-2} & 7.39 & 2.69\\ 
  \end{tabular}};
    \node (ctilde2) [c-rectangle2, below of=min_depth2] {\begin{tabular}{p{1.2cm}l|rr}
    \multirow{3}{\linewidth}{\textbf{solving $\boldsymbol{u} H = \boldsymbol{q}$}} & \textbf{p} & \textcolor[HTML]{FFFFFF} 
     { \cellcolor[HTML]{808B96 }\textbf{CPU}} &  \cellcolor[HTML]{808B96 }\textcolor[HTML]{FFFFFF}{\textbf{GPU}} \\ \cline{2-4}
&\cellcolor[HTML]{EAECEE }\textbf{0-1} & 20.14 & 15.49\\ 

&\cellcolor[HTML]{EAECEE }\textbf{1-2} & \textcolor{red}{159.02} & \textcolor{forest_green}{\underline{80.71}}\\ 
  \end{tabular}};
    \node (bc2) [c-rectangle2, below of=ctilde2] {\begin{tabular}{p{1.2cm}l|rr}
    \multirow{3}{\linewidth}{\textbf{BC comput.}} &  \textbf{p}  & \textcolor[HTML]{FFFFFF} 
     { \cellcolor[HTML]{808B96 }\textbf{CPU}} &  \cellcolor[HTML]{808B96 }\textcolor[HTML]{FFFFFF}{\textbf{GPU}} \\ \cline{2-4}
&\cellcolor[HTML]{EAECEE }\textbf{0-1} & 0.64 & 0.87\\ 

&\cellcolor[HTML]{EAECEE }\textbf{1-2} & 1.04 & 1.51\\ 
  \end{tabular}};
    \node (end2) [circle_large, below of=bc2, yshift=-2.45cm] {end substep (total) \begin{tabular}{l|rr}
      \textbf{p} & \textcolor[HTML]{FFFFFF} 
     { \cellcolor[HTML]{808B96 }\textbf{CPU}} &  \cellcolor[HTML]{808B96 }\textcolor[HTML]{FFFFFF}{\textbf{GPU}} \\ \hline
\cellcolor[HTML]{EAECEE }\textbf{0-1} & 129.47 & 118.87\\ 

\cellcolor[HTML]{EAECEE }\textbf{1-2} & 517.70 & 587.11\\  
  \end{tabular}};
    \draw[arrow] (start2) -- (edge2);
    \draw[arrow] (edge2) -- (el_rhs2);
    \draw[arrow] (el_rhs2) -- (rk2);
    \draw[arrow] (rk2) -- (min_depth2);
    \draw[arrow] (min_depth2) -- (ctilde2);
    \draw[arrow] (ctilde2) -- (bc2);
    \draw[arrow] (bc2) -- (end2);

\node (headline3) [rectangle_white, xshift=11cm] {\textbf{dynamically p-adaptive}};
\node (start3) [circle, below of=headline3, yshift=1cm] {start substep };
    \node (edge3) [c-rectangle2, below of=start3, yshift=0.3cm] { \begin{tabular}{p{1.2cm}l|rr}
    \multirow{3}{\linewidth}{\textbf{edge comput.}} & \textbf{p} &\textcolor[HTML]{FFFFFF} 
     { \cellcolor[HTML]{808B96 }\textbf{CPU}} &  \cellcolor[HTML]{808B96 }\textcolor[HTML]{FFFFFF}{\textbf{GPU}} \\ \cline{2-4}
&\cellcolor[HTML]{EAECEE }\textbf{0-1} & 39.41 & 27.14\\ 

&\cellcolor[HTML]{EAECEE }\textbf{1-2} & \textcolor{red}{143.37} & \textcolor{forest_green}{\underline{81.59}}\\  
      \end{tabular}};
    \node (el_rhs3) [c-rectangle2, below of=edge3] { \begin{tabular}{p{1.2cm}l|rr}
    \multirow{3}{\linewidth}{\textbf{element \& RHS comput.}} &\textbf{p} &\textcolor[HTML]{FFFFFF} 
     { \cellcolor[HTML]{808B96 }\textbf{CPU}} &  \cellcolor[HTML]{808B96 }\textcolor[HTML]{FFFFFF}{\textbf{GPU}} \\ \cline{2-4}
&\cellcolor[HTML]{EAECEE }\textbf{0-1} & 22.34 & 15.17\\ 

&\cellcolor[HTML]{EAECEE }\textbf{1-2} & \textcolor{forest_green}{\underline{86.78}} & \textcolor{red}{135.49}\\ 
 \end{tabular}};
    \node (rk3) [c-rectangle2,below of=el_rhs3] { \begin{tabular}{p{1.2cm}l|rr}
    \multirow{3}{\linewidth}{\textbf{RK substep}} &\textbf{p} & \textcolor[HTML]{FFFFFF} 
     { \cellcolor[HTML]{808B96 }\textbf{CPU}} &  \cellcolor[HTML]{808B96 }\textcolor[HTML]{FFFFFF}{\textbf{GPU}} \\ \cline{2-4}
&\cellcolor[HTML]{EAECEE }\textbf{0-1} & \textcolor{red}{8.06} & \textcolor{forest_green}{\underline{4.38}}\\ 

&\cellcolor[HTML]{EAECEE }\textbf{1-2} & \textcolor{red}{26.06} & \textcolor{forest_green}{\underline{11.80}}\\
 \end{tabular}};
    \node (min_depth3) [c-rectangle2, below of=rk3] { \begin{tabular}{p{1.2cm}l|rr}
    \multirow{3}{\linewidth}{\textbf{min depth}} & \textbf{p}  & \textcolor[HTML]{FFFFFF} 
     { \cellcolor[HTML]{808B96 }\textbf{CPU}} &  \cellcolor[HTML]{808B96 }\textcolor[HTML]{FFFFFF}{\textbf{GPU}} \\ \cline{2-4}
&\cellcolor[HTML]{EAECEE }\textbf{0-1} & 3.49 & 1.28\\ 

&\cellcolor[HTML]{EAECEE }\textbf{1-2} & 4.61 & 1.65\\ 
  \end{tabular}};
    \node (ctilde3) [c-rectangle2, below of=min_depth3] {\begin{tabular}{p{1.2cm}l|rr}
    \multirow{3}{\linewidth}{\textbf{solving $\boldsymbol{u} H = \boldsymbol{q}$}} & \textbf{p} & \textcolor[HTML]{FFFFFF} 
     { \cellcolor[HTML]{808B96 }\textbf{CPU}} &  \cellcolor[HTML]{808B96 }\textcolor[HTML]{FFFFFF}{\textbf{GPU}} \\ \cline{2-4}
&\cellcolor[HTML]{EAECEE }\textbf{0-1} & \textcolor{red}{15.84} & \textcolor{forest_green}{\underline{4.66}}\\ 

&\cellcolor[HTML]{EAECEE }\textbf{1-2} & \textcolor{red}{119.72} & \textcolor{forest_green}{\underline{29.85}}\\ 
  \end{tabular}};
    \node (bc3) [c-rectangle2, below of=ctilde3] {\begin{tabular}{p{1.2cm}l|rr}
    \multirow{3}{\linewidth}{\textbf{BC comput.}} &  \textbf{p}  & \textcolor[HTML]{FFFFFF} 
     { \cellcolor[HTML]{808B96 }\textbf{CPU}} &  \cellcolor[HTML]{808B96 }\textcolor[HTML]{FFFFFF}{\textbf{GPU}} \\ \cline{2-4}
&\cellcolor[HTML]{EAECEE }\textbf{0-1} & 0.44 & 0.54\\ 

&\cellcolor[HTML]{EAECEE }\textbf{1-2} & 0.71 & 0.88\\ 

  \end{tabular}};
    \node (ind3) [c-rectangle2, below of=bc3] {\begin{tabular}{p{1.3cm}l|rr}
    \multirow{3}{\linewidth}{\textbf{adap\-tivity indicator}} &  \textbf{p}  & \textcolor[HTML]{FFFFFF} 
     { \cellcolor[HTML]{808B96 }\textbf{CPU}} &  \cellcolor[HTML]{808B96 }\textcolor[HTML]{FFFFFF}{\textbf{GPU}} \\ \cline{2-4}
&\cellcolor[HTML]{EAECEE }\textbf{0-1} & \textcolor{red}{20.84} & \textcolor{forest_green}{\underline{7.34}}\\ 

&\cellcolor[HTML]{EAECEE }\textbf{1-2} & \textcolor{red}{33.79} & \textcolor{forest_green}{\underline{10.70}}\\ 

  \end{tabular}};
    \node (end3) [circle_large, below of=ind3, yshift=-0.4cm] {end substep (total) \begin{tabular}{l|rr}
      \textbf{p} & \textcolor[HTML]{FFFFFF} 
     { \cellcolor[HTML]{808B96 }\textbf{CPU}} &  \cellcolor[HTML]{808B96 }\textcolor[HTML]{FFFFFF}{\textbf{GPU}} \\ \hline
\cellcolor[HTML]{EAECEE }\textbf{0-1} & \textcolor{red}{110.92} & \textcolor{forest_green}{\underline{59.80}}\\ 

\cellcolor[HTML]{EAECEE }\textbf{1-2} & \textcolor{red}{416.16} & \textcolor{forest_green}{\underline{270.92}}\\  
  \end{tabular}};
    \draw[arrow] (start3) -- (edge3);
    \draw[arrow] (edge3) -- (el_rhs3);
    \draw[arrow] (el_rhs3) -- (rk3);
    \draw[arrow] (rk3) -- (min_depth3);
    \draw[arrow] (min_depth3) -- (ctilde3);
    \draw[arrow] (ctilde3) -- (bc3);
    \draw[arrow] (bc3) -- (ind3);
    \draw[arrow] (ind3) -- (end3);
\end{tikzpicture}
\caption{Radial dam break: Data flow and kernel execution times (in ms) on the ARM-AGX platform for the unseparated setup. Piecewise constant, linear, and quadratic solutions (left), statically adaptive p0-1 and p1-2 solutions (middle), dynamically p-adaptive p0-1 and p1-2 solutions (right). We highlight significantly faster execution times (green, underlined) with a difference of more than 1/3 with respect to the slower ones (red).}
\label{flow_non-separated}
\end{figure}
\
Next, we turn on our new separation approach and consider the execution times of all kernels on the CPU and GPU (see Fig.~\ref{flow_adaptive-separated} "homog." rows). To exploit further parallelism, we utilize these results to distribute the kernels of the separated algorithm between the GPU and CPU (see Fig.~\ref{flow_adaptive-separated} "heterog." rows). The resulting optimal distribution assigns the correction computation to the CPU, whereas the base computation and the remaining kernels, except for the BC computation, are run on the GPU. This leads to approx. 22\,\% speedup compared to the fastest pure (on either CPU or GPU) separated computation and is approx. 11\,\% faster than the fastest unseparated one.

\setlength{\tabcolsep}{3pt}
\tikzset{c-rectangle2/.style={rectangle,  thick, rounded corners, minimum
width=2cm, minimum height=1cm,text centered, text width = 4cm, draw=black,
fill=white},
circle/.style={ellipse,  thick, minimum
width=2cm, minimum height=0.5cm,text centered, text width = 1.7cm, draw=black,
fill=white},
circle_large/.style={ellipse,  thick, minimum
width=2cm, minimum height=0.5cm,text centered, text width = 3.6cm, draw=black,
fill=white},
arrow/.style={ thick,->,>=stealth}}
\begin{figure}[b!]
\centering
    \centering
\begin{tikzpicture}[node distance = 2.8cm]
\footnotesize
	\node (start) [circle] {start substep };
    \node (edge_base) [c-rectangle2, below of=start, xshift=-6.6cm, yshift=0.7cm] {\textbf{edge base computation}\\\begin{tabular}{llrr} \hline
       \textbf{p} & \textbf{distrib.} & \textcolor[HTML]{FFFFFF} 
     {\cellcolor[HTML]{808B96 }\textbf{CPU}} &  \cellcolor[HTML]{808B96 }\textcolor[HTML]{FFFFFF}{\textbf{GPU}} \\ \hline
\multirow{2}{*}{\textbf{0-1}} &homog. & \textcolor{red}{21.19} & \textcolor{forest_green}{\underline{17.89}}\\ 

&heterog. &  - & 18.30\\ 
\hline 

\multirow{2}{*}{\textbf{1-2}} &homog. & \textcolor{red}{87.58} & \textcolor{forest_green}{\underline{62.57}}\\ 

&heterog. &  - & 63.05\\ 

\end{tabular}};
    \node (edge_cor) [c-rectangle2, below of=start, xshift=-2.2cm, yshift=0.7cm] {\textbf{edge correction computation}\\\begin{tabular}{llrr}
    \hline
       \textbf{p} & \textbf{distrib.} & \textcolor[HTML]{FFFFFF} 
     {\cellcolor[HTML]{808B96 }\textbf{CPU}} &  \cellcolor[HTML]{808B96 }\textcolor[HTML]{FFFFFF}{\textbf{GPU}} \\ \hline
\multirow{2}{*}{\textbf{0-1}} &homog. & \textcolor{forest_green}{\underline{43.52}} & \textcolor{red}{50.92}\\ 

&heterog. & 64.19 & - \\ 
\hline 

\multirow{2}{*}{\textbf{1-2}} &homog. & \textcolor{forest_green}{\underline{88.32}} & \textcolor{red}{127.89}\\  

&heterog. & 155.37 & - \\ 
   \end{tabular}};
     \node (el_rhs_base) [c-rectangle2, below of=start, xshift=2.2cm, yshift=0.7cm] {\textbf{element \& RHS base comp.}\\\begin{tabular}{llrr}
     \hline
       \textbf{p} & \textbf{distrib.} & \textcolor[HTML]{FFFFFF} 
     {\cellcolor[HTML]{808B96 }\textbf{CPU}} &  \cellcolor[HTML]{808B96 }\textcolor[HTML]{FFFFFF}{\textbf{GPU}} \\ \hline
\multirow{2}{*}{\textbf{0-1}} &homog. & \textcolor{red}{9.18} & \textcolor{forest_green}{\underline{6.64}}\\ 

&heterog. &  - & 6.66\\ 
\hline 

\multirow{2}{*}{\textbf{1-2}} &homog. & \textcolor{forest_green}{\underline{44.11}} & \textcolor{red}{97.81}\\ 

&heterog. &  - & 97.85\\ 
\end{tabular}};
    \node (el_rhs_cor) [c-rectangle2, below of=start, xshift=6.6cm, yshift=0.7cm] {\textbf{element \& RHS \\ correction computation}\\\begin{tabular}{llrr}
    \hline
       \textbf{p} & \textbf{distrib.} & \textcolor[HTML]{FFFFFF} 
     {\cellcolor[HTML]{808B96 }\textbf{CPU}} &  \cellcolor[HTML]{808B96 }\textcolor[HTML]{FFFFFF}{\textbf{GPU}} \\ \hline
\multirow{2}{*}{\textbf{0-1}} &homog. & \textcolor{forest_green}{\underline{14.41}} & \textcolor{red}{48.45}\\ 

&heterog. & 20.04 & - \\ 
\hline 

\multirow{2}{*}{\textbf{1-2}} &homog. & \textcolor{forest_green}{\underline{42.34}} & \textcolor{red}{340.42}\\ 

&heterog. & 51.11 & - \\
   \end{tabular}};
    \node (rk_add) [c-rectangle2,below of=start,yshift=-2cm ] {\textbf{RK substep \& additions}\\\begin{tabular}{llrr}
    \hline
       \textbf{p} & \textbf{distrib.} & \textcolor[HTML]{FFFFFF} 
     {\cellcolor[HTML]{808B96 }\textbf{CPU}} &  \cellcolor[HTML]{808B96 }\textcolor[HTML]{FFFFFF}{\textbf{GPU}} \\ \hline
\multirow{2}{*}{\textbf{0-1}} &homog. & \textcolor{red}{56.20} & \textcolor{forest_green}{\underline{17.95}}\\ 

&heterog. &  - & 19.22\\ 
\hline 

\multirow{2}{*}{\textbf{1-2}} &homog. & \textcolor{red}{168.07} & \textcolor{forest_green}{\underline{60.23}}\\ 

&heterog. &  - & 61.28\\ 
\end{tabular}};
    \node (min_depth) [c-rectangle2, below of=rk_add] {\textbf{min depth}\\\begin{tabular}{llrr}
    \hline
       \textbf{p} & \textbf{distrib.} & \textcolor[HTML]{FFFFFF} 
     {\cellcolor[HTML]{808B96 }\textbf{CPU}} &  \cellcolor[HTML]{808B96 }\textcolor[HTML]{FFFFFF}{\textbf{GPU}} \\ \hline
\multirow{2}{*}{\textbf{0-1}} &homog. & \textcolor{red}{3.49} & \textcolor{forest_green}{\underline{1.84}}\\ 

&heterog. &  - & 1.85\\ 
\hline 

\multirow{2}{*}{\textbf{1-2}} &homog. & \textcolor{red}{7.02} & \textcolor{forest_green}{\underline{2.69}}\\ 

&heterog. &  - & 2.68\\ 
\end{tabular}};
    \node (ctilde) [c-rectangle2, below of=min_depth] {\textbf{solving $\boldsymbol{u} H = \boldsymbol{q}$}\\\begin{tabular}{llrr}
    \hline
       \textbf{p} & \textbf{distrib.} & \textcolor[HTML]{FFFFFF} 
     {\cellcolor[HTML]{808B96 }\textbf{CPU}} &  \cellcolor[HTML]{808B96 }\textcolor[HTML]{FFFFFF}{\textbf{GPU}} \\ \hline
\multirow{2}{*}{\textbf{0-1}} &homog. & \textcolor{red}{21.70} & \textcolor{forest_green}{\underline{15.47}}\\ 

&heterog. &  - & 15.42\\ 
\hline 

\multirow{2}{*}{\textbf{1-2}} &homog. & \textcolor{red}{150.02} & \textcolor{forest_green}{\underline{80.70}}\\ 

&heterog. &  - & 80.73\\ 
   \end{tabular}};
    \node (bc_tot) [c-rectangle2, below of=ctilde] {\textbf{BC computation}\\\begin{tabular}{llrr}
    \hline
       \textbf{p} & \textbf{distrib.} & \textcolor[HTML]{FFFFFF} 
     {\cellcolor[HTML]{808B96 }\textbf{CPU}} &  \cellcolor[HTML]{808B96 }\textcolor[HTML]{FFFFFF}{\textbf{GPU}} \\ \hline
\multirow{2}{*}{\textbf{0-1}} &homog. & \textcolor{forest_green}{\underline{0.62}} & \textcolor{red}{0.77}\\ 

&heterog. & 0.98 & - \\ 
\hline 

\multirow{2}{*}{\textbf{1-2}} &homog. & \textcolor{forest_green}{\underline{1.09}} & \textcolor{red}{1.44}\\ 

&heterog. & 1.94 & - \\ 
\end{tabular}};
    \node (end) [circle_large, below of=bc_tot, yshift=-0.4cm] {end substep (total) \begin{tabular}{llrr}
      \textbf{p} & & \textcolor[HTML]{FFFFFF} 
     { \cellcolor[HTML]{808B96 }\textbf{CPU}} &  \cellcolor[HTML]{808B96 }\textcolor[HTML]{FFFFFF}{\textbf{GPU}} \\ \hline
\multirow{2}{*}{\textbf{0-1}} &homog. & \textcolor{red}{170.86} & 144.55\\ 

&heterog. & \multicolumn{2}{c}{\textcolor{forest_green}{\underline{127.40}}} \\ 
\hline 

\multirow{2}{*}{\textbf{1-2}} &homog. & 588.95 & \textcolor{red}{714.25}\\ 

&heterog. & \multicolumn{2}{c}{\textcolor{forest_green}{\underline{460.01}}}\\ 
  \end{tabular}};
    \draw[arrow] (start) -| (edge_base);
    \draw[arrow] (start) -- (edge_cor);
    \draw[arrow] (start) -- (el_rhs_base);
    \draw[arrow] (start) -| (el_rhs_cor);
    \draw[arrow] (edge_base) |- (rk_add);
    \draw[arrow] (edge_cor) -- (rk_add);
    \draw[arrow] (el_rhs_base) -- (rk_add);
    \draw[arrow] (el_rhs_cor) |- (rk_add);
    \draw[arrow] (rk_add) -- (min_depth);
    \draw[arrow] (min_depth) -- (ctilde);
    \draw[arrow] (ctilde) -- (bc_tot);
    \draw[arrow] (bc_tot) -- (end);
\end{tikzpicture}
\caption{Radial dam break: Data flow and kernel execution times (in ms) on the ARM-AGX platform for the separated statically adaptive p0-1 and p1-2 solution. The faster and slower execution times are highlighted in green (underlined) and red, respectively, to substantiate the decision on the heterogeneous kernel distribution.}
\label{flow_adaptive-separated}
\end{figure}
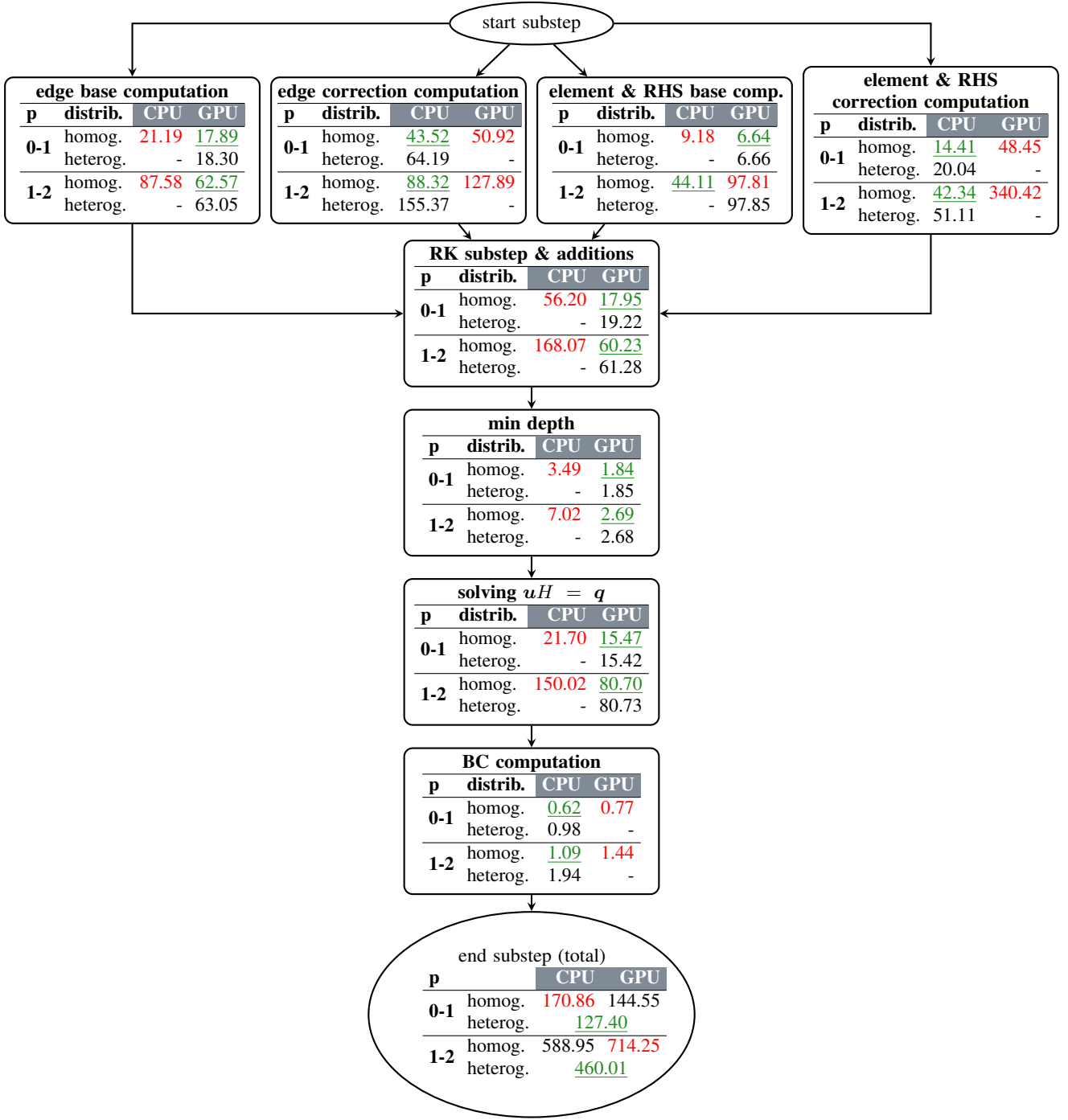

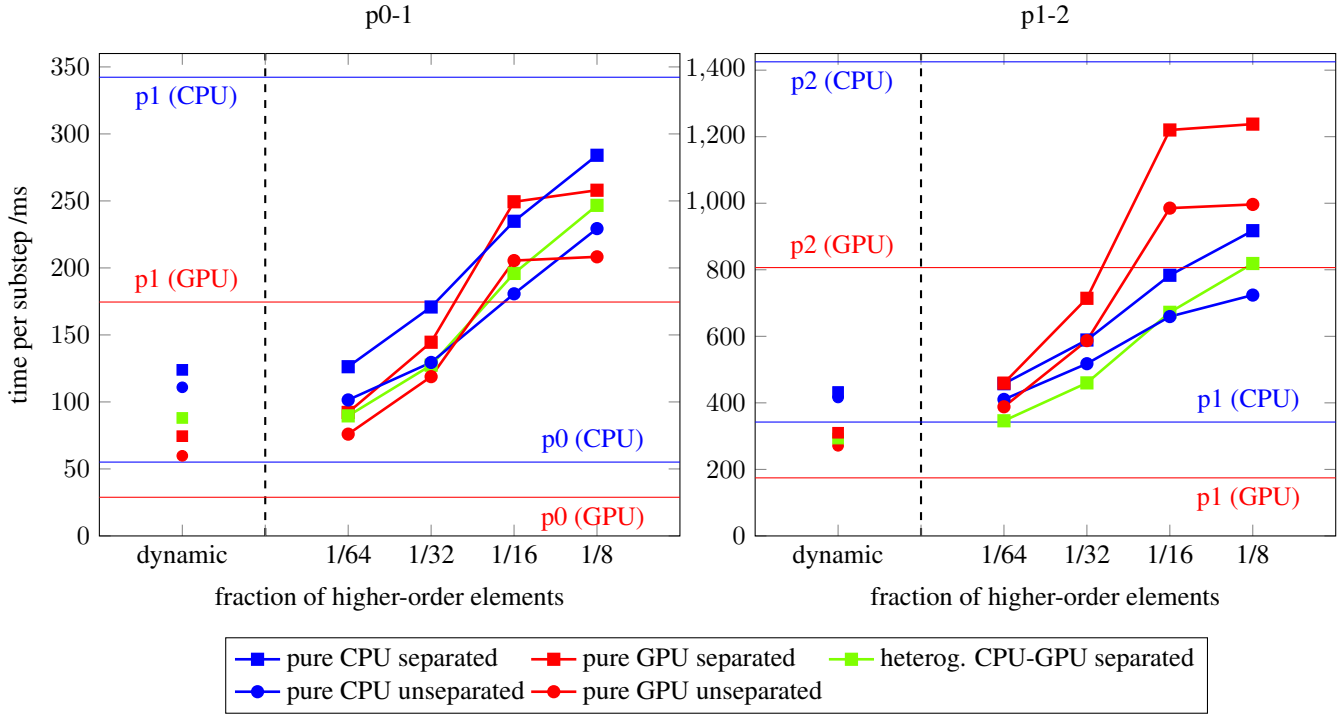
\begin{figure}[h!]
\centering
\pgfplotsset{compat=newest, width=0.53\columnwidth}
\begin{tikzpicture}
\begin{axis}[
name=ax1,
height = 8cm,
width = 9.3cm,
title= {p0-1},
symbolic x coords={0,dynamic,1/256,1/64,1/32,1/16, 1/8,1},
xtick={dynamic,1/64,1/32,1/16, 1/8},
extra x ticks={1/256},
extra tick style={grid=major,major grid style={black,thick, dashed}},
ymin =0,
ymax =360,
ytick={0, 50, 100, 150, 200, 250, 300, 350},
xmin =0,
xmax=1,
ylabel={time per substep /ms },
xlabel={fraction of higher-order elements},
extra x tick labels=\raisebox{-5pt}{}
]

\addplot[red,mark=square*, line width=1pt] coordinates {
(1/64,92.18) (1/32,144.55) (1/16,249.37) (1/8,257.98) };

\addplot[lime,mark=square*, line width=1pt] coordinates {
(1/64,89.60) (1/32,127.40) (1/16,195.91) (1/8,246.70) };

\addplot[blue,mark=square*, line width=1pt] coordinates {
(1/64,126.28) (1/32,170.86) (1/16,234.83) (1/8,284.07) };

\addplot[blue,mark=*, line width=1pt] coordinates {
(1/64,101.50) (1/32,129.47) (1/16,180.69) (1/8,229.35) };

\addplot[red,mark=*, line width=1pt] coordinates {
(1/64,75.97) (1/32,118.87) (1/16,205.49) (1/8,208.29) };

\draw [draw=blue]   (axis cs: 0,55.11) -- (axis cs: 1,55.11)
        node[pos=0.85, above] {\color{blue}{p0 (CPU)}};
\draw [draw=red]   (axis cs: 0,28.86) -- (axis cs: 1,28.86)
        node[pos=0.85, below] {\color{red}{p0 (GPU)}};
\draw [draw=blue]   (axis cs: 0,342.31) -- (axis cs: 1,342.31)
        node[pos=0.15, below] {\color{blue}{p1 (CPU)}};
\draw [draw=red]   (axis cs: 0,174.55) -- (axis cs: 1,174.55)
        node[pos=0.15, above] {\color{red}{p1 (GPU)}};

\addplot[blue,mark=*] coordinates { (dynamic,110.92)};
\addplot[red,mark=*] coordinates { (dynamic,59.8)};
\addplot[lime,mark=square*] coordinates { (dynamic,88.05)};
\addplot[blue,mark=square*] coordinates { (dynamic,123.94)};
\addplot[red,mark=square*] coordinates { (dynamic,74.44)};
\end{axis}
\begin{axis}[
at={(ax1.south east)},
xshift=1cm,
height = 8cm,
width = 9.3cm,
title= {p1-2},
symbolic x coords={0,dynamic,1/128,1/64,1/32,1/16, 1/8,1},
xtick={dynamic,1/64,1/32,1/16, 1/8},
extra x ticks={1/128},
extra tick style={grid=major,major grid style={black,thick, dashed}},
ymin =0,
ymax =1450,
ytick={0, 200, 400, 600, 800, 1000, 1200,1400},
xmin =0,
xmax=1,
legend cell align=left,
xlabel={fraction of higher-order elements},
extra x tick labels=\raisebox{-5pt}{},
legend style={at={(0.78,-0.21)},anchor=north east,legend columns=3}
]

\addplot[blue,mark=square*, line width=1pt] coordinates {
(1/64,457.52) (1/32,588.95) (1/16,783.51) (1/8,917.53) };

\addplot[red,mark=square*, line width=1pt] coordinates {
(1/64,459.34) (1/32,714.25) (1/16,1220.16) (1/8,1237.78) };

\addplot[lime,mark=square*, line width=1pt] coordinates {
(1/64,346.12) (1/32,460.01) (1/16,672.17) (1/8,818.82) };

\addplot[blue,mark=*, line width=1pt] coordinates {
(1/64,410.31) (1/32,517.70) (1/16,659.38) (1/8,723.95) };

\addplot[red,mark=*, line width=1pt] coordinates {
(1/64,388.20) (1/32,587.11) (1/16,985.30) (1/8,996.37) };

\draw [draw=blue]   (axis cs: 0,342.31) -- (axis cs: 1,342.31)
        node[pos=0.85, above] {\color{blue}{p1 (CPU)}};
\draw [draw=red]   (axis cs: 0,174.55) -- (axis cs: 1,174.55)
        node[pos=0.85, below] {\color{red}{p1 (GPU)}};
\draw [draw=blue]   (axis cs: 0,1425.03) -- (axis cs: 1,1425.03)
        node[pos=0.15, below] {\color{blue}{p2 (CPU)}};
\draw [draw=red]   (axis cs: 0,806.55) -- (axis cs: 1,806.55)
        node[pos=0.15, above] {\color{red}{p2 (GPU)}};

\addplot[blue,mark=*] coordinates { (dynamic,416.16)};
\addplot[red,mark=*] coordinates { (dynamic,270.92)};
\addplot[lime,mark=square*] coordinates { (dynamic,292.65)};
\addplot[blue,mark=square*] coordinates { (dynamic,432.64)};
\addplot[red,mark=square*] coordinates { (dynamic,310.16)};
\legend{pure CPU separated,pure GPU separated,heterog. CPU-GPU separated,pure CPU unseparated,pure GPU unseparated}
\end{axis}
\end{tikzpicture}
\captionof{figure}{Radial dam break: ARM-AGX total execution time for non-adaptive, statically adaptive with different fractions of higher-order elements, and dynamically p-adaptive setups. For p0-1, in the latter case, on average (over the whole simulation) approx. 1/482 of the elements use the higher order and, for p1-2, the average fraction of higher-order elements is 1/172. The horizontal lines mark the non-adaptive (p0, p1, and p2) execution times. Constant-linear (left) and linear-quadratic (right) approximation.}
     \label{fig:line_plot}
\end{figure}

In Fig.~\ref{fig:line_plot}, we show the substep execution times for statically adaptive scenarios with different ratios of higher-order elements, which are fixed during the run. Here, we compare the unseparated and separated schemes run on the CPU, the GPU, or with the optimal heterogeneous distribution of kernels between the CPU and GPU. First, the overhead caused by separation is now easy to quantify in each configuration. Furthermore, one can conclude that the CPU clearly outperforms the GPU if approx. 1/32 or more elements use the higher-order approximation, whereas the GPU is faster if the fraction of higher-order elements is small. Additionally, we observe that all adaptive computations are faster than the corresponding non-adaptive higher-order computations on the CPU. For the non-adaptive GPU execution times, this is also the case as long as 1/32 or fewer elements use the higher order. 
In the heterogeneous case, for p1-2 for some fractions of higher-order elements (i.e., 1/32 and 1/64), we achieve more than 10\,\% speedup compared to the fastest homogeneous version.

The values left of the vertical dashed line in Fig.~\ref{fig:line_plot} show the substep execution times for the dynamically p-adaptive case (cf., Fig. \ref{fig:dambreak}) with the solution accuracy enforced to be similar to that of the full higher-order solution, cf.\ \citep{FaghihNainiA2022}. For p0-1, on average about 1/482 of the elements use the higher order and for p1-2, the portion is 1/172. Since the dynamically p-adaptive runs are more than twice as fast as the higher-order runs (p1 in Fig.~\ref{fig:line_plot} (left) and p2 in Fig.~\ref{fig:line_plot} (right)), these measurements confirm the advantages of p-adaptivity in general. The heterogeneous version for p1-2 is faster than the separated homogeneous ones but not faster than the unseparated GPU version. When comparing the p0-1 to the p1-2 versions, one must note that the performance difference between constant and linear computations is smaller than between linear and quadratic computations. Additionally, the overhead caused by separating the element and edge computations is, in some cases, so significant that it cannot be compensated by distributing the kernels and doing the computations in parallel. Here, flexible code generation can be used to the best advantage by easily generating configurations which provide the best performance depending on the problem setup and the hardware configuration.

Achieving efficient heterogeneous kernel distribution is made possible due to the CPU and the GPU sharing the memory on our ARM-AGX SoC architecture. For the conventional hardware with a discrete GPU (AMD-RTX in this case), memory transfers between the CPU and GPU are a significant bottleneck difficult to amortize by any performance benefits arising from heterogeneous kernel parallelism. For the separated versions and p0-1, we get reasonable substep execution times of 24.9\,ms on the CPU and 16.1 ms on the GPU but 101.0\,ms for the heterogeneous setup. This is similar for the p1-2 case where we get 68.1\,ms, 72.4\,ms, and 222.2\,ms, respectively. For detailed execution times on the AMD-RTX architecture, we refer the reader to the last column of Tab.~\ref{tab:timings} in Appendix~\ref{sec:app_A}.

\subsection{Tidal flow at Bahamas with water hump}
\label{bahamas}
\begin{figure}[h!]
\centering
     \includegraphics[width=0.30\textwidth, trim=220 100 760 240, clip]{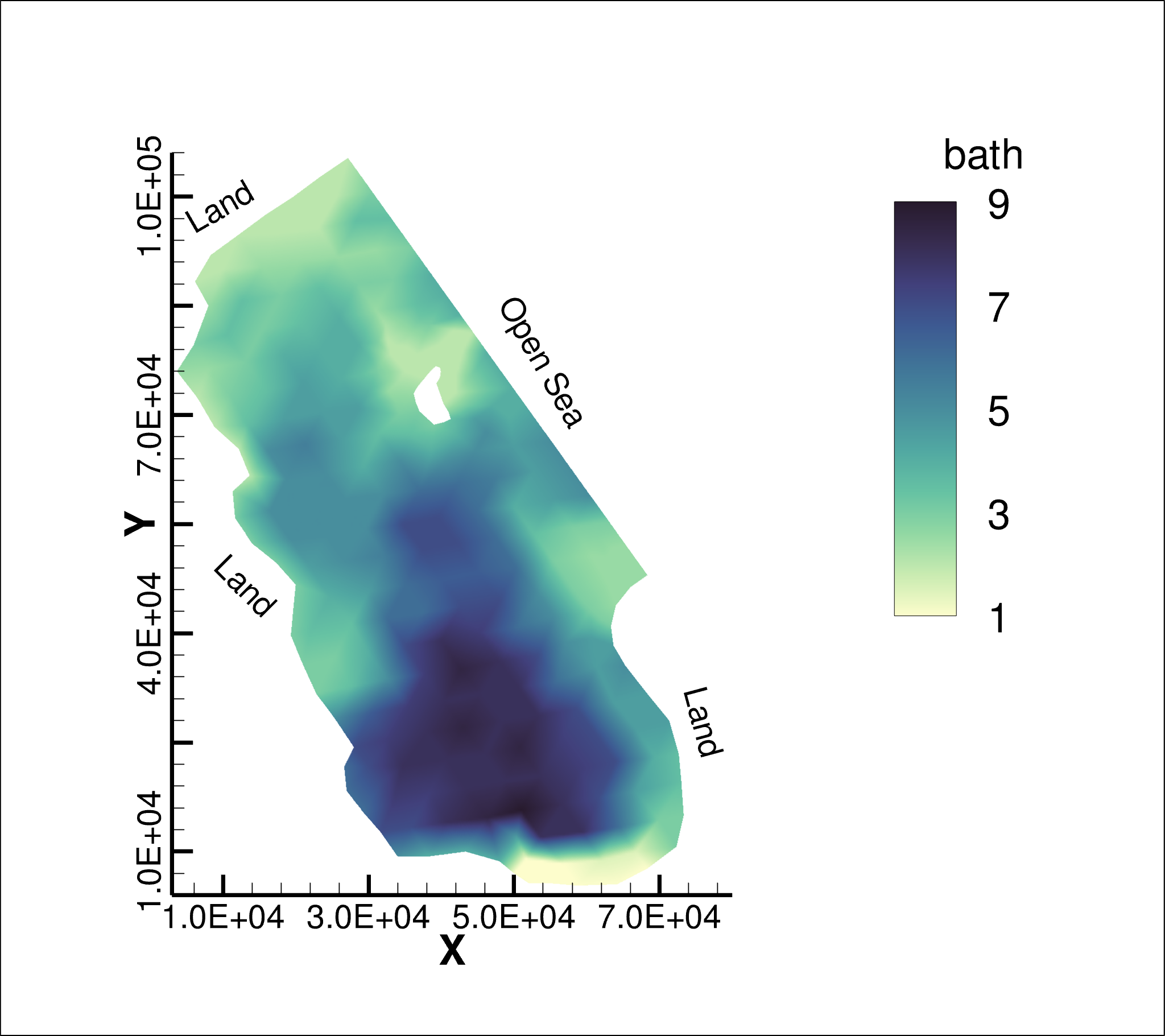}
                    \includegraphics[width=0.065\textwidth, trim=1570 200 230 240, clip]{figures/bahamas/bahamas_bath.png}
          \includegraphics[width=0.60\textwidth, trim=5 180 5 70, clip]{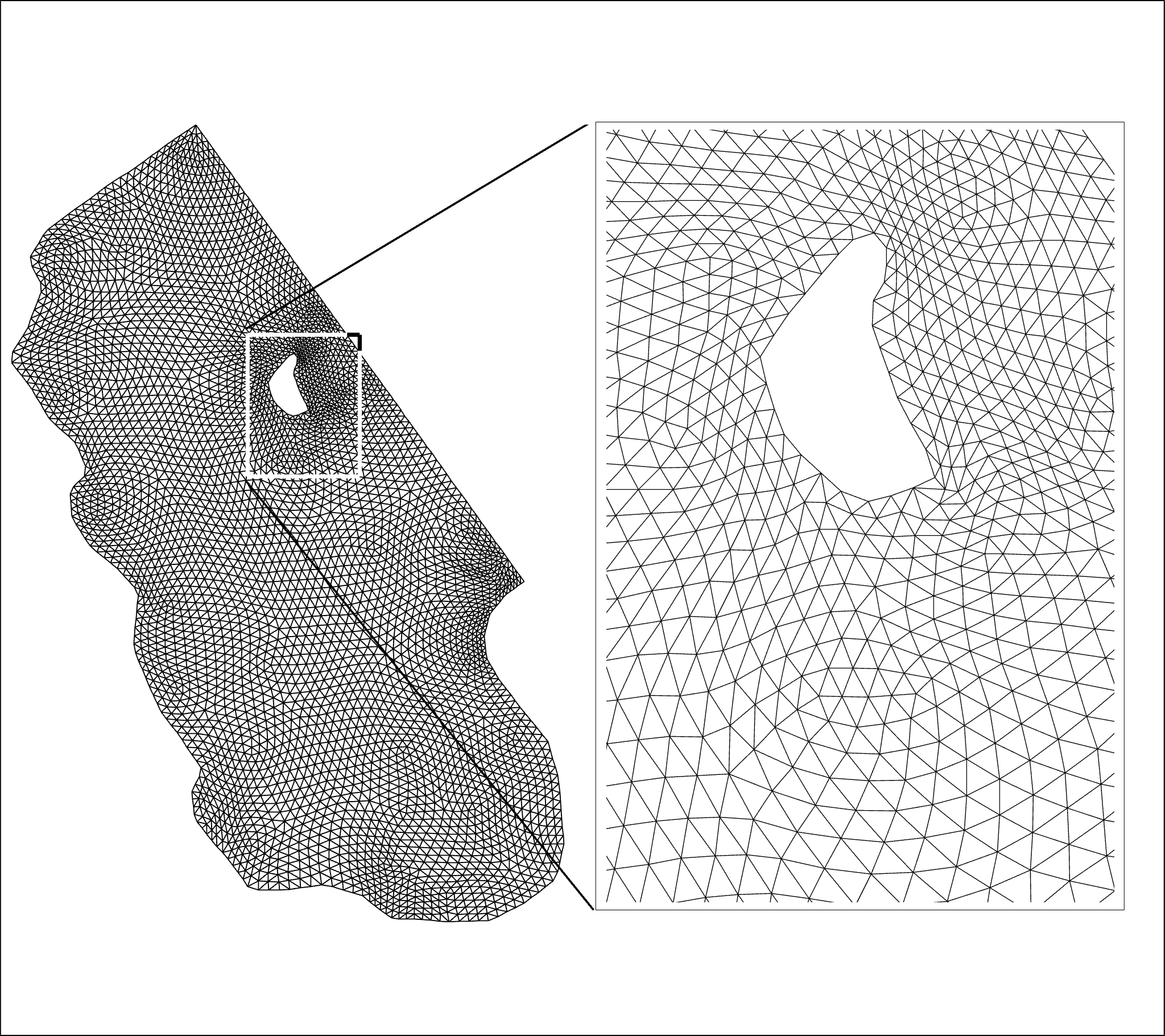}
  \caption{Tidal flow at Bahamas: Bathymetry (left) and block-structured grid with 256 blocks of 32 elements each. The grid used for the computations was uniformly refined four times, i.e., contains 8192 elements per block. All units are meters.}
  \label{fig:bahamas_bath_grid}
\end{figure}

Next, we consider a~tide-driven flow scenario in the Bight of Abaco (Bahamas). The simulations were started from the lake-at-rest initial conditions with an added water column of 2 meters height -- a prototypical tsunami simulation without wetting and drying -- and run for 50 minutes driven by the tidal surface elevation at the open sea boundary. 
We imposed no normal flow boundary conditions at the land boundaries (see~\citep{FaghihNainiKAZGK2020} for more details on the tidal problem setup). 
Fig.~\ref{fig:bahamas_bath_grid} shows the bathymetry (left) and the block-structured grid (BSG). 
The displayed BSG contains 256 blocks with only 32 elements each for better visualization. In the computations, a four times uniformly refined (via bisecting each edge) BSG with 8192 elements per block was used, that is, the total number of elements was the same as in the uniform dam break examples. The simulations were run for 12\,000 time steps with $\Delta t=0.25\,s$.

\begin{figure}[h!]
\centering
     \includegraphics[width=0.32\textwidth, trim=150 190 210 180, clip]{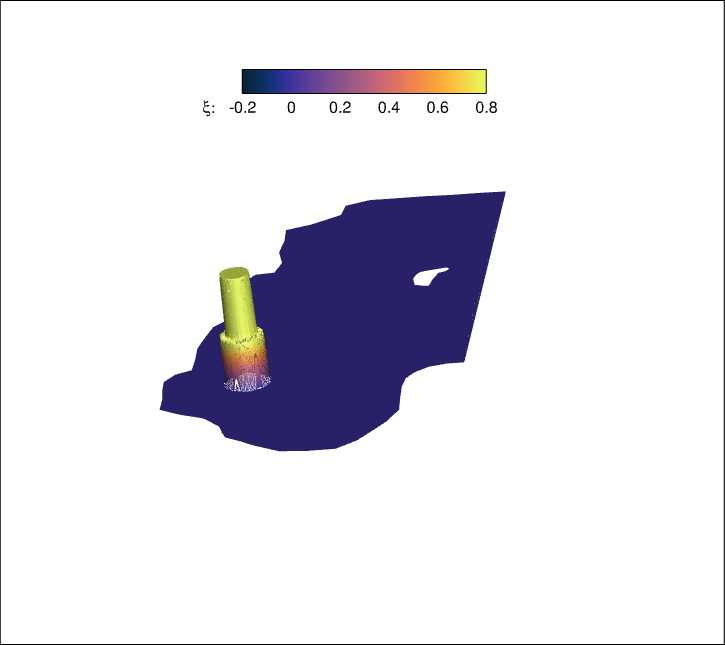}
      \includegraphics[width=0.32\textwidth, trim=150 190 210 180, clip]{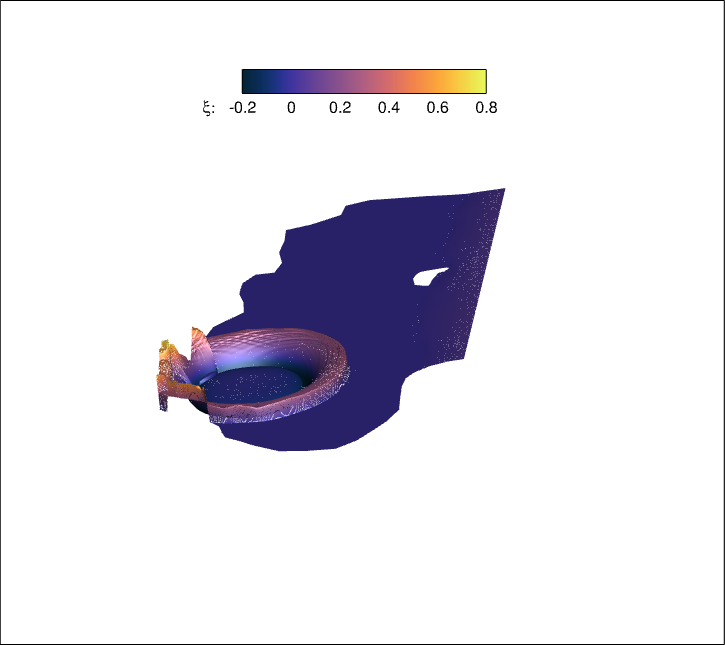}
      \includegraphics[width=0.32\textwidth, trim=150 190 210 180, clip]{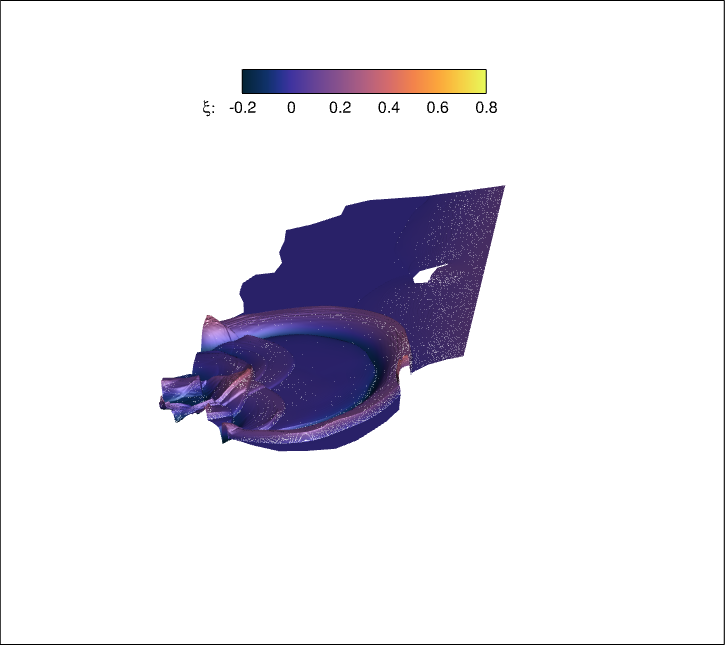} 
       \includegraphics[width=0.32\textwidth, trim=150 190 210 180, clip]{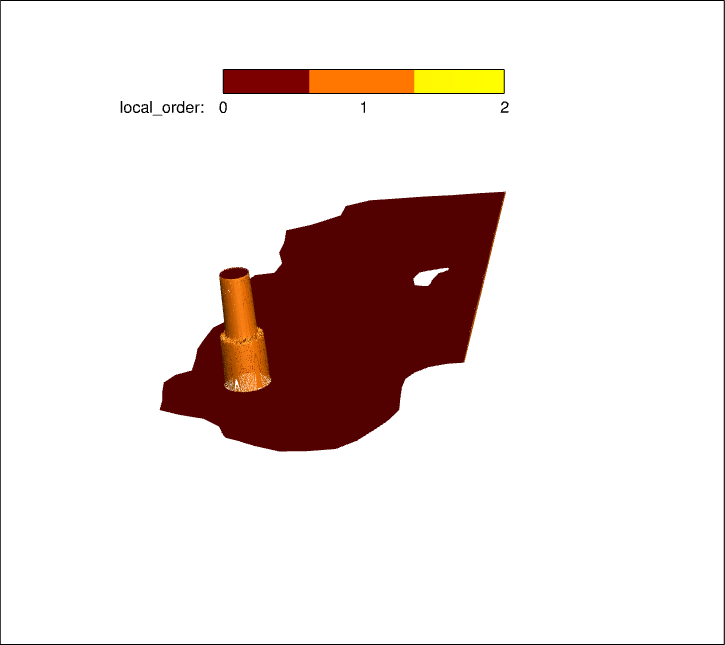}
       \includegraphics[width=0.32\textwidth, trim=150 190 210 180, clip]{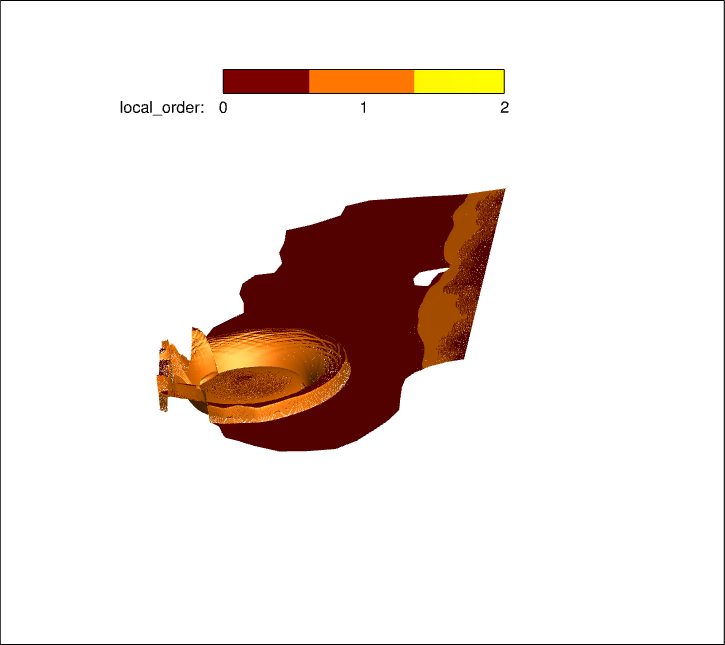}
       \includegraphics[width=0.32\textwidth, trim=150 190 210 180, clip]{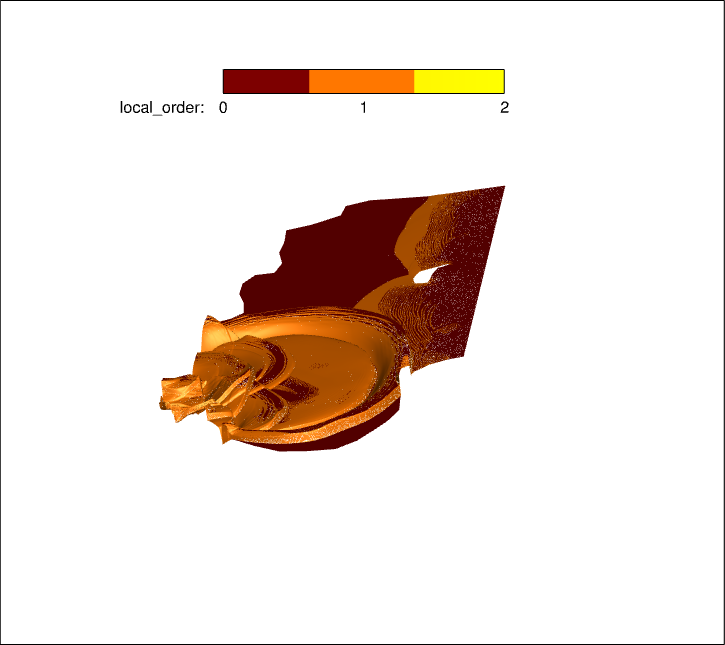} 
       \includegraphics[width=0.32\textwidth, trim=150 190 210 180, clip]{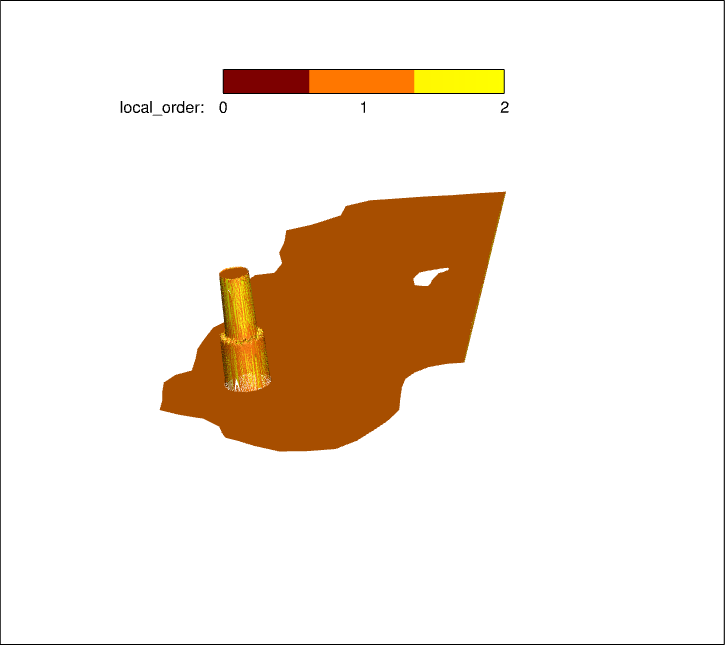}
       \includegraphics[width=0.32\textwidth, trim=150 190 210 180, clip]{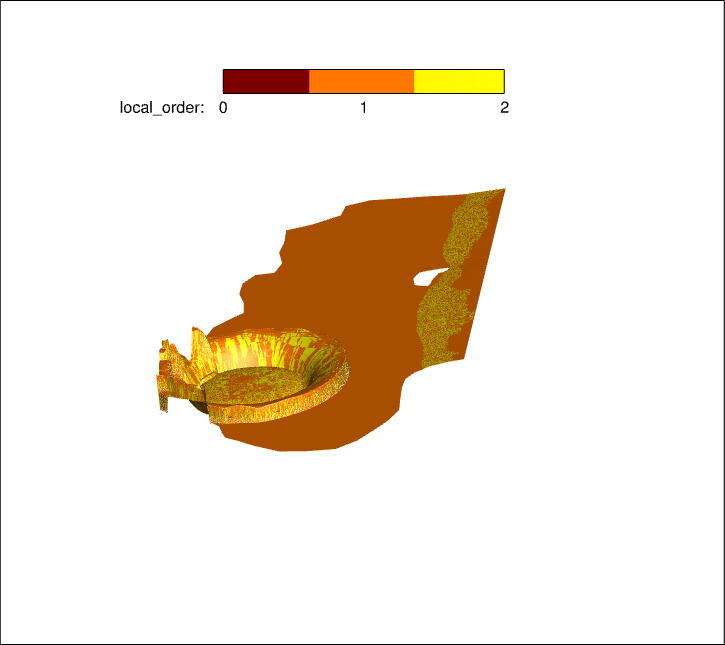}
       \includegraphics[width=0.32\textwidth, trim=150 190 210 180, clip]{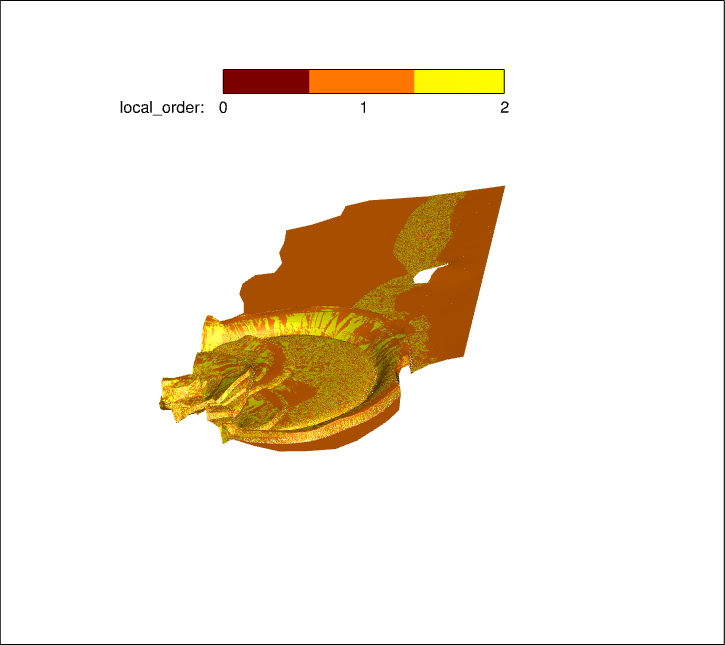} 
       \includegraphics[width=0.34\textwidth, trim=550 1475 630  130, clip]{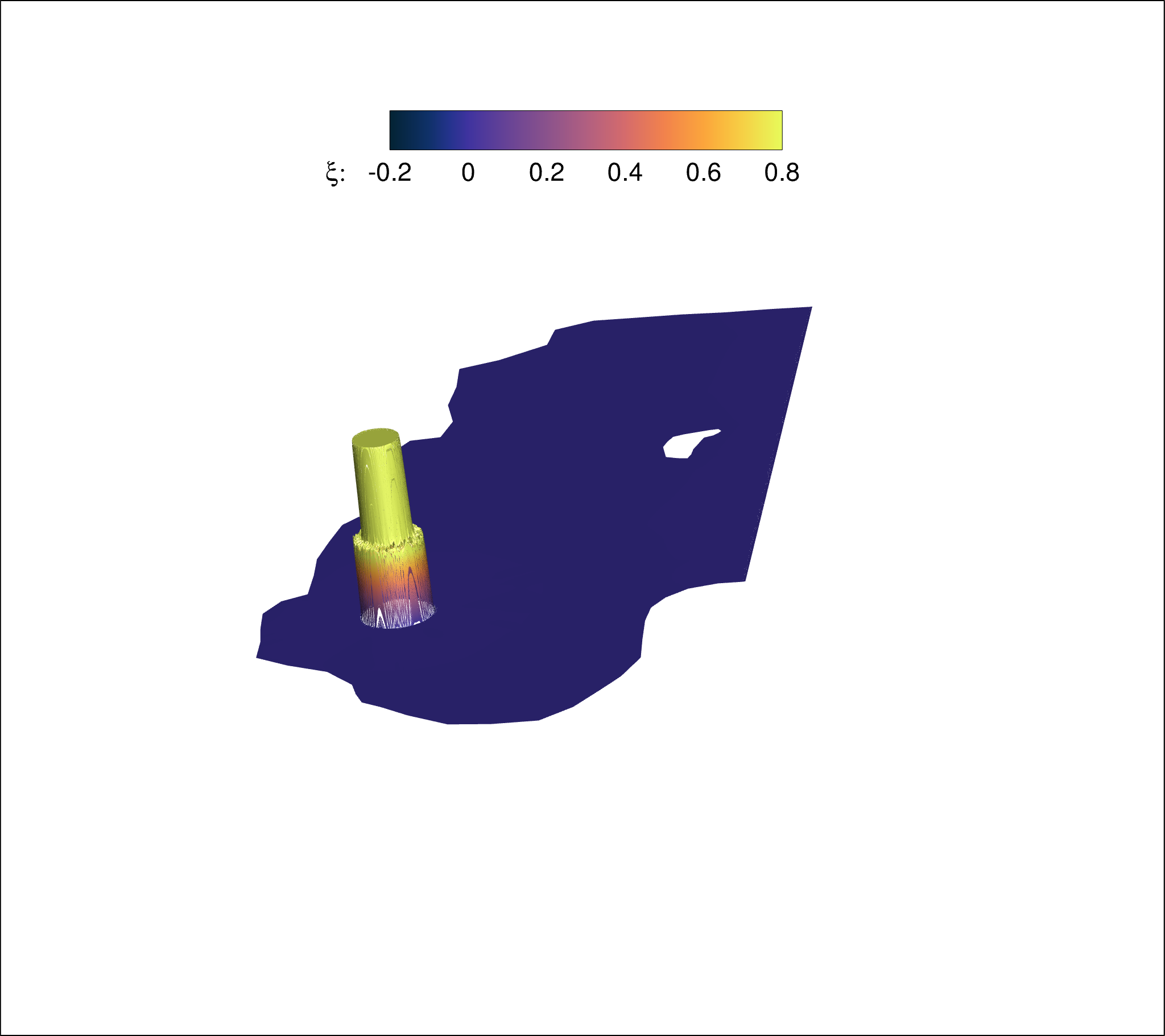} \hspace{1.5cm}
	\includegraphics[width=0.34\textwidth, trim=1090 1300 150  300, clip]{figures/dambreak/dambreak_eta_p_l8_o0-1_1_hr_colorbar.png} 
  \caption{Tidal flow at Bahamas: surface elevation (top row) and local approximation orders for p0-1 (middle row) and p1-2 (bottom row) at $t=1\,\mathrm{s}$ (left), $t=25.0\,\mathrm{s}$ (middle) and $t=50\,\mathrm{s}$ (right). The z-axis is scaled up by factor 10\,000.}
  \label{fig:bahamas}
\end{figure}

Fig.~\ref{fig:bahamas} illustrates the surface elevation at different times for the constant-linear approximation (top) along with the corresponding local approximation orders for the constant-linear (middle) and the linear-quadratic (bottom) discretization.
In the p0-1 case, about 24.5\,\% of the elements use order 1, 
and, in the p1-2 case, about 12.1\,\% of the elements use order 2. 

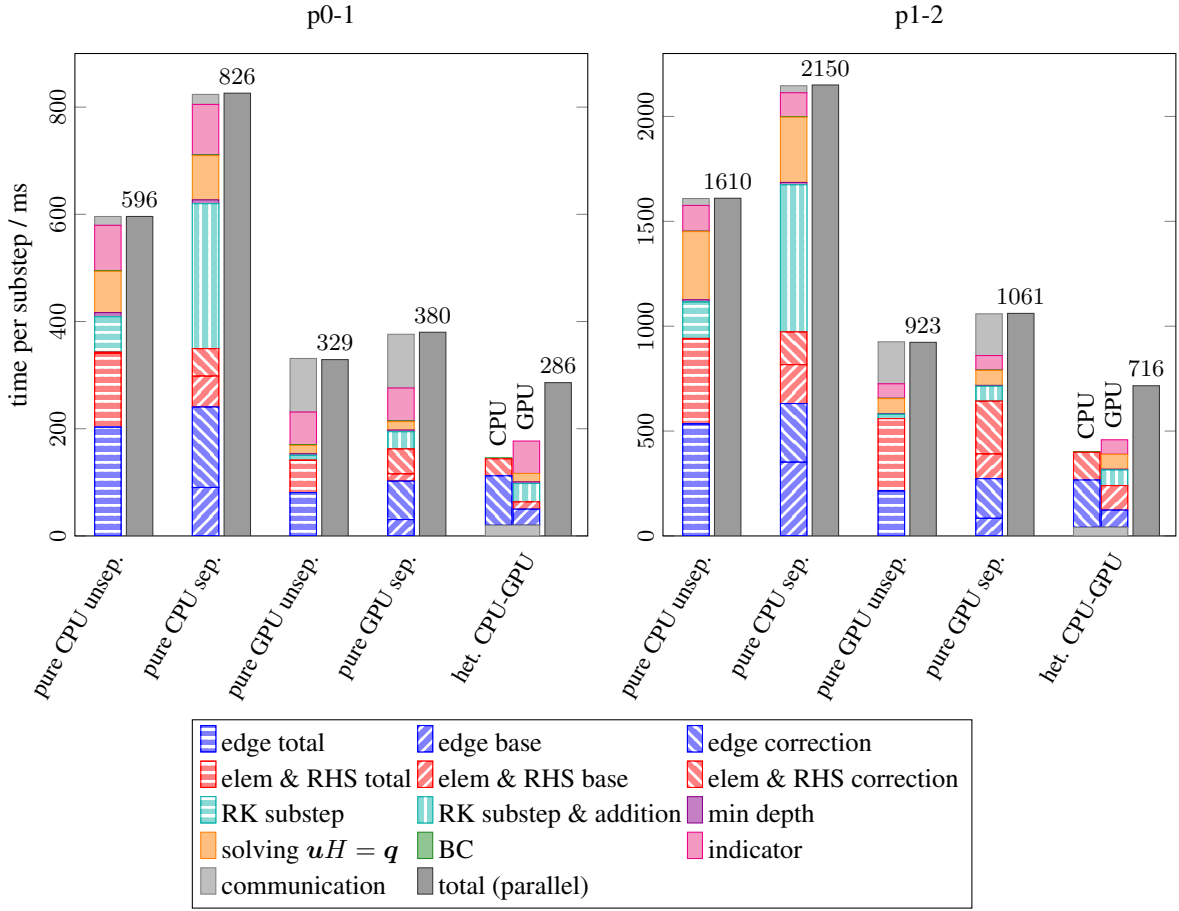
\begin{figure}[]
  \centering
  \sisetup{group-separator = {\,}}
\begin{tikzpicture}[
  every axis/.style={ 
  		nodes near coords ybar stacked configuration/.style={},
  		/pgf/number format/1000 sep={},
    		ybar stacked,
        width=8.4cm, 
		height = 8cm,
		xtick={1,2,3,4,5.2},
        x tick style={draw=none},
        xticklabels={pure CPU unsep.,pure CPU sep.,pure GPU unsep.,pure GPU sep.,het. CPU-GPU},
        x tick label style={rotate=60,anchor=east,font=\small},
        y tick label style={rotate=90,font=\small},
        ymin=0,
        xmin=0.5,
        xmax= 5.75,
        ylabel style={font=\small},
		y tick label style={font=\small},
        ybar=0pt,
         totals/.style={nodes near coords align={anchor=south}},
        every axis plot/.append style={
          bar width=10,
          legend image code/.code={
        \draw [#1] (0cm,-0.1cm) rectangle (0.2cm,0.25cm);},
  },
  }
]

\begin{axis}[
name=ax1,
ymax= 900,
title= {p0-1},
ylabel=time per substep / ms,
every axis y label/.style={at={(ticklabel cs:0.5)},rotate=90,anchor=near ticklabel},
legend style={cells={anchor=west}},
		legend entries={edge total, edge base, edge correction,  elem \& RHS total, elem \& RHS base, elem \& RHS correction, RK substep, RK substep \& addition,  min depth, solving $\boldsymbol{u} H = \boldsymbol{q}$, BC, indicator, communication, total (parallel)},
		legend style={at={(1.75,-0.38)},anchor=north east,legend columns=3}]
	\addplot+[bar shift=-6pt, visualization depends on={value \thisrow{Ax} \as \xdelta},
              visualization depends on={value \thisrow{Ay} \as \ydelta},
              every node near coord/.append style={xshift=\xdelta,yshift=\ydelta},ybar=0pt,draw =blue,fill=white!50!blue,postaction={pattern=horizontal hatch}, pattern color=white, hatch distance=1.5mm, hatch thickness=1pt] table [x={x}, y={A}] {
      x A Ax Ay 
      1 202.66 0  0  
      2 0 0  0  
      3 80.90 0  0  
      4 0 0  0 
    };
   \addplot+[bar shift=-6pt, visualization depends on={value \thisrow{Ax} \as \xdelta},
              visualization depends on={value \thisrow{Ay} \as \ydelta},
              every node near coord/.append style={xshift=\xdelta,yshift=\ydelta},ybar=0pt,draw =blue,fill=white!50!blue,postaction={pattern=north east hatch}, pattern color=white, hatch distance=2mm, hatch thickness=1pt] table [x={x}, y={A}] {
      x A Ax Ay 
      1 0 0  0  
      2 90.39 0  0  
      3 0 0  0  
      4 30.57 0  0 
    };
   \addplot+[bar shift=-6pt, visualization depends on={value \thisrow{Ax} \as \xdelta},
              visualization depends on={value \thisrow{Ay} \as \ydelta},
              every node near coord/.append style={xshift=\xdelta,yshift=\ydelta},ybar=0pt,draw =blue,fill=white!50!blue,postaction={pattern=north west hatch}, pattern color=white, hatch distance=2mm, hatch thickness=1pt] table [x={x}, y={A}] {
      x A Ax Ay 
      1 0 0  0  
      2 150.05 0  0  
      3 0 0  0  
      4 71.62 0  0 
    };
   \addplot+[bar shift=-6pt, visualization depends on={value \thisrow{Ax} \as \xdelta},
              visualization depends on={value \thisrow{Ay} \as \ydelta},
              every node near coord/.append style={xshift=\xdelta,yshift=\ydelta},ybar=0pt,draw =red,fill=white!50!red, postaction={pattern=horizontal hatch}, pattern color=white, hatch distance=1.5mm, hatch thickness=1pt] table [x={x}, y={A}] {
      x A Ax Ay 
      1 138.22 0  0  
      2 0 0  0  
      3 60.07 0  0  
      4 0 0  0 
    };
   \addplot+[bar shift=-6pt, visualization depends on={value \thisrow{Ax} \as \xdelta},
              visualization depends on={value \thisrow{Ay} \as \ydelta},
              every node near coord/.append style={xshift=\xdelta,yshift=\ydelta},ybar=0pt,draw =red,fill=white!50!red,postaction={pattern=north east hatch}, pattern color=white, hatch distance=2mm, hatch thickness=1pt] table [x={x}, y={A}] {
      x A Ax Ay 
      1 0 0  0  
      2 57.61 0  0  
      3 0 0  0  
      4 13.55 0  0 
    };
      \addplot+[bar shift=-6pt, visualization depends on={value \thisrow{Ax} \as \xdelta},
              visualization depends on={value \thisrow{Ay} \as \ydelta},
              every node near coord/.append style={xshift=\xdelta,yshift=\ydelta},ybar=0pt,draw =red,fill=white!50!red,postaction={pattern=north west hatch}, pattern color=white, hatch distance=2mm, hatch thickness=1pt] table [x={x}, y={A}] {
      x A Ax Ay 
      1 2 0  0  
      2 51.40 0  0  
      3 0 0  0  
      4 46.56 0  0 
    };
      \addplot+[bar shift=-6pt, visualization depends on={value \thisrow{Ax} \as \xdelta},
              visualization depends on={value \thisrow{Ay} \as \ydelta},
              every node near coord/.append style={xshift=\xdelta,yshift=\ydelta},ybar=0pt,draw=aquamarine,fill=white!50!aquamarine,postaction={pattern=horizontal hatch}, pattern color=white, hatch distance=1.5mm, hatch thickness=1pt] table [x={x}, y={A}] {
      x A Ax Ay 
      1 66.23 0  0  
      2 0 0  0  
      3 9.31 0  0  
      4 0 0  0 
    };
   \addplot+[bar shift=-6pt, visualization depends on={value \thisrow{Ax} \as \xdelta},
              visualization depends on={value \thisrow{Ay} \as \ydelta},
              every node near coord/.append style={xshift=\xdelta,yshift=\ydelta},ybar=0pt,draw=aquamarine,fill=white!50!aquamarine,postaction={pattern=vertical hatch}, pattern color=white, hatch distance=1.5mm, hatch thickness=1pt] table [x={x}, y={A}] {
      x A Ax Ay 
      1 0 0  0  
      2 270.44 0  0  
      3 0 0  0  
      4 32.18 0  0 
    };
   \addplot+[bar shift=-6pt, visualization depends on={value \thisrow{Ax} \as \xdelta},
              visualization depends on={value \thisrow{Ay} \as \ydelta},
              every node near coord/.append style={xshift=\xdelta,yshift=\ydelta},ybar=0pt,draw=dark_magenta,fill=white!50!dark_magenta] table [x={x}, y={A}] {
      x A Ax Ay 
      1 7.54 0  0  
      2 7.13 0  0  
      3 3.08 0  0  
      4 3.09 0  0 
    };
	\addplot+[bar shift=-6pt, visualization depends on={value \thisrow{Ax} \as \xdelta},
              visualization depends on={value \thisrow{Ay} \as \ydelta},
              every node near coord/.append style={xshift=\xdelta,yshift=\ydelta},ybar=0pt,draw=orange,fill=white!50!orange] table [x={x}, y={A}] {
      x A Ax Ay 
      1 77.09 0  0  
      2 82.49 0  0  
      3 15.68 0  0  
      4 15.70 0  0 
    };
	\addplot+[bar shift=-6pt, visualization depends on={value \thisrow{Ax} \as \xdelta},
              visualization depends on={value \thisrow{Ay} \as \ydelta},
              every node near coord/.append style={xshift=\xdelta,yshift=\ydelta},ybar=0pt,draw=forest_green,fill=white!50!forest_green] table [x={x}, y={A}] {
      x A Ax Ay 
      1 1.30 0  0  
      2 1.59 0  0  
      3 1.46 0  0  
      4 1.47 0  0 
    };
	\addplot+[bar shift=-6pt, visualization depends on={value \thisrow{Ax} \as \xdelta},
              visualization depends on={value \thisrow{Ay} \as \ydelta},
              every node near coord/.append style={xshift=\xdelta,yshift=\ydelta},ybar=0pt,draw=magenta,fill=white!50!magenta] table [x={x}, y={A}] {
      x A Ax Ay 
      1 84.37 0  0  
      2 93.90 0  0  
      3 60.60 0  0  
      4 61.19 0  0 
    };
	\addplot+[bar shift=-6pt, visualization depends on={value \thisrow{Ax} \as \xdelta},
              visualization depends on={value \thisrow{Ay} \as \ydelta},
              every node near coord/.append style={xshift=\xdelta,yshift=\ydelta},ybar=0pt,draw=gray,fill=white!50!gray] table [x={x}, y={A}] {
      x A Ax Ay 
      1 16.54 0  0  
      2 18.77 0  0  
      3 99.99 0  0  
      4 100.33 0  0 
    };

\resetstackedplots
\addplot+[bar shift=6pt,draw=dark_gray,fill=white!50!dark_gray, nodes near coords, text=black, font=\small]
     table [x={x}, y={A}] {
      x A  Ax Ay    
      1 596  0  0     
      2 826  0  0     
      3 329  0  0   
      4 380 0  0 
    };
    
    \resetstackedplots
	\addplot+[bar width=20.5,bar shift=-0.75pt, visualization depends on={value \thisrow{Ax} \as \xdelta},
              visualization depends on={value \thisrow{Ay} \as \ydelta},
              every node near coord/.append style={xshift=\xdelta,yshift=\ydelta},ybar=0pt,draw=gray,fill=white!50!gray] table [x={x}, y={A}] {
      x A Ax Ay 
      5 20.09 0  0 
    };
   \addplot+[bar shift=-6pt, visualization depends on={value \thisrow{Ax} \as \xdelta},
              visualization depends on={value \thisrow{Ay} \as \ydelta},
              every node near coord/.append style={xshift=\xdelta,yshift=\ydelta},ybar=0pt,draw =blue,fill=white!50!blue,postaction={pattern=north west hatch}, pattern color=white, hatch distance=2mm, hatch thickness=1pt] table [x={x}, y={A}] {
      x A Ax Ay 
      5 92.14 0  0 
    };
      \addplot+[bar shift=-6pt, visualization depends on={value \thisrow{Ax} \as \xdelta},
              visualization depends on={value \thisrow{Ay} \as \ydelta},
              every node near coord/.append style={xshift=\xdelta,yshift=\ydelta},ybar=0pt,draw =red,fill=white!50!red,postaction={pattern=north west hatch}, pattern color=white, hatch distance=2mm, hatch thickness=1pt] table [x={x}, y={A}] {
      x A Ax Ay 
      5 31.73 0  0 
    };
	\addplot+[bar shift=-6pt, visualization depends on={value \thisrow{Ax} \as \xdelta},
              visualization depends on={value \thisrow{Ay} \as \ydelta},
              every node near coord/.append style={xshift=\xdelta,yshift=\ydelta},ybar=0pt,draw=forest_green,fill=white!50!forest_green] table [x={x}, y={A}] {
      x A Ax Ay 
      5 1.83 0  0 
    };
\resetstackedplots
	\addplot+[bar width=20.5, bar shift=-0.75pt, visualization depends on={value \thisrow{Ax} \as \xdelta},
              visualization depends on={value \thisrow{Ay} \as \ydelta},
              every node near coord/.append style={xshift=\xdelta,yshift=\ydelta},ybar=0pt,draw=gray,fill=white!50!gray] table [x={x}, y={A}] {
      x A Ax Ay 
      5 20.09 0  0 
    };
   \addplot+[bar shift=4.5pt, visualization depends on={value \thisrow{Ax} \as \xdelta},
              visualization depends on={value \thisrow{Ay} \as \ydelta},
              every node near coord/.append style={xshift=\xdelta,yshift=\ydelta},ybar=0pt,draw =blue,fill=white!50!blue,postaction={pattern=north east hatch}, pattern color=white, hatch distance=2mm, hatch thickness=1pt] table [x={x}, y={A}] {
      x A Ax Ay 
      5 29.77 0  0 
    };
   \addplot+[bar shift=4.5pt, visualization depends on={value \thisrow{Ax} \as \xdelta},
              visualization depends on={value \thisrow{Ay} \as \ydelta},
              every node near coord/.append style={xshift=\xdelta,yshift=\ydelta},ybar=0pt,draw =red,fill=white!50!red,postaction={pattern=north east hatch}, pattern color=white, hatch distance=2mm, hatch thickness=1pt] table [x={x}, y={A}] {
      x A Ax Ay 
      5 13.83 0  0 
    };
   \addplot+[bar shift=4.5pt, visualization depends on={value \thisrow{Ax} \as \xdelta},
              visualization depends on={value \thisrow{Ay} \as \ydelta},
              every node near coord/.append style={xshift=\xdelta,yshift=\ydelta},ybar=0pt,draw=aquamarine,fill=white!50!aquamarine,postaction={pattern=vertical hatch}, pattern color=white, hatch distance=1.5mm, hatch thickness=1pt] table [x={x}, y={A}] {
      x A Ax Ay 
      5 34.13 0  0 
    };
   \addplot+[bar shift=4.5pt, visualization depends on={value \thisrow{Ax} \as \xdelta},
              visualization depends on={value \thisrow{Ay} \as \ydelta},
              every node near coord/.append style={xshift=\xdelta,yshift=\ydelta},ybar=0pt,draw=dark_magenta,fill=white!50!dark_magenta] table [x={x}, y={A}] {
      x A Ax Ay 
      5 3.07 0  0 
    };
	\addplot+[bar shift=4.5pt, visualization depends on={value \thisrow{Ax} \as \xdelta},
              visualization depends on={value \thisrow{Ay} \as \ydelta},
              every node near coord/.append style={xshift=\xdelta,yshift=\ydelta},ybar=0pt,draw=orange,fill=white!50!orange] table [x={x}, y={A}] {
      x A Ax Ay 
      5 15.61 0  0 
    };
	\addplot+[bar shift=4.5pt, visualization depends on={value \thisrow{Ax} \as \xdelta},
              visualization depends on={value \thisrow{Ay} \as \ydelta},
              every node near coord/.append style={xshift=\xdelta,yshift=\ydelta},ybar=0pt,draw=magenta,fill=white!50!magenta] table [x={x}, y={A}] {
      x A Ax Ay 
      5 60.53 0  0 
    };
    \resetstackedplots
    \addplot+[bar shift=16.5pt,draw=dark_gray,fill=white!50!dark_gray, nodes near coords, text=black, font=\small]
     table [x={x}, y={A}] {
      x A  Ax Ay    
      5 286  0  0      
    };

\node (CPU) at (433,150) [rotate=90, anchor=west, font=\small] {CPU};    
\node (GPU) at (461,180) [rotate=90, anchor=west, font=\small] {GPU};  
  
\end{axis}

\begin{axis}[
title= {p1-2},
ymax= 2300,
at={(ax1.south east)},
xshift=1cm]
	\addplot+[bar shift=-6pt, visualization depends on={value \thisrow{Ax} \as \xdelta},
              visualization depends on={value \thisrow{Ay} \as \ydelta},
              every node near coord/.append style={xshift=\xdelta,yshift=\ydelta},ybar=0pt,draw =blue,fill=white!50!blue,postaction={pattern=horizontal hatch}, pattern color=white, hatch distance=1.5mm, hatch thickness=1pt] table [x={x}, y={A}] {
      x A Ax Ay 
      1 533.79 0  0  
      2 0 0  0  
      3 214.41 0  0  
      4 0 0  0 
    };
   \addplot+[bar shift=-6pt, visualization depends on={value \thisrow{Ax} \as \xdelta},
              visualization depends on={value \thisrow{Ay} \as \ydelta},
              every node near coord/.append style={xshift=\xdelta,yshift=\ydelta},ybar=0pt,draw =blue,fill=white!50!blue,postaction={pattern=north east hatch}, pattern color=white, hatch distance=2mm, hatch thickness=1pt] table [x={x}, y={A}] {
      x A Ax Ay 
      1 0 0  0  
      2 351.31 0  0  
      3 0 0  0  
      4 83.77 0  0 
    };
   \addplot+[bar shift=-6pt, visualization depends on={value \thisrow{Ax} \as \xdelta},
              visualization depends on={value \thisrow{Ay} \as \ydelta},
              every node near coord/.append style={xshift=\xdelta,yshift=\ydelta},ybar=0pt,draw =blue,fill=white!50!blue,postaction={pattern=north west hatch}, pattern color=white, hatch distance=2mm, hatch thickness=1pt] table [x={x}, y={A}] {
      x A Ax Ay 
      1 0 0  0  
      2 279.91 0  0  
      3 0 0  0  
      4 189.47 0  0 
    };
   \addplot+[bar shift=-6pt, visualization depends on={value \thisrow{Ax} \as \xdelta},
              visualization depends on={value \thisrow{Ay} \as \ydelta},
              every node near coord/.append style={xshift=\xdelta,yshift=\ydelta},ybar=0pt,draw =red,fill=white!50!red, postaction={pattern=horizontal hatch}, pattern color=white, hatch distance=1.5mm, hatch thickness=1pt] table [x={x}, y={A}] {
      x A Ax Ay 
      1 405.78 0  0  
      2 0 0  0  
      3 344.57 0  0  
      4 0 0  0 
    };
   \addplot+[bar shift=-6pt, visualization depends on={value \thisrow{Ax} \as \xdelta},
              visualization depends on={value \thisrow{Ay} \as \ydelta},
              every node near coord/.append style={xshift=\xdelta,yshift=\ydelta},ybar=0pt,draw =red,fill=white!50!red,postaction={pattern=north east hatch}, pattern color=white, hatch distance=2mm, hatch thickness=1pt] table [x={x}, y={A}] {
      x A Ax Ay 
      1 0 0  0  
      2 185.54 0  0  
      3 0 0  0  
      4 117.48 0  0 
    };
      \addplot+[bar shift=-6pt, visualization depends on={value \thisrow{Ax} \as \xdelta},
              visualization depends on={value \thisrow{Ay} \as \ydelta},
              every node near coord/.append style={xshift=\xdelta,yshift=\ydelta},ybar=0pt,draw =red,fill=white!50!red,postaction={pattern=north west hatch}, pattern color=white, hatch distance=2mm, hatch thickness=1pt] table [x={x}, y={A}] {
      x A Ax Ay 
      1 2 0  0  
      2 155.80 0  0  
      3 0 0  0  
      4 251.78 0  0 
    };
      \addplot+[bar shift=-6pt, visualization depends on={value \thisrow{Ax} \as \xdelta},
              visualization depends on={value \thisrow{Ay} \as \ydelta},
              every node near coord/.append style={xshift=\xdelta,yshift=\ydelta},ybar=0pt,draw=aquamarine,fill=white!50!aquamarine,postaction={pattern=horizontal hatch}, pattern color=white, hatch distance=1.5mm, hatch thickness=1pt] table [x={x}, y={A}] {
      x A Ax Ay 
      1 173.95 0  0  
      2 0 0  0  
      3 20.66 0  0  
      4 0 0  0 
    };
   \addplot+[bar shift=-6pt, visualization depends on={value \thisrow{Ax} \as \xdelta},
              visualization depends on={value \thisrow{Ay} \as \ydelta},
              every node near coord/.append style={xshift=\xdelta,yshift=\ydelta},ybar=0pt,draw=aquamarine,fill=white!50!aquamarine,postaction={pattern=vertical hatch}, pattern color=white, hatch distance=1.5mm, hatch thickness=1pt] table [x={x}, y={A}] {
      x A Ax Ay 
      1 0 0  0  
      2 701.75 0  0  
      3 0 0  0  
      4 71.23 0  0 
    };
   \addplot+[bar shift=-6pt, visualization depends on={value \thisrow{Ax} \as \xdelta},
              visualization depends on={value \thisrow{Ay} \as \ydelta},
              every node near coord/.append style={xshift=\xdelta,yshift=\ydelta},ybar=0pt,draw=dark_magenta,fill=white!50!dark_magenta] table [x={x}, y={A}] {
      x A Ax Ay 
      1 10.56 0  0  
      2 11.04 0  0  
      3 3.95 0  0  
      4 4.02 0  0 
    };
	\addplot+[bar shift=-6pt, visualization depends on={value \thisrow{Ax} \as \xdelta},
              visualization depends on={value \thisrow{Ay} \as \ydelta},
              every node near coord/.append style={xshift=\xdelta,yshift=\ydelta},ybar=0pt,draw=orange,fill=white!50!orange] table [x={x}, y={A}] {
      x A Ax Ay 
      1 324.97 0  0  
      2 311.35 0  0  
      3 71.25 0  0  
      4 71.34 0  0 
    };
	\addplot+[bar shift=-6pt, visualization depends on={value \thisrow{Ax} \as \xdelta},
              visualization depends on={value \thisrow{Ay} \as \ydelta},
              every node near coord/.append style={xshift=\xdelta,yshift=\ydelta},ybar=0pt,draw=forest_green,fill=white!50!forest_green] table [x={x}, y={A}] {
      x A Ax Ay 
      1 2.12 0  0  
      2 2.30 0  0  
      3 1.72 0  0  
      4 1.75 0  0 
    };
	\addplot+[bar shift=-6pt, visualization depends on={value \thisrow{Ax} \as \xdelta},
              visualization depends on={value \thisrow{Ay} \as \ydelta},
              every node near coord/.append style={xshift=\xdelta,yshift=\ydelta},ybar=0pt,draw=magenta,fill=white!50!magenta] table [x={x}, y={A}] {
      x A Ax Ay 
      1 122.38 0  0  
      2 113.93 0  0  
      3 68.80 0  0  
      4 69.20 0  0 
    };
	\addplot+[bar shift=-6pt, visualization depends on={value \thisrow{Ax} \as \xdelta},
              visualization depends on={value \thisrow{Ay} \as \ydelta},
              every node near coord/.append style={xshift=\xdelta,yshift=\ydelta},ybar=0pt,draw=gray,fill=white!50!gray] table [x={x}, y={A}] {
      x A Ax Ay 
      1 33.10 0  0  
      2 33.38 0  0  
      3 200.23 0  0  
      4 198.97 0  0 
      };
      
\resetstackedplots
\addplot+[bar shift=6pt,draw=dark_gray,fill=white!50!dark_gray, nodes near coords, text=black, font=\small]
     table [x={x}, y={A}] {
      x A  Ax Ay    
      1 1610  0  0     
      2 2150  0  0     
      3 923  0  0   
      4 1061 0  0      
    };

    \resetstackedplots
	\addplot+[bar width=20.5, bar shift=-0.75pt, visualization depends on={value \thisrow{Ax} \as \xdelta},
              visualization depends on={value \thisrow{Ay} \as \ydelta},
              every node near coord/.append style={xshift=\xdelta,yshift=\ydelta},ybar=0pt,draw=gray,fill=white!50!gray] table [x={x}, y={A}] {
      x A Ax Ay 
      5 41.44 0  0
    };
   \addplot+[bar shift=-6pt, visualization depends on={value \thisrow{Ax} \as \xdelta},
              visualization depends on={value \thisrow{Ay} \as \ydelta},
              every node near coord/.append style={xshift=\xdelta,yshift=\ydelta},ybar=0pt,draw =blue,fill=white!50!blue,postaction={pattern=north west hatch}, pattern color=white, hatch distance=2mm, hatch thickness=1pt] table [x={x}, y={A}] {
      x A Ax Ay 
      5 225.40 0  0 
    };
      \addplot+[bar shift=-6pt, visualization depends on={value \thisrow{Ax} \as \xdelta},
              visualization depends on={value \thisrow{Ay} \as \ydelta},
              every node near coord/.append style={xshift=\xdelta,yshift=\ydelta},ybar=0pt,draw =red,fill=white!50!red,postaction={pattern=north west hatch}, pattern color=white, hatch distance=2mm, hatch thickness=1pt] table [x={x}, y={A}] {
      x A Ax Ay 
      5 131.79 0  0 
    };
	\addplot+[bar shift=-6pt, visualization depends on={value \thisrow{Ax} \as \xdelta},
              visualization depends on={value \thisrow{Ay} \as \ydelta},
              every node near coord/.append style={xshift=\xdelta,yshift=\ydelta},ybar=0pt,draw=forest_green,fill=white!50!forest_green] table [x={x}, y={A}] {
      x A Ax Ay 
      5 2.46 0  0 
    };
    \resetstackedplots
	\addplot+[bar width=20.5, bar shift=-0.75pt, visualization depends on={value \thisrow{Ax} \as \xdelta},
              visualization depends on={value \thisrow{Ay} \as \ydelta},
              every node near coord/.append style={xshift=\xdelta,yshift=\ydelta},ybar=0pt,draw=gray,fill=white!50!gray] table [x={x}, y={A}] {
      x A Ax Ay 
      5 41.44 0  0 
    };
   \addplot+[bar shift=4.5pt, visualization depends on={value \thisrow{Ax} \as \xdelta},
              visualization depends on={value \thisrow{Ay} \as \ydelta},
              every node near coord/.append style={xshift=\xdelta,yshift=\ydelta},ybar=0pt,draw =blue,fill=white!50!blue,postaction={pattern=north east hatch}, pattern color=white, hatch distance=2mm, hatch thickness=1pt] table [x={x}, y={A}] {
      x A Ax Ay 
      5 81.54 0  0 
    };
   \addplot+[bar shift=4.5pt, visualization depends on={value \thisrow{Ax} \as \xdelta},
              visualization depends on={value \thisrow{Ay} \as \ydelta},
              every node near coord/.append style={xshift=\xdelta,yshift=\ydelta},ybar=0pt,draw =red,fill=white!50!red,postaction={pattern=north east hatch}, pattern color=white, hatch distance=2mm, hatch thickness=1pt] table [x={x}, y={A}] {
      x A Ax Ay 
      5 117.08 0  0 
    };
   \addplot+[bar shift=4.5pt, visualization depends on={value \thisrow{Ax} \as \xdelta},
              visualization depends on={value \thisrow{Ay} \as \ydelta},
              every node near coord/.append style={xshift=\xdelta,yshift=\ydelta},ybar=0pt,draw=aquamarine,fill=white!50!aquamarine,postaction={pattern=vertical hatch}, pattern color=white, hatch distance=1.5mm, hatch thickness=1pt] table [x={x}, y={A}] {
      x A Ax Ay 
      5 74.23 0  0 
    };
   \addplot+[bar shift=4.5pt, visualization depends on={value \thisrow{Ax} \as \xdelta},
              visualization depends on={value \thisrow{Ay} \as \ydelta},
              every node near coord/.append style={xshift=\xdelta,yshift=\ydelta},ybar=0pt,draw=dark_magenta,fill=white!50!dark_magenta] table [x={x}, y={A}] {
      x A Ax Ay 
      5 3.95 0  0 
    };
	\addplot+[bar shift=4.5pt, visualization depends on={value \thisrow{Ax} \as \xdelta},
              visualization depends on={value \thisrow{Ay} \as \ydelta},
              every node near coord/.append style={xshift=\xdelta,yshift=\ydelta},ybar=0pt,draw=orange,fill=white!50!orange] table [x={x}, y={A}] {
      x A Ax Ay 
      5 71.29 0  0 
    };
	\addplot+[bar shift=4.5pt, visualization depends on={value \thisrow{Ax} \as \xdelta},
              visualization depends on={value \thisrow{Ay} \as \ydelta},
              every node near coord/.append style={xshift=\xdelta,yshift=\ydelta},ybar=0pt,draw=magenta,fill=white!50!magenta] table [x={x}, y={A}] {
      x A Ax Ay 
      5 68.93 0  0 
    };
    \resetstackedplots
\addplot+[bar shift=16.5pt,draw=dark_gray,fill=white!50!dark_gray, nodes near coords, text=black, font=\small]
     table [x={x}, y={A}] {
      x A  Ax Ay    
      5 716  0  0      
    };

\node (CPU) at (433,42) [rotate=90, anchor=west, font=\small] {CPU}; 
\node (GPU) at (461,47) [rotate=90, anchor=west, font=\small] {GPU};  
 
\end{axis}

\end{tikzpicture}
  \caption{Tidal flow at Bahamas: detailed kernel and total execution times for dynamically p-adaptive simulation with unseparated and separated setup on the CPU and the GPU as well as with the optimal heterogeneous distribution.}
  \label{fig:bahamas_bar}
\end{figure}

Fig.~\ref{fig:bahamas_bar} details the kernel execution times for the unseparated and separated setups on the CPU, the GPU, as well as for the heterogeneous kernel distribution. In the heterogeneous case, the CPU-kernels are on the left and the GPU ones are on the right of the corresponding bar. The total execution time is larger than the individual ones because not all kernels can be overlapped due to data dependencies. We use the heterogeneous kernel distribution derived in the previous section, i.e., the correction computations are performed on the CPU, whereas the base computations and the remaining kernels, except for the BC computation, are done on the GPU. Therefore, the heterogeneous bar (the rightmost bar of the corresponding subplots of Fig.~\ref{fig:bahamas_bar}) mostly consists of faster kernels (i.e., smaller blocks) out of separated CPU and GPU execution times plotted in the corresponding bars. When using the CPU and the GPU in parallel, we obtain approx. 13\,\% speedup for the constant-linear approximation and approx. 22\,\% speedup for the linear-quadratic one. These results show that our approach also works well in realistic simulation scenarios.

\section{Conclusions and outlook}
\label{conclusion}
In this work, we proposed and tested a~specially re-designed p-adaptive discontinuous Galerkin scheme for the shallow water equations. Using a hierarchical modal basis, our approach separates the lower-order degrees of freedom computations from the rest of the discretization. Furthermore, by exploiting automatic code generation, we distribute the computational kernels between the CPU and the GPU based on kernel performance evaluation for specific hardware. Performance measurements demonstrated that this approach can lead  to significant performance improvements for certain simulation scenarios if used on a hardware where the CPU and the GPU share memory.
Since integrated architectures such as the SoC used as a test platform in our work are often also particularly energy efficient, this is a promising approach for the future HPC applications.

A further improvement of our p-adaptive approach may include an~online performance measurement system which, e.g., could evaluate the kernel execution times at certain time points during the simulation run and automatically re-distribute the kernels as needed. Also porting our implementation to other types of SoC (e.g., integrated Intel GPUs or the NVIDIA Grace Hopper Superchip) and comparing its performance in the energy-to-solution metric to traditional CPU and GPU realizations of the same numerical scheme could generate interesting insights.

\section*{Acknowledgements}
The authors gratefully acknowledge the scientific support and HPC resources provided by the Erlangen National High Performance Computing Center (NHR@FAU) of the Friedrich-Alexander-Universit\"at Erlangen-N\"urnberg (FAU), in particular support from Dominik Ernst and Jan Laukemann. The hardware is funded by the German Research Foundation (DFG). The authors are very grateful to Prof. Stefan Turek for granting access to the ICARUS cluster at the TU Dortmund and to Markus Geveler and Dominik M\"utter for technical support on this cluster. ICARUS hardware is financed by MIWF NRW under the lead of MERCUR. We also thank Dinesh Parthasarathy for the initial setup of the performance measurement script. The work in this paper was supported in part by the DFG through grant AI 117/6-1 'Performance optimized software strategies for unstructured-mesh applications in ocean modeling'.

\bibliographystyle{plainnat}
\bibliography{bibliography}  

\begin{thebibliography}{27}
\providecommand{\natexlab}[1]{#1}
\providecommand{\url}[1]{\texttt{#1}}
\expandafter\ifx\csname urlstyle\endcsname\relax
  \providecommand{\doi}[1]{doi: #1}\else
  \providecommand{\doi}{doi: \begingroup \urlstyle{rm}\Url}\fi

\bibitem[Aizinger and Dawson(2002)]{AizingerDawson2002}
V.~Aizinger and C.~Dawson.
\newblock A discontinuous {G}alerkin method for two-dimensional flow and
  transport in shallow water.
\newblock \emph{Advances in Water Resources}, 25\penalty0 (1):\penalty0 67--84,
  2002.
\newblock \doi{10.1016/S0309-1708(01)00019-7}.

\bibitem[Aizinger et~al.(2013)Aizinger, Proft, Dawson, Pothina, and
  Negusse]{AizingerPDPN2013}
V.~Aizinger, J.~Proft, C.~Dawson, D.~Pothina, and S.~Negusse.
\newblock A three-dimensional discontinuous galerkin model applied to the
  baroclinic simulation of corpus christi bay.
\newblock \emph{Ocean Dynamics}, 63\penalty0 (1):\penalty0 89--113, 2013.
\newblock \doi{10.1007/s10236-012-0579-8}.

\bibitem[Alt et~al.(2023)Alt, Kenter, Faghih-Naini, Faj, Opdenh{\"o}vel,
  Plessl, Aizinger, H{\"o}nig, and K{\"o}stler]{AltKFFOPAHK2023}
C.~Alt, T.~Kenter, S.~Faghih-Naini, J.~Faj, J.-O. Opdenh{\"o}vel, C.~Plessl,
  V.~Aizinger, J.~H{\"o}nig, and H.~K{\"o}stler.
\newblock {Shallow Water DG Simulations on FPGAs: Design and Comparison of a
  Novel Code Generation Pipeline}.
\newblock In A.~Bhatele, J.~Hammond, M.~Baboulin, and C.~Kruse, editors,
  \emph{High Performance Computing}, pages 86--105, Cham, 2023. Springer Nature
  Switzerland.
\newblock ISBN 978-3-031-32041-5.
\newblock \doi{10.1007/978-3-031-32041-5{\_}5}.

\bibitem[Baiges and Bayona(2017)]{Baiges2017}
J.~Baiges and C.~Bayona.
\newblock {Refficientlib: An Efficient Load-Rebalanced Adaptive Mesh Refinement
  Algorithm for High-Performance Computational Physics Meshes}.
\newblock \emph{SIAM Journal on Scientific Computing}, 39\penalty0
  (2):\penalty0 C65--C95, 2017.
\newblock \doi{10.1137/15M105330X}.

\bibitem[Biswas et~al.(2000)Biswas, Das, Harvey, and Oliker]{Biswas2000}
R.~Biswas, S.~K. Das, D.~Harvey, and L.~Oliker.
\newblock Parallel dynamic load balancing strategies for adaptive irregular
  applications.
\newblock \emph{Applied Mathematical Modelling}, 25\penalty0 (2):\penalty0
  109--122, 2000.
\newblock \doi{10.1016/S0307-904X(00)00040-8}.

\bibitem[Bosilca et~al.(2013)Bosilca, Bouteiller, Danalis, Faverge, Herault,
  and Dongarra]{Bosilca2013}
G.~Bosilca, A.~Bouteiller, A.~Danalis, M.~Faverge, Th. Herault, and
  J.~Dongarra.
\newblock {PaRSEC: Exploiting Heterogeneity to Enhance Scalability}.
\newblock \emph{Computing in Science \& Engineering}, 15\penalty0 (6):\penalty0
  36--45, 2013.
\newblock \doi{10.1109/MCSE.2013.98}.

\bibitem[Chaplygin et~al.(2022)Chaplygin, Gusev, and Diansky]{Chaplygin2022}
A.~Chaplygin, A.~Gusev, and N.~Diansky.
\newblock High-performance shallow water model for use on massively parallel
  and heterogeneous computing systems.
\newblock \emph{Supercomputing Frontiers and Innovations}, 8, 02 2022.
\newblock \doi{10.14529/jsfi210407}.

\bibitem[Dawson and Aizinger(2005)]{DawsonAizinger2005}
C.~Dawson and V.~Aizinger.
\newblock A discontinuous galerkin method for three-dimensional shallow water
  equations.
\newblock \emph{Journal of Scientific Computing}, 22\penalty0 (1-3):\penalty0
  245--267, 2005.
\newblock \doi{10.1007/s10915-004-4139-3}.

\bibitem[Echeverribar et~al.(2020)Echeverribar, Morales-Hern\'andez, Brufau,
  and Garc\'ia-Navarro]{Echeverribar2020}
I.~Echeverribar, M.~Morales-Hern\'andez, P.~Brufau, and P.~Garc\'ia-Navarro.
\newblock {Analysis of the performance of a hybrid CPU/GPU 1D2D coupled model
  for real flood cases}.
\newblock \emph{Journal of Hydroinformatics}, 22\penalty0 (5):\penalty0
  1198--1216, 07 2020.
\newblock ISSN 1464-7141.
\newblock \doi{10.2166/hydro.2020.032}.

\bibitem[Faghih-Naini and Aizinger(2022)]{FaghihNainiA2022}
S.~Faghih-Naini and V.~Aizinger.
\newblock {p-adaptive discontinuous Galerkin method for the shallow water
  equations with a parameter-free error indicator}.
\newblock \emph{International Journal on Geomathematics}, 13\penalty0 (18), 10
  2022.
\newblock ISSN 0022-1481.
\newblock \doi{10.1007/s13137-022-00208-3}.

\bibitem[Faghih-Naini et~al.(2020)Faghih-Naini, Kuckuk, Aizinger, Zint, Grosso,
  and K\"ostler]{FaghihNainiKAZGK2020}
S.~Faghih-Naini, S.~Kuckuk, V.~Aizinger, D.~Zint, R.~Grosso, and H.~K\"ostler.
\newblock Quadrature-free discontinuous {{Galerkin}} method with code
  generation features for shallow water equations on automatically generated
  block-structured meshes.
\newblock \emph{Advances in Water Resources}, 138:\penalty0 103552, 2020.
\newblock \doi{10.1016/j.advwatres.2020.103552}.

\bibitem[Faghih-Naini et~al.(2023)Faghih-Naini, Kuckuk, Zint, Kemmler,
  K{\"o}stler, and Aizinger]{FaghihNainiKZKA2022}
S.~Faghih-Naini, S.~Kuckuk, D.~Zint, S.~Kemmler, H.~K{\"o}stler, and
  V.~Aizinger.
\newblock Discontinuous {G}alerkin method for the shallow water equations on
  complex domains using masked block-structured grids.
\newblock \emph{Advances in Water Resources}, page 104584, 2023.
\newblock ISSN 0309-1708.
\newblock \doi{10.1016/j.advwatres.2023.104584}.

\bibitem[Flynn(1972)]{Flynn1972}
M.~Flynn.
\newblock Some computer organizations and their effectiveness. ieee trans
  comput c-21:948.
\newblock \emph{Computers, IEEE Transactions on}, C-21:\penalty0 948 -- 960, 10
  1972.
\newblock \doi{10.1109/TC.1972.5009071}.

\bibitem[Fu et~al.(2017)Fu, Gan, Yang, Xue, Wang, Wang, Huang, and
  Yang]{Fu2017}
H.~Fu, L.~Gan, C.~Yang, W.~Xue, L.~Wang, X.~Wang, X.~Huang, and G.~Yang.
\newblock Solving global shallow water equations on heterogeneous
  supercomputers.
\newblock \emph{PLOS ONE}, 12:\penalty0 e0172583, 03 2017.
\newblock \doi{10.1371/journal.pone.0172583}.

\bibitem[Garcia-Gasulla et~al.(2019)Garcia-Gasulla, Houzeaux, Ferrer, Artigues,
  López, Labarta, and Vázquez]{Garcia-Gasulla2019}
M.~Garcia-Gasulla, G.~Houzeaux, R.~Ferrer, A.~Artigues, V.~López, J.~Labarta,
  and M.~Vázquez.
\newblock {MPI+X:} task-based parallelisation and dynamic load balance of
  finite element assembly.
\newblock \emph{International Journal of Computational Fluid Dynamics},
  33\penalty0 (3):\penalty0 115--136, 2019.
\newblock \doi{10.1080/10618562.2019.1617856}.

\bibitem[Geveler et~al.(2016)Geveler, Reuter, Aizinger, G{\"o}ddeke, and
  Turek]{GevelerRAGT2016}
M.~Geveler, B.~Reuter, V.~Aizinger, D.~G{\"o}ddeke, and S.~Turek.
\newblock Energy efficiency of the simulation of three-dimensional coastal
  ocean circulation on modern commodity and mobile processors.
\newblock \emph{Computer Science : Research + Development}, 31\penalty0
  (4):\penalty0 225--234, 2016.
\newblock ISSN 2524-8529.
\newblock \doi{10.1007/s00450-016-0324-5}.

\bibitem[Gottlieb and Shu(1998)]{GottliebShu1998}
S.~Gottlieb and C.-W. Shu.
\newblock {Strong Stability-Preserving High-Order Time Discretization Methods}.
\newblock \emph{Math. Comp.}, 67\penalty0 (221):\penalty0 73--85, 1998.
\newblock \doi{10.1090/S0025-5718-98-00913-2}.

\bibitem[Hajduk(2021)]{Hajduk2021}
H.~Hajduk.
\newblock Monolithic convex limiting in discontinuous {G}alerkin
  discretizations of hyperbolic conservation laws.
\newblock \emph{Computers \& Mathematics with Applications}, 87:\penalty0
  120--138, 04 2021.
\newblock \doi{10.1016/j.camwa.2021.02.012}.

\bibitem[Hajduk et~al.(2018)Hajduk, Hodges, Aizinger, and
  Reuter]{HajdukHAR2018}
H.~Hajduk, B.~R. Hodges, V.~Aizinger, and B.~Reuter.
\newblock Locally {F}iltered {T}ransport for computational efficiency in
  multi-component advection-reaction models.
\newblock \emph{Environmental Modelling \& Software}, 102:\penalty0 185--198,
  2018.
\newblock \doi{10.1016/j.envsoft.2018.01.003}.

\bibitem[Hendrickson and Devine(2000)]{Hendrickson2000}
B.~Hendrickson and K.~Devine.
\newblock Dynamic load balancing in computational mechanics.
\newblock \emph{Computer Methods in Applied Mechanics and Engineering},
  184\penalty0 (2):\penalty0 485--500, 2000.
\newblock ISSN 0045-7825.
\newblock \doi{10.1016/S0045-7825(99)00241-8}.

\bibitem[Kronawitter and Lengauer(2018)]{KronawitterLengauer2018}
S.~Kronawitter and C.~Lengauer.
\newblock Polyhedral search space exploration in the exastencils code
  generator.
\newblock \emph{ACM Transactions on Architecture and Code Optimization},
  15\penalty0 (4), 2018.
\newblock \doi{10.1145/3274653}.

\bibitem[Kubatko et~al.(2009)Kubatko, Bunya, Dawson, and
  Westerink]{Kubatko2009}
E.~J. Kubatko, S.~Bunya, C.~Dawson, and J.~J. Westerink.
\newblock Dynamic p-adaptive {R}unge-{K}utta discontinuous {G}alerkin methods
  for the shallow water equations.
\newblock \emph{Computer Methods in Applied Mechanics and Engineering},
  198\penalty0 (21):\penalty0 1766--1774, 2009.
\newblock ISSN 0045-7825.
\newblock \doi{10.1016/j.cma.2009.01.007}.
\newblock Advances in Simulation-Based Engineering Sciences -- Honoring J.
  Tinsley Oden.

\bibitem[Lengauer et~al.(2020)Lengauer, Apel, Bolten, Chiba, R{\"u}de, Teich,
  Gr{\"o}{\ss}linger, Hannig, K{\"o}stler, Claus, Grebhahn, Groth, Kronawitter,
  Kuckuk, Rittich, Schmitt, and Schmitt]{LengauerABCRTGHKCGGKKRSS2020}
C.~Lengauer, S.~Apel, M.~Bolten, S.~Chiba, U.~R{\"u}de, J.~Teich,
  A.~Gr{\"o}{\ss}linger, F.~Hannig, H.~K{\"o}stler, L.~Claus, A.~Grebhahn,
  S.~Groth, S.~Kronawitter, S.~Kuckuk, H.~Rittich, C.~Schmitt, and J.~Schmitt.
\newblock Exastencils: Advanced multigrid solver generation.
\newblock In H.-J. Bungartz, S.~Reiz, B.~Uekermann, P.~Neumann, and W.~E.
  Nagel, editors, \emph{Software for Exascale Computing - SPPEXA 2016-2019},
  pages 405--452, Cham, 2020. Springer International Publishing.
\newblock ISBN 978-3-030-47956-5.

\bibitem[LeVeque(2002)]{Leveque2002}
R.~J. LeVeque.
\newblock \emph{Finite Volume Methods for Hyperbolic Problems}.
\newblock Cambridge Texts in Applied Mathematics. Cambridge University Press,
  2002.
\newblock \doi{10.1017/CBO9780511791253}.

\bibitem[Meurer et~al.(2017)Meurer, Smith, Paprocki, \v{C}ert\'{i}k, Kirpichev,
  Rocklin, Kumar, Ivanov, Moore, Singh, Rathnayake, Vig, Granger, Muller,
  Bonazzi, Gupta, Vats, Johansson, Pedregosa, Curry, Terrel, Rou\v{c}ka, Saboo,
  Fernando, Kulal, Cimrman, and Scopatz]{MeurerSPCKRKIMSRVGMBGVJPCTRSFKCS2017}
A.~Meurer, C.~P. Smith, M.~Paprocki, O.~\v{C}ert\'{i}k, S.~B. Kirpichev,
  M.~Rocklin, A.~Kumar, S.~Ivanov, J.~K. Moore, S.~Singh, T.~Rathnayake,
  S.~Vig, B.~E. Granger, R.~P. Muller, F.~Bonazzi, H.~Gupta, S.~Vats,
  F.~Johansson, F.~Pedregosa, M.~J. Curry, A.~R. Terrel, \v{S}. Rou\v{c}ka,
  A.~Saboo, I.~Fernando, S.~Kulal, R.~Cimrman, and A.~Scopatz.
\newblock {SymPy: symbolic computing in Python}.
\newblock \emph{PeerJ Computer Science}, 3:\penalty0 e103, 2017.
\newblock ISSN 2376-5992.
\newblock \doi{10.7717/peerj-cs.103}.

\bibitem[Reuter et~al.(2016)Reuter, Aizinger, Wieland, Frank, and
  Knabner]{ReuterAWFK2016}
B.~Reuter, V.~Aizinger, M.~Wieland, F.~Frank, and P.~Knabner.
\newblock {FESTUNG}: A {MATLAB}/{GNU} {O}ctave toolbox for the discontinuous
  {G}alerkin method, {P}art {II}: {A}dvection operator and slope limiting.
\newblock \emph{Computers and Mathematics with Applications}, 72\penalty0
  (7):\penalty0 1896--1925, 2016.
\newblock \doi{10.1016/j.camwa.2016.08.006}.

\bibitem[Teresco et~al.(2006)Teresco, Devine, and Flaherty]{TerescoDF2006}
J.~D. Teresco, K.~D. Devine, and J.~E. Flaherty.
\newblock Partitioning and dynamic load balancing for the numerical solution of
  partial differential equations.
\newblock In A.~M. Bruaset and A.~Tveito, editors, \emph{Numerical Solution of
  Partial Differential Equations on Parallel Computers}, pages 55--88, Berlin,
  Heidelberg, 2006. Springer Berlin Heidelberg.
\newblock ISBN 978-3-540-31619-0.

\end{thebibliography}

\appendix
\section{Detailed performance results}\label{sec:app_A}
The kernel execution times for all presented adaptive measurements are detailed in Tab.~\ref{tab:timings}.
\begin{table}[h!]
	\centering
	\renewcommand{\arraystretch}{1.1}
	\setlength{\tabcolsep}{2.5pt}
	\footnotesize
	\sisetup{table-format=2.1,table-text-alignment=right, table-comparator,round-mode=places}
	\begin{tabular}{@{}p{1.5cm}ll|S[round-precision=1]S[round-precision=1]S[round-precision=1]S[round-precision=1]S[round-precision=1]S[round-precision=1]S[round-precision=1]S[round-precision=1]S[round-precision=1]S[round-precision=1]S[round-precision=1]S[round-precision=1]S[round-precision=1]S[round-precision=1]} 
\multicolumn{3}{l|}{\textbf{test scenario}} & \multicolumn{2}{c|}{\textbf{dam break}}  & \multicolumn{2}{c|}{\textbf{dam break}} & \multicolumn{2}{c|}{\textbf{dam break}} & \multicolumn{2}{c|}{\textbf{dam break}} &\multicolumn{2}{c|}{\textbf{dam break}} & \multicolumn{2}{c|}{\textbf{Bahamas}}& \multicolumn{2}{c}{\textbf{dam break}}\\ 
\multicolumn{3}{l|}{\textbf{adaptivity strategy}} & \multicolumn{2}{c|}{\textbf{static, 1/8}}  & \multicolumn{2}{c|}{\textbf{static, 1/16}} & \multicolumn{2}{c|}{\textbf{static, 1/32}} & \multicolumn{2}{c|}{\textbf{static, 1/64}} &\multicolumn{2}{c|}{\textbf{dynamic}} & \multicolumn{2}{c|}{\textbf{dynamic}}& \multicolumn{2}{c}{\textbf{static, 1/32}}\\ 
\multicolumn{3}{l|}{\textbf{hardware platform}}  & \multicolumn{2}{c|}{\textbf{ARM-AGX}}  & \multicolumn{2}{c|}{\textbf{ARM-AGX}} & \multicolumn{2}{c|}{\textbf{ARM-AGX}} & \multicolumn{2}{c|}{\textbf{ARM-AGX}} &\multicolumn{2}{c|}{\textbf{ARM-AGX}} & \multicolumn{2}{c|}{\textbf{ARM-AGX}}& \multicolumn{2}{c}{\textbf{AMD-RTX}}\\ \hline
       \textbf{kernel}&\textbf{p} & \textbf{distrib.} & \textcolor[HTML]{FFFFFF} 
     {\cellcolor[HTML]{808B96}\textbf{CPU}} &  \cellcolor[HTML]{808B96}\textcolor[HTML]{FFFFFF}{\textbf{GPU}}& \textcolor[HTML]{FFFFFF}
     {\cellcolor[HTML]{808B96}\textbf{CPU}} &  \cellcolor[HTML]{808B96}\textcolor[HTML]{FFFFFF}{\textbf{GPU}}& \textcolor[HTML]{FFFFFF} 
     {\cellcolor[HTML]{808B96}\textbf{CPU}} &  \cellcolor[HTML]{808B96}\textcolor[HTML]{FFFFFF}{\textbf{GPU}}& \textcolor[HTML]{FFFFFF} 
     {\cellcolor[HTML]{808B96}\textbf{CPU}} &  \cellcolor[HTML]{808B96}\textcolor[HTML]{FFFFFF}{\textbf{GPU}}& \textcolor[HTML]{FFFFFF} 
     {\cellcolor[HTML]{808B96}\textbf{CPU}} &  \cellcolor[HTML]{808B96}\textcolor[HTML]{FFFFFF}{\textbf{GPU}}& \textcolor[HTML]{FFFFFF} 
     {\cellcolor[HTML]{808B96}\textbf{CPU}} &  \cellcolor[HTML]{808B96}\textcolor[HTML]{FFFFFF}{\textbf{GPU}}& \textcolor[HTML]{FFFFFF} 
     {\cellcolor[HTML]{808B96}\textbf{CPU}} &  \cellcolor[HTML]{808B96}\textcolor[HTML]{FFFFFF}{\textbf{GPU}}\\ \hline
\multirow{4}{\linewidth}{\textbf{edge base computation}} &\multirow{2}{*}{\textbf{0-1}}                                                                                                                                 
&homog.                                                                                       & 21.64 & 17.60     & 21.59 & 17.90    & 21.19 & 17.89    & 21.41 & 17.59   & 20.66 & 17.85    & 90.39 & 30.57    & 4.06 & 1.70            \\
& &heterog.                                                                                   & \textemdash   & 18.31     & \textemdash   & 18.31    & \textemdash   & 18.30    & \textemdash   & 18.29   & \textemdash   & 18.30    & \textemdash   & 29.77    & \textemdash   & 8.40          \\
\arrayrulecolor{light_gray}\cline{2-17}\arrayrulecolor{black}                                                                                                                                                                                                  
&\multirow{2}{*}{\textbf{1-2}} &homog.                                                        & 89.05 & 62.59     & 85.51 & 62.59    & 87.58 & 62.57    & 90.85 & 62.17   & 83.72 & 61.73    & 351.31 & 83.77   & 14.70 & 5.79         \\
&&heterog.                                                                                    & \textemdash   & 63.07     & \textemdash   & 63.05    & \textemdash   & 63.05    & \textemdash   & 63.08   & \textemdash   & 63.08    & \textemdash   & 81.54    & \textemdash   & 23.09         \\
\hline                                                                                                                                                                                      
\multirow{4}{\linewidth}{\textbf{edge correction computation}} &\multirow{2}{*}{\textbf{0-1}}                                                                                                                                       
&homog.                                                                                       & 86.03 & 99.21     & 69.87 & 98.74    & 43.52 & 50.92    & 24.64 & 26.74   & 16.16 & 13.85    & 150.05 & 71.622  & 4.30 & 4.82             \\
& &heterog.                                                                                   & 140.56 & \textemdash       & 109.73 & \textemdash      & 64.19 & \textemdash       & 37.34 & \textemdash      & 32.54 & \textemdash       & 92.14 & \textemdash       & 9.45 & \textemdash               \\
\arrayrulecolor{light_gray}\cline{2-17}\arrayrulecolor{black}                                                                                                                                                                                                  
&\multirow{2}{*}{\textbf{1-2}} &homog.                                                        & 196.04 & 253.97   & 161.36 & 251.87  & 88.32 & 127.89   & 48.68 & 65.45   & 20.48 & 23.46    & 279.91 & 189.47  & 12.33 & 11.97             \\
&&heterog.                                                                                    & 411.00 & \textemdash       & 281.67 & \textemdash      & 155.37 & \textemdash      & 83.43 & \textemdash      & 38.26 & \textemdash       & 225.40 & \textemdash      & 22.92 & \textemdash              \\
\hline                                                                                                                                                                                      
\multirow{4}{\linewidth}{\textbf{elem \& RHS base computation}} &\multirow{2}{*}{\textbf{0-1}}                                                                                                                                       
&homog.                                                                                       & 9.21 & 6.58       & 9.16 & 6.64      & 9.18 & 6.64      & 9.26 & 6.58     & 8.08 & 6.63      & 57.61 & 13.55    & 1.95 & 0.77            \\ 
& &heterog.                                                                                   & \textemdash  & 6.66       & \textemdash  & 6.66      & \textemdash  & 6.66      & \textemdash  & 6.66     & \textemdash  & 6.66      & \textemdash  & 13.83     & \textemdash  & 0.84            \\ 
\arrayrulecolor{light_gray}\cline{2-17}\arrayrulecolor{black}                                                                                                                                                                                
&\multirow{2}{*}{\textbf{1-2}} &homog.                                                        & 43.72 & 97.84     & 43.89 & 97.81    & 44.11 & 97.81    & 44.02 & 97.75   & 42.14 & 97.82    & 185.54 & 117.48  & 4.95 & 9.04            \\ 
&&heterog.                                                                                    & \textemdash  & 97.86      & \textemdash  & 97.86     & \textemdash  & 97.85     & \textemdash  & 97.86    & \textemdash  & 97.95     & \textemdash  & 117.08    & \textemdash  & 9.14            \\ 
\hline                                                                                                                                                                                      
\multirow{4}{\linewidth}{\textbf{elem \& RHS correction computation}} &\multirow{2}{*}{\textbf{0-1}}                                                                                                                              
&homog.                                                                                       & 26.29 & 96.17     & 21.94 & 96.17    & 14.41 & 48.45    & 8.33 & 24.46    & 5.96 & 7.17      & 51.40 & 46.56    & 1.81 & 4.49             \\ 
& &heterog.                                                                                   & 42.02 & \textemdash        & 30.02 & \textemdash       & 20.04 & \textemdash       & 13.69 & \textemdash      & 11.46 & \textemdash       & 31.73 & \textemdash       & 1.79 & \textemdash                \\ 
\arrayrulecolor{light_gray}\cline{2-17}\arrayrulecolor{black}                                                                                                                                                                                
&\multirow{2}{*}{\textbf{1-2}} &homog.                                                        & 105.28 & 680.47   & 75.65 & 679.73   & 42.34 & 340.42   & 22.53 & 170.70  & 8.80 & 36.87     & 155.80 & 251.78  & 3.70 & 31.50              \\ 
&&heterog.                                                                                    & 127.50 & \textemdash       & 92.76 & \textemdash       & 51.11 & \textemdash       & 29.28 & \textemdash      & 12.52 & \textemdash       & 131.79 & \textemdash      & 3.88 & \textemdash                \\ 
\hline                                                                                                                                                                                        
\multirow{4}{\linewidth}{\textbf{RK substep \& addition}} &\multirow{2}{*}{\textbf{0-1}}                                                                                                                                  
&homog.                                                                                       & 97.67 & 32.05     & 76.55 & 25.29    & 56.20 & 17.95    & 42.30 & 14.07   & 31.69 & 16.33    & 270.44 & 32.18   & 9.88 & 4.32               \\ 
& &heterog.                                                                                   & \textemdash   & 32.77     & \textemdash   & 26.16    & \textemdash   & 19.22    & \textemdash   & 15.47   & \textemdash   & 17.71    & \textemdash   & 34.13    & \textemdash   & 59.42          \\ 
\arrayrulecolor{light_gray}\cline{2-17}\arrayrulecolor{black}
&\multirow{2}{*}{\textbf{1-2}} &homog.                                                        & 259.72 & 96.51    & 211.31 & 85.64   & 168.07 & 60.23   & 139.30 & 45.01  & 123.78 & 48.61   & 701.75 & 71.23   & 23.11 & 11.60           \\ 
&&heterog.                                                                                    & \textemdash   & 95.90     & \textemdash   & 86.05    & \textemdash   & 61.28    & \textemdash   & 46.41   & \textemdash   & 50.20    & \textemdash   & 74.23    & \textemdash   & 118.08         \\ 
\hline                                                                                                                                                                                      
\multirow{4}{\linewidth}{\textbf{min depth}} &\multirow{2}{*}{\textbf{0-1}}                                                                                                                                                        
&homog.                                                                                       & 3.55 & 2.45       & 3.67 & 2.45      & 3.49 & 1.84      & 3.53 & 1.52     & 2.93 & 1.29      & 7.13 & 3.09      & 0.49 & 0.19             \\ 
& &heterog.                                                                                   & \textemdash  & 2.48       & \textemdash  & 2.47      & \textemdash  & 1.85      & \textemdash  & 1.54     & \textemdash  & 1.30      & \textemdash  & 3.07      & \textemdash  & 0.19             \\ 
\arrayrulecolor{light_gray}\cline{2-17}\arrayrulecolor{black}                                                                                                                                                                               
&\multirow{2}{*}{\textbf{1-2}} &homog.                                                        & 8.86 & 3.81       & 8.10 & 3.81      & 7.02 & 2.69      & 5.64 & 2.12     & 4.73 & 1.65      & 11.04 & 4.02     & 0.89 & 0.27             \\ 
&&heterog.                                                                                    & \textemdash  & 3.82       & \textemdash  & 3.81      & \textemdash  & 2.68      & \textemdash  & 2.13     & \textemdash  & 1.68      & \textemdash  & 3.95      & \textemdash  & 0.28             \\ 
\hline                                                                                                                                                                                                  
\multirow{4}{\linewidth}{\textbf{solving $\boldsymbol{u} H = \boldsymbol{q}$}} &\multirow{2}{*}{\textbf{0-1}}                                                                                                                  
&homog.                                                                                       & 38.59 & 26.78     & 30.89 & 26.92    & 21.70 & 15.47    & 15.60 & 9.68    & 16.35 & 4.67     & 82.49 & 15.70    & 0.97 & 1.92               \\ 
& &heterog.                                                                                   & \textemdash   & 26.81     & \textemdash   & 26.88    & \textemdash   & 15.42    & \textemdash   & 9.70    & \textemdash   & 4.76     & \textemdash   & 15.61    & \textemdash   & 4.74       \\ 
\arrayrulecolor{light_gray}\cline{2-17}\arrayrulecolor{black}                                                                                                                                                                                
&\multirow{2}{*}{\textbf{1-2}} &homog.                                                        & 213.75 & 143.60   & 196.37 & 138.10  & 150.02 & 80.70   & 104.96 & 52.32  & 112.54 & 29.87   & 311.35 & 71.34   & 5.78 & 11.65           \\ 
&&heterog.                                                                                    & \textemdash   & 143.19    & \textemdash   & 138.29   & \textemdash   & 80.73    & \textemdash   & 52.47   & \textemdash   & 29.95    & \textemdash   & 71.29    & \textemdash   & 14.51      \\ 
\hline                                                                                                                                                                                      
\multirow{4}{\linewidth}{\textbf{BC computation}} &\multirow{2}{*}{\textbf{0-1}}                                                                                                                                                
&homog.                                                                                       & 0.63 & 0.38       & 0.64 & 0.79      & 0.62 & 0.77      & 0.62 & 0.37     & 0.44 & 0.53      & 1.59 & 1.47      & 0.94 & 0.13            \\ 
& &heterog.                                                                                   & 1.03 & \textemdash         & 0.99 & \textemdash        & 0.98 & \textemdash        & 0.96 & \textemdash       & 0.75 & \textemdash        & 1.83 & \textemdash        & 12.58 & \textemdash          \\ 
\arrayrulecolor{light_gray}\cline{2-17}\arrayrulecolor{black}                                                                                                                                                                                                  
&\multirow{2}{*}{\textbf{1-2}} &homog.                                                        & 1.20 & 1.51       & 1.16 & 1.46      & 1.09 & 1.44      & 1.10 & 0.79     & 0.65 & 0.86      & 2.30 & 1.75      & 1.57 & 0.16             \\ 
&&heterog.                                                                                    & 2.17 & \textemdash         & 2.07 & \textemdash        & 1.94 & \textemdash        & 1.92 & \textemdash       & 1.21 & \textemdash        & 2.46 & \textemdash        & 25.63 & \textemdash          \\ 
\hline                                                                                                                                    
\multirow{4}{\linewidth}{\textbf{indicator}} &\multirow{2}{*}{\textbf{0-1}}                                                                                                                                             
&homog.                                                                                       & \textemdash  & \textemdash         & \textemdash  & \textemdash        & \textemdash  & \textemdash        & \textemdash  & \textemdash       & 21.14 & 7.35     & 93.90 & 61.19    & \textemdash  & \textemdash              \\
& &heterog.                                                                                   & \textemdash  & \textemdash         & \textemdash  & \textemdash        & \textemdash  & \textemdash        & \textemdash  & \textemdash       & \textemdash   & 8.05     & \textemdash   & 60.53    & \textemdash  & \textemdash              \\
\arrayrulecolor{light_gray}\cline{2-17}\arrayrulecolor{black}                                                                                                                                                                                                    
&\multirow{2}{*}{\textbf{1-2}} &homog.                                                        & \textemdash  & \textemdash         & \textemdash  & \textemdash        & \textemdash  & \textemdash        & \textemdash  & \textemdash       & 34.68 & 10.69    & 113.93 & 69.20   & \textemdash  & \textemdash                \\
&&heterog.                                                                                    & \textemdash  & \textemdash         & \textemdash  & \textemdash        & \textemdash  & \textemdash        & \textemdash  & \textemdash       & \textemdash   & 11.28    & \textemdash   & 68.93    & \textemdash  & \textemdash              \\
\hline \hline  
\multirow{4}{\linewidth}{\textbf{total (parallel)}} &\multirow{2}{*}{\textbf{0-1}}                                                                             
&homog.                                 & 284.07 & 257.98                  & 234.83 & 249.37               & 170.86 & 144.55                 & 126.28 & 92.18                   & 123.94 & 74.44                  & 826.17 & 379.73                & 24.85 & 16.12                       \\
& &heterog.                             &  \multicolumn{2}{c}{246.7}     &  \multicolumn{2}{c}{195.9}  &  \multicolumn{2}{c}{127.4}    &  \multicolumn{2}{c}{89.6}      &  \multicolumn{2}{c}{88.1}     &  \multicolumn{2}{c}{286.5}   &  \multicolumn{2}{c}{101.0}       \\
\arrayrulecolor{light_gray}\cline{2-17}\arrayrulecolor{black}                                                                                                                                                                                                                                           
&\multirow{2}{*}{\textbf{1-2}} &homog.  & 917.53 & 1237.78                 & 783.51 & 1220.16              & 588.95 & 714.25                 & 457.52 & 459.34                  & 432.64 & 310.16                 & 2149.52 & 1061.00              & 68.05 & 72.39                       \\
&&heterog.                              &  \multicolumn{2}{c}{818.8}     &  \multicolumn{2}{c}{672.2}  &  \multicolumn{2}{c}{460.1}    &  \multicolumn{2}{c}{346.1}     &  \multicolumn{2}{c}{292.7}    &  \multicolumn{2}{c}{716.1}   &  \multicolumn{2}{c}{222.2}      \\
\end{tabular}                                                                                                     
	\normalsize                                                                                                 
	\caption{Detailed kernel execution times (in ms) for different scenarios. The partial execution times were measured without overlap, i.e., with synchronization after the kernel calls and therefore their sums do not not always match the total execution times.}                                                     
	\label{tab:timings}
\end{table}

\end{document}